%% file: ver_28.tex
\documentclass[a4paper,11pt]{article}
\pdfoutput=1 

\usepackage{jheppub} 
\usepackage{graphicx,color}
\usepackage{amsmath}
\usepackage{slashed}
\usepackage{multirow}
\usepackage{autobreak}
\usepackage[normalem]{ulem}
\usepackage{eufrak}
\usepackage{cleveref}
\usepackage{amsfonts}
\SetSymbolFont{largesymbols}{bold}{OMX}{txex}{b}{n}

\DeclareMathAlphabet{\mathpzc}{OT1}{pzc}{m}{it}

\newcommand{\minitab}[2][l]{\begin{tabular}{#1}#2\end{tabular}}

\def\vvevof#1{\left\langle\!\!\!\left\langle#1\right\rangle\!\!\!\right\rangle}
\def\op{\mathpzc{o}}
\def\Op{\mathpzc{O}}

\def\fft{{\mathfrak{p}}_{\tt theo}}
\def\FFt{{\mathfrak{F}}_{\tt theo}}

\def\Nt{\Sigma_{\tt theo}}

\def\lum{{\mathfrak{L}}_{\tt int} }
\def\ert{{\mathfrak N}_{\tt theo}}

\def\sw{s_{\tt w}}

\def\stw{s_{2\tt w}}
\def\ctw{c_{2\tt w}}
\def\mz{ m_{\tt z}}
\def\sg{\sin\gamma}
\def\cg{\cos\gamma}

\def\ttg{\tan{2\gamma}}
\def\etm{E_{\tt T}^{\tt miss}}

\def\vvevof#1{\left\lgroup\!#1\!\right\rgroup}

\def\fft{{\mathfrak{f}}_{\tt theo}}

\def\FFt{{\mathfrak{F}}_{\tt theo}}

\input mylatex

\title{\boldmath Probing heavy charged fermions at $e^+e^-$ collider using the Optimal Observable Technique}

\author[a]{Subhaditya Bhattacharya,}
\author[a]{Sahabub Jahedi,}
\author[b]{Jose Wudka}

\affiliation[a]{Department of Physics, Indian Institute of Technology Guwahati, Assam 781039, India}
\affiliation[b]{ Department of Physics and Astrophysics, University of California, Riverside, USA}
\emailAdd{subhab@iitg.ac.in}
\emailAdd{sahabub@iitg.ac.in}
\emailAdd{jose.wudka@ucr.edu}

\abstract{
In this work we study the production of color-neutral and singly-charged heavy leptons at the proposed International Linear Collider. We use the optimal observable technique to determine the statistical accuracy to which the coupling of such fermions to the $Z$ gauge boson (vector, axial or chiral) can be measured. We also consider a UV-complete model that contains these particles as well as a dark matter candidate, and consider some observable effects involving both; the correspondence to chargino production in supersymmetric models with heavy sleptons is briefly discussed.}

\keywords{Beyond Standard Model, $e^+e^-$ Experiments, Particle and resonance production}

\begin{document} 
\maketitle
\flushbottom

\section{Introduction}
\label{sec:intro}
The expected presence of physics beyond the Standard Model (BSM) is motivated by the observation of non-vanishing neutrino masses,  the overwhelming evidence for dark matter (DM), and the need for an effective mechanism to explain the baryon asymmetry; in addition, theoretical issues such as the stabilization of Higgs vacuum, also point towards the presence of new physics (NP). The search for such new particles and interactions is one of the central programs at the Large Hadron Collider (LHC). Despite this effort, and excepting the discovery of the long sought-after Higgs boson in 2012 \cite{Aad:2012tfa,Chatrchyan:2012ufa}, no direct observation of new physics at the LHC (or other experiments) has been confirmed, though there are significant hints \cite{Lees:2012xj,Aaij:2017tyk,Lees:2013uzd,Aaij:2017deq,Aaij:2021vac,Bennett:2002jb,Bennett:2006fi,Aoyama:2012wk,Aoyama:2020ynm,Abi:2021gix}.

One major challenge facing the search for NP at the LHC is its large QCD background that makes the detection of possible weakly-coupled BSM physics difficult. Yet this type of NP is expected in several scenarios ({\it e.g.} many DM and neutrino mass generation paradigms), which makes the prospect of an electron-positron ($ e^+e^-$) collider, such as the International Linear Collider (ILC) \cite{Behnke:2013xla}, a very attractive possibility for probing a variety of BSM physics. It is the goal of this paper to study some of the capabilities of the ILC and determine its sensitivity to simple NP extensions of the SM: we consider first the sensitivity of the ILC to an additional heavy vector-like fermion, and then to a SM extension that includes, in addition, a viable DM candidate. The discussion here presented can be easily extended to other proposed $ e^+e^-$ colliders and to a diversity of other types of hypothesized BSM physics.
 
The current bounds on heavy lepton masses depend on their nature (chiral or vector-like) and dominating decay channels. For example, LEP put a bound on the mass of $101.2\,\gev$ (95\% CL) on the mass of heavy, unstable, singly-charged fermion  \cite{Achard:2001qw} when its main decay channel is $\nu W^\pm$, while the bound is $102.6\,\gev$ (95\% CL) if it is stable \cite{Tanabashi:2018oca}. Searches at LHC have been so far in three main directions: {\it(i)} triplet leptons associated with type III seesaw mechanism for neutrino mass generation \cite{Sirunyan:2017qkz}; {\it(ii)} stable or long-lived charged leptons; and {\it(iii)} superpartners of the SM gauge bosons (neutralino and chargino). In the first case, CMS has put a ($3\sigma$) bound of $840\,\gev$ \cite{Sirunyan:2017qkz} (using $137\,\text{fb}^{-1}$ of data at $\sqrt s=13\,\tev$). The current bound for a long-lived singly charged fermion is $\sim$ 574 GeV \cite{Chatrchyan:2013oca} (CMS, using $18.8\,\text{fb}^{-1}$ of data at $\sqrt s=8\,\tev$). The limit on chargino mass in supersymmetric theories from production of chargino pairs \cite{Aad:2019vnb}  is $\sim 400\,\gev$ when the neutralino mass is zero, and $\sim 250\,\gev$ from chargino-neutralino pair production \cite{Aad:2019qnd} (both obtained at $13\,\tev$ CM energy). 

In our discussion below we will first study the detectability of a singly-charged lepton with mass of either $150\,\gev$ or $ 245\,\gev$ at the ILC, with a center-of-mass (CM) energy of $500\,\gev$ and determine the optimal statistical precision to which its couplings to the $Z$ boson can be measured using the optimal-observable technique (OOT) \cite{Atwood:1991ka, Davier:1992nw, Diehl:1993br, Gunion:1996vv}. Charged fermion pair production in the context of type-III seesaw framework has been studied in literature \cite{Das:2020gnt, Das:2020uer}, but no study has been done yet using the optimal observable approach.  We will then consider this particle in the context of a specific NP model and provide an event-level collider simulation of its dominating decay channel; this model has the added feature of containing a viable dark matter candidate, some of whose effects at the ILC will also be considered. We will discuss the effects of beam polarization and the extent to which the conclusions drawn for these specific cases can be generalized. 

The OOT has been used previous in a variety of studies, including the estimation of the uncertainty of the Higgs couplings \cite{Gunion:1996vv,Hagiwara:2000tk,Dutta:2008bh} and top-quark couplings at $e^+e^-$ colliders \cite{Grzadkowski:1996pc,Grzadkowski:1997cj,Grzadkowski:1998bh,Grzadkowski:1999kx,Grzadkowski:2000nx}, of the top-quark interactions in a $\gamma\gamma$ collider \cite{Grzadkowski:2003tf,Grzadkowski:2004iw,Grzadkowski:2005ye}, of the CP properties of Higgs boson at a muon collider \cite{Hioki:2007jc}, and of possible non-standard top-quark couplings at LHC \cite{Gunion:1998hm,Hioki:2012vn, Hioki:2014eca}; other studies using this technique include estimating the sensitivity to NP effects in flavor physics  \cite{Bhattacharya:2015ida, Calcuttawala:2017usw, Calcuttawala:2018wgo} and NP searches in top-quark production at $e\gamma$ colliders \cite{Cao:2006pu}. 

Our paper is organized as follows: the OOT is described in section \ref{sec:method}; the phenomenological model that we will use to study the $Z$ couplings of a heavy charged lepton is presented in  \ref{sec:model}; sections \ref{sec:cross-section} and \ref{sec:analysis} discuss the relevant cross-section calculations and OOT for this model; the UV-complete model and associated collider signals are examined in section \ref{sec:DMmodel}; with section \ref{sec:conclusion} containing parting comments and conclusions. 

\section{Optimal uncertainties}
\label{sec:method}
This section contains a summary of several results concerning the statistical uncertainty of experimental observables. These results have appeared previously (see, {\it e.g.}, \cite{Diehl:1993br}); they are included here for convenience and to ensure uniformity of notation.

We consider models where the SM has been complemented by some type of new physics; the (theoretical) differential cross section for any given collider process involving the production of new particles can be written in the form
\beq
\mathcal{O}=\frac{d\sigma_{\tt theo}}{d\phi} = \sum_i c_i f_i(\phi) \,,
\label{eq:expnd1}
\eeq
where $ \phi $ denotes the appropriate  phase-space coordinates and the coefficients $c_i $, composed of (sums of products of) coupling and numerical constants, parametrize the process in terms of the linearly-independent functions $ f_i $. In the following, we will discuss 2 $\rightarrow$ 2 scattering process for which there is a single phase-space variable, that we take as the CM scattering angle; naturally, $\phi$ changes according to the process under consideration and experimental convenience. The separation of coefficients $c_i$ and functions $f_i $ is not unique -- we will comment on this below.

The goal is now to determine the coefficients $c_i$ as accurately as possible. If one assumes a constant event rate together with the fact that an experiment occurs over a finite time, the event number follows a Poisson distribution, then the optimal covariance matrix becomes 
\beq
V_{ij} =\frac{ M_{ij}^{-1} \sigma_T}{N}= \inv\lum  M_{ij}^{-1}\, ,
\label{eq:covmat2}
\eeq
with
 \beq
M_{ij} =\int \frac{f_i(\phi) f_j(\phi)}{\mathcal{O}(\phi)}d\phi\, ,
\label{eq:mij}
\eeq
where $\sigma_T=\int \mathcal{O}(\phi) d\phi$ and N is total number of events ($N=\sigma_T \lum $). $\lum $ denotes the integrated luminosity over this period. The detailed 
derivation of the covariance matrix as in Eq.~\eqref{eq:covmat2} has been furnished in the Appendix \ref{sec:covmat} for the convenience of the readers. This can also be 
achieved by choosing a weighting function $w_i(\phi)$ such that $c_i=\int w_i(\phi) \mathcal{O}(\phi)d\phi$ \cite{Gunion:1996vv}, where the expression of $w_i(\phi)$ is given by,
	
\beq
w_i(\phi)=\frac{\sum_{j}M_{ij}^{-1}f_{j}(\phi)}{\mathcal{O}(\phi)}.
\eeq	

$V$ can be used to estimate the width of the distribution of the $c_i$ as follows. We {\em assume} that these parameters have average (or `seed') values $ c_i^0 $ and define ($ \epsilon$ is an efficiency factor discussed below)
\beq
\chi^2= \epsilon \sum_{\{i,j\}=1}^{n} \delta c_i \, \delta c_j \left( V_0^{-1}\right)_{ij}, \qquad \delta c_i=c_i-c_i^0, \qquad V_0 = \left. V\right|_{ c=c^0 }\,;
\label{chi2}
\eeq
In practice the optimal observable technique (OOT) consists in  using this covariance matrix to determine statistical uncertainties and correlations between the coefficients $ c_i $.

Regarding $ \chi^2$ as a random variable, one can determine the probability ${\sf p}_n(\ell)$ (often termed the confidence level, C.L.) for $ \chi^2 \le \ell $ to occur. If the $c_i$ are normally distributed this is given by the usual $ \chi^2$ distribution with $n$ degrees of freedom: ${\sf p}_n(\ell) = 1- \Gamma(n/2,\ell/2)/\Gamma(n/2) $. In particular, ${\sf p}_2(1) = 39.3\%$, and ${\sf p}_3(1) = 19.8\%$ so that for $n=2,\,3$ the $ \ell=1 $ C.L.  is relatively low;  a $68\%$ C.L. requires $ \ell = 2.3$ for $n=2$ and $ \ell=3.5$ for $n=3$. In the discussion below we will be mainly concerned with the regions determined by $ \chi^2 \le \ell $ for a given $ \ell$, referring to them as  the $ \sqrt \ell$-$\sigma$ regions; this can be contrasted to the common usage of ``1-$\sigma$ standard deviation" referred to $\chi^2=2.3$ for 2 parameter space and $\chi^2=3.5$ for 3 parameter space. An illustration on how the $ 1$-$\sigma$ regions change when we use a given C.L. is discussed in Appendix \ref{sec:68cl}.

In the following we will consider the (electron-positron) collider production of new physics (NP) which in turn decays to SM states; symbolically, $e^+ e^- \to$NP$\to$SM. We denote the `hard' cross section for $ e^+ e^- \to$NP production by $\sigma^{\tt NP}$, and by $\sigma^{\tt FS}$ the final-state cross section $ e^+ e^- \to$NP$\to$SM, including all event selection cuts aimed at reducing and SM background and enhancing the NP contribution. The efficiency factor $\epsilon$ in Eq.~\eqref{chi2} is then defined by the ratio 
\beq
\epsilon=\frac{\sigma^{\tt FS}}{\sigma^{\tt NP}}~, 
\label{eq:eps.def}
\eeq

We would also like to note further that the statistical analysis done in section \ref{sec:model} is based on the NP signal process without including the effects of 
SM backgrounds, since this requires a specific model for a detailed characterization of the final state events (we return to this in section \ref{sec:DMmodel}). 
However, the efficiency $\epsilon$ in Eq.~\eqref{chi2} includes not only the branching ratio of NP$\to$ SM final state, but also the effects of event selection cuts 
that suppress the SM background contamination. The values of $ \epsilon $ must be then estimated using a complete model of NP production and decay; in the 
next section  we will assume $\epsilon=0.001$ and $ 0.005 $, justified by the analysis of the specific model of section \ref{sec:DMmodel}. The use of $ \epsilon$ 
to include these effects is, of course, an approximation; it is appropriate for the type of situations we consider: the resonant production of new particles which then 
decay into a SM final state. This approximation would not be appropriate in processes where on-shell NP particle is similar in mass and spin to that of a SM particle 
leading to same signal and providing large interference (for example, a new $Z^{'}$ boson having similar mass to SM $Z$ boson), or when the new particle contribution 
to the signal is virtual or in narrow-width s-channel resonance. In all such cases the cross section in Eq.~\eqref{eq:expnd1} receives also a SM contribution, and the corresponding 
OOT must be modified ({\it cf. e.g.} \cite{Diehl:1993br}); we will return to this issue in a future publication. However, the procedure as adopted here, will be less 
conclusive, given a large irreducible SM background contribution and the estimation of $\epsilon$ will be limited in this case. 

As noted earlier the choice of $c_i $ and $f_i $ is not unique; in practice one uses a separation that is convenient computationally and, if possible, has some physical motivation. The final results are independent of this choice in the sense that, if we use different functions and coefficients, $f_i  = \sum u_{ij} \tilde f_j $  and  $ \tilde c_i = \sum u_{ji} c_j $, where $u_{ij}$ is a constant invertible matrix, $ \chi^2 $ in Eq.~\eqref{chi2} is invariant.

The covariance matrix $V$ depends on the physical process under consideration and on the experimental parameters such as collider energy and luminosity. Therefore the above expression can also be used to determine the (minimal) collider properties that are required to obtain a given desired statistical uncertainty.

The seed coefficients $ c_i^0 $ take different values depending on the type of new physics being considered. One can then take a different approach and regard Eq.~\eqref{eq:expnd1}  as a generic expansion of the cross section under consideration in a convenient basis of functions $ f_i$. If a model has parameters $ p_a $, then the $c_i  = c_i(p_a)$ and $ \delta c_i  = c_i(p_a^0 + \delta p_a) - c_i(p_a^0)$; from which the statistical uncertainties and correlations of the $  p_a$ can be readily extracted; an example of this procedure when is presented in the next section. The case where the $c_i$ are linear combinations of the $p_a$ is considered in appendix \ref{sec:ellipses}. The number of parameters $p_a$ can be larger than the number of coefficients $c_i $; in which case the measurements under consideration provide a consistency test of the model. 

\section{Phenomenological framework}
\label{sec:model}

\begin{figure}[htb!]
	$$
	\includegraphics[scale=0.27]{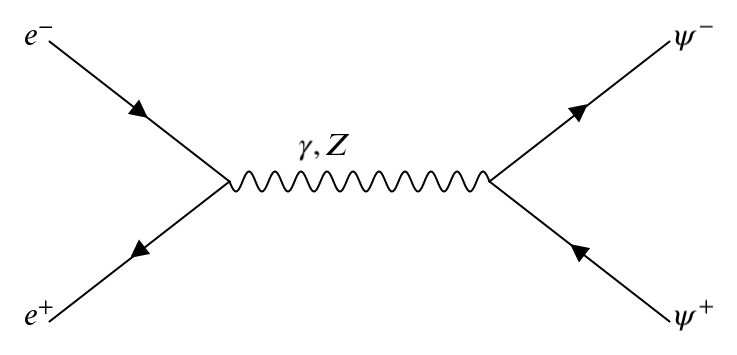}
	$$
	\caption{Production of heavy charged fermions ($\psi^+\psi^-$) at $e^-e^+$ collision (ILC).}
	\label{fig:productn}
\end{figure}

In this and the following sections we will use the OOT to determine the accuracy to which the parameters of a simple model of BSM physics can be measured at the projected International Linear Collider (ILC). The model we consider is a simple extension of the SM by the addition of a heavy charged fermion $ \psi^\pm $, that can be produced by $Z$ and photon exchange (Fig.~\ref{fig:productn}). We will discuss the precision to which the OOT allows the determination of the $\psi $ couplings to the $Z$ at an $e^+e^-$ collider. 

This type of heavy fermion appears in various extensions of SM; {\it e.g.} those containing a fermion isodoublet $\left(\psi^0,\psi^-\right)$ with hypercharge $Y_\psi=-1$; we  elaborate upon a possible model framework below (Sect. \ref{sec:DMmodel}). Here we adopt a purely phenomenological approach,  allowing $ \psi^\pm $ to have general chiral couplings to the $Z$ boson\footnote {We postpone any constraints coming from chiral anomalies to our discussion of a specific model.}:
\beq
\psi^+\psi^-Z:- ~\frac{i e_0}\stw \gamma^\mu\left(a+b\gamma^5\right)~,
\label{eq:vertex1}
\eeq
(where $e_0=U(1)_{\tt em}$ coupling, and $\stw = \sin(2 \theta_{\tt w})$; $\theta_{\tt w}$ is the weak-mixing angle)  assuming for simplicity~\footnote{It is worth noting that in weakly-coupled theories modifications to the photon minimal coupling are generated at 1 or higher loops and are correspondingly suppressed.} that it has the usual minimal coupling to the photon:
\beq
\psi^+\psi^-\gamma:- ~ie_0\gamma^\mu \,.
\label{eq:vertex2}
\eeq
The paramters $a,\,b$ correspond to the $ p_a $ discussed briefly at the end of section \ref{sec:method}.

We will call any specific choice of $a,\,b$ a {\em hypothesis} and the corresponding parameters as {\em seed parameters}, of which we will consider the following:
\begin{itemize}
	\item $a=\pm1$, $b=0$ (pure vector coupling).
	\item $a=0$, $b=\pm 1$ (pure axial vector coupling).
	\item $a=\pm1$, $b=\pm 1$ (chiral coupling).
\end{itemize}
We note that, for this simple model, $ b \to - b $ under a parity transformation, so we need to consider only $ b\ge 0 $. 

Using the couplings in Eqs.~\eqref{eq:vertex1} and \eqref{eq:vertex2}, we can evaluate $ d\sigma_{\tt theo} $ and, upon selection of the $f_i $, extract the coefficients $ c_i = c_i(a,b) $;  the hypothesis $ a=a^0,\,b=b^0$ corresponds to assuming that these coefficients have seed values $ c_i^0 = c_i(a^0,b^0)$ ({\it cf.} comments at the end of sect. \ref{sec:method}). We then use Eq.~\eqref{eq:covmat2} to compute the covariance matrix $ V$ and corresponding $ \chi^2$; the regions $ \chi^2 < $const. determine the optimal statistical uncertainties \cite{Gunion:1996vv, Calcuttawala:2017usw}, and the accuracy to which different hypotheses can be differentiated.  

For the calculations below we will assume the following collider parameters:
\beq
m_{\psi^{\pm}} = 150\,\gev\,,~\text{or}~ 245\,\gev \mcr \,, \qquad \sqrt{s} = 500 \, \gev\,; \quad \lum =567 \, \text{fb}^{-1}\,,
\label{eq:params}
\eeq
where $\sqrt{s}$ is the CM energy of the collider and integrated luminosity $ \lum $, whose values were taken from the ILC design parameters~\cite{Behnke:2013xla}. The lower value of $ m_{\psi^\pm} $ is chosen above the current collider limit of $O( 100 )\,\gev $ \cite{Tanabashi:2018oca} (Sect. \ref{sec:intro})~\footnote{This limit is obtained using the $ \psi^\pm \to W^\pm+$neutral decay, which naturally occurs in the simplest models containing a $ \psi^\pm $; see Sect. \ref{sec:DMmodel}.}; the higher value is chosen to be close to threshold.

Our analysis is carried out for an $ e^+ e^- $ linear collider because {\it(i)} it provides a much cleaner platform where QCD processes are suppressed, and so provides much better opportunity for the precision measurements we consider here; {\it(ii)} the expected availability of (partially) polarized beams allows a better probe of the new physics we are considering; and {\it(iii)} the construction of the covariance matrix and $\chi^2$ can be done analytically, avoiding insertion of the quark distribution functions that are unavoidable in a hadron collider.

\subsection{The $\psi^+\psi^-$ production cross section at an $e^+e^-$ collider}
\label{sec:cross-section}

The amplitude for the process $ e^+e^- \to \psi^+ \psi^-$, which we denote by $\mcal(\lambda_{e^-},\,\lambda_{e^+},\,\lambda_\psi,\,\lambda_{\bar\psi} )$ (where $\lambda_i = \pm1$ denotes the helicity of particle $i$), is easily calculated  \cite{Vega:1995cc}:
\bal
\mcal(\lambda_{e^-},\,- \lambda_{e^-},\,\lambda_\psi,\,-\lambda_\psi  ) &= - e e_0 \left(\lambda_{e^-} \lambda_\psi + \cos\theta \right) \left[ 1 + \xi \left(a + b \lambda_\psi \beta_\psi \right) \right] \,; \quad \xi = \xi_1+\lambda_{e^-} \xi_2\,,\mcr
\mcal(\lambda_{e^-},\,- \lambda_{e^-},\,\lambda_\psi,\,\lambda_\psi ) &= - e e_0\left( \frac{2 m_{\psi^{\pm}} \lambda_\psi \sin\theta}{\sqrt{s}} \right) \left( 1 + \xi a \right) \,,
\label{eq:amplitude}
\end{align}
where $e$ is the electron charge, $\sqrt{s}$  the CM energy, $ \beta_\psi = \sqrt{1 - 4 m_{\psi^{\pm}}^2/s}$, and
\beq
\xi_1=\frac{C_v}{\stw^2 (1 - \mz^2/s)}\,, \qquad  \xi_2=\frac{C_a}{\stw^2 (1 - \mz^2/s)} \,,
\eeq
with $C_v = ( 4 \sw^2-1)/2,\, C_a=1/2$,  the vector and axial couplings of the electron to the $Z$, respectively (and $ \sw = \sin\theta_{\tt w}$). If $ \hat\pp_{e^-}$ and $ \hat\pp_{\psi^-} $ are unit vectors parallel to the corresponding momenta, then the scattering angle $ \theta $ is defined by $ \cos\theta = \hat\pp_{e^-} \cdot \hat\pp_{\psi^-} $.

\begin{figure}[htb!]
$$
\includegraphics[scale=0.36]{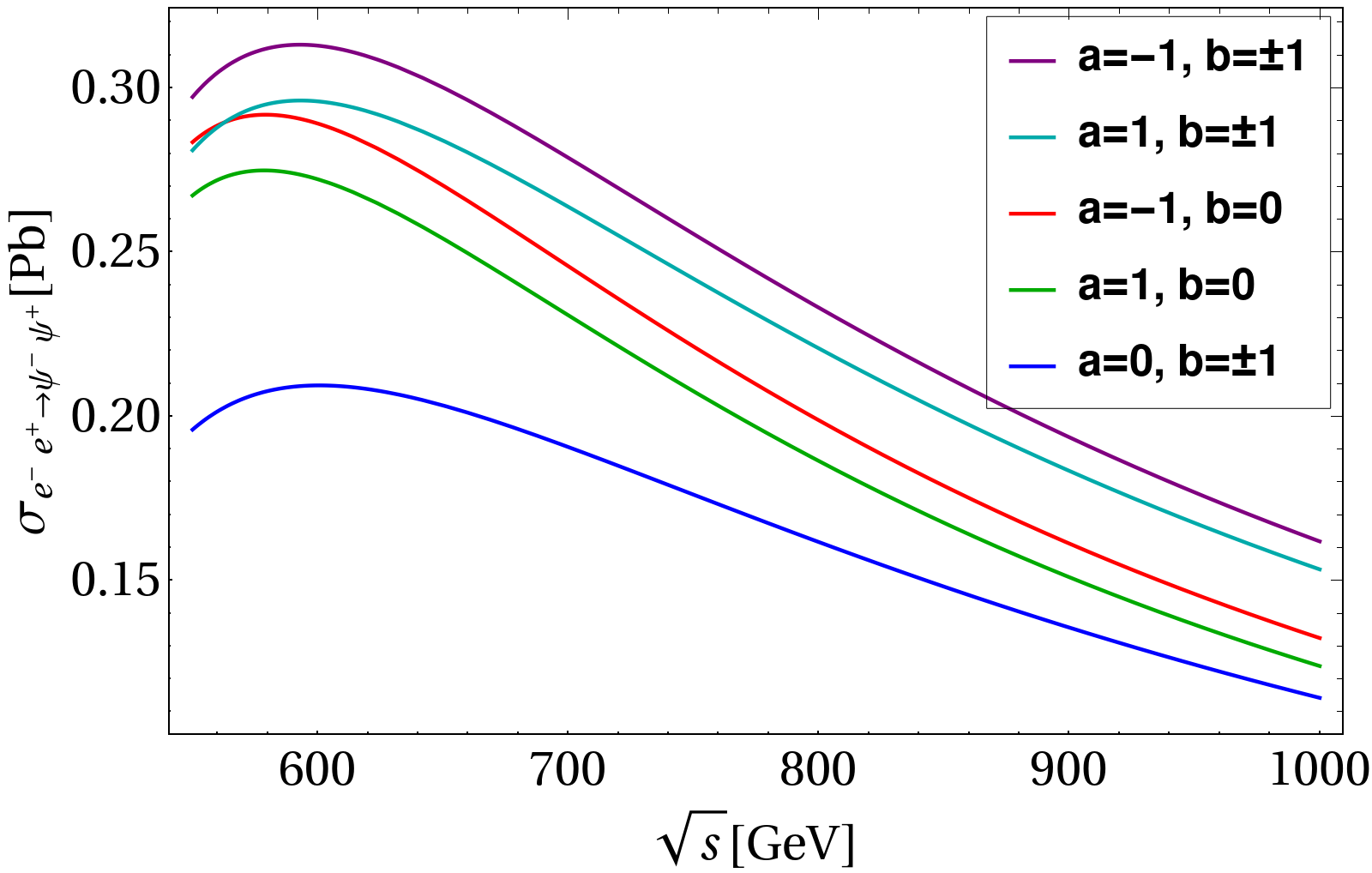}
\includegraphics[scale=0.5]{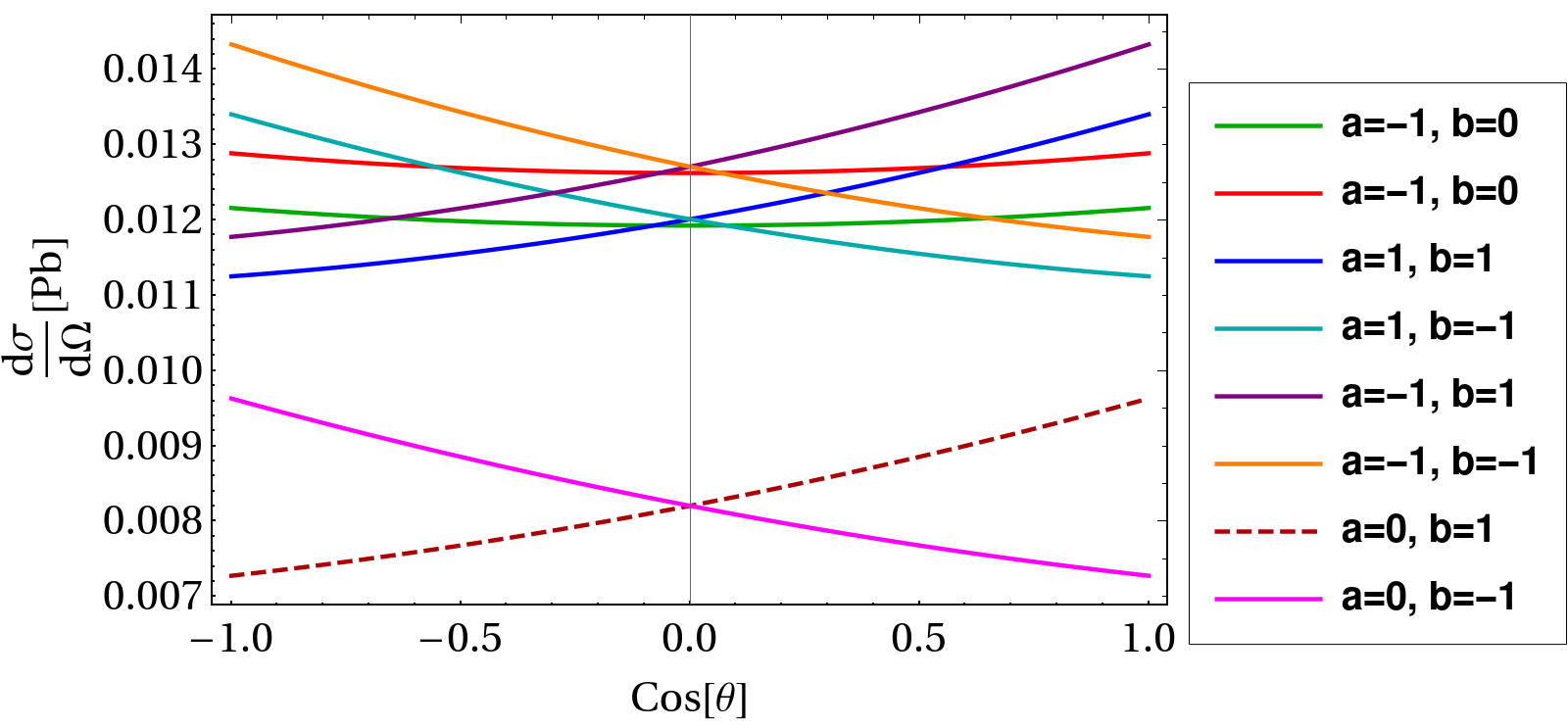}
$$
\caption{Left: Total spin-averaged cross-section for $e^+e^-\to \psi^+\psi^-$ as a function of the CM energy $\sqrt{s}$; right:  differential  spin-averaged cross-section as a function of the scattering angle for c.o.m energy ($\sqrt s$) $=500$ GeV. We took $m_{\psi^\pm} =245$ GeV and the collider parameters of Eq.~(\ref{eq:params}).}
\label{fig:cs}
\end{figure}

Using Eq.~(\ref{eq:amplitude}), the cross-section when the $ e^\pm$ beams have partial polarizations $P_{e^\pm} $ (with $ -1\le P_{e^\pm}\le1$) is given by
\bal
\label{eq:poldiffcrosssection}
\frac{d\sigma(P_{e^+},\,P_{e^-})}{d\Omega} &= \frac{(1-P_{e^-})(1+P_{e^+})}4 \left( \frac{d\sigma}{d\Omega}\right)_{\lambda_{e^-}=-1} + \frac{(1+P_	{e^-})(1-P_{e^+})}4 \left( \frac{d\sigma}{d\Omega}\right)_{\lambda_{e^-}=1}\,, \mcr
&= \sum c_i f_i\,,
\end{align}
where we choose \footnote{It is straightforward to verify that these functions are linearly independent}
\beq
\{f_1,\,f_2,\,f_3\} = \frac{ \beta_\psi }{2 s} \left\{  (2 - \beta_\psi^2),\,  \beta_\psi   \cos \theta,\,\beta_\psi^2  \cos^2\theta \right\}\,,
\label{fi}
\eeq
and
\begin{align}
\frac{c_1}{\alpha \alpha_0}  =&\frac{1-P_{e^-}P_{e^+}}{2}\bigg[ 1+2\xi_1 a+(\xi_1^2+\xi_2^2) \left( a^2+\frac{\beta_{\psi}^2}{2-\beta_{\psi}^2} b^2 \right) -2P_{\tt eff}\bigg\{\xi_2 a+\xi_1\xi_2 a^2+\frac{\beta_{\psi}^2}{2-\beta_{\psi}^2}\xi_1\xi_2b^2\bigg\}\bigg];\mcr
\frac{c_2}{\alpha \alpha_0}  =&\frac{1-P_{e^-}P_{e^+}}{2}\bigg[2\xi_2b+4\xi_1\xi_2ab-P_{\tt eff}\bigg\{2\xi_1b+(\xi_1^2+\xi_2^2)ab\bigg\}\bigg];\mcr
\frac{c_3}{\alpha \alpha_0}  =&\frac{1-P_{e^-}P_{e^+}}{2}\bigg[ 1+2\xi_1a+(\xi_1^2+\xi_2^2)(a^2+b^2) -2P_{\tt eff}\bigg\{\xi_2a+\xi_1\xi_2(a^2+b^2)\bigg\}\bigg];
\label{ci}
\end{align}
We defined
\beq
P_{\tt eff} = \frac{P_{e^-}-P_{e^+}}{1-P_{e^-}P_{e^+}}.
\eeq
and  $ \alpha_0 = e_0^2/(4\pi)$, while  $\alpha = e^2/(4\pi) $ is the usual fine-structure constant.  It is also useful to note that $ d\sigma/d\Omega $ is invariant under 
$ b \to - b$ and $ \theta \to \pi - \theta $, a consequence of the invariance of Eqs.~(\ref{eq:vertex1}) and (\ref{eq:vertex2}) under CP.

The spin-averaged total and differential cross-sections (corresponding to $ P_{e^\pm} =0 $) for different seed values of $\{a,b\}$ are plotted in Fig.~\ref{fig:cs}.  It is worth noting that the total cross sections exhibit the same behavior for the values of $a$ and $b$ considered, especially at large $s$; this is due to a combination of two effects. First, since $ \sw^2 \simeq1/4 $, $ \xi_1 \sim 0 $; second, for large $s$, $ \beta_\psi \simeq 1$; it follows that the average cross section $ \sigma \propto 1+ \xi_2^2 (a^2 + b^2) $, explicitly displaying its dependence on $a$ and $b$. In contrast the unpolarized {\em differential} cross section depends on $ c_2 \simeq \xi_2 b $ and will have a very different behavior depending on the values of $b$. As expected from unitarity, the total cross section drops with increasing CM energy.

\begin{figure}[htb!]
$$
\includegraphics[scale=0.28]{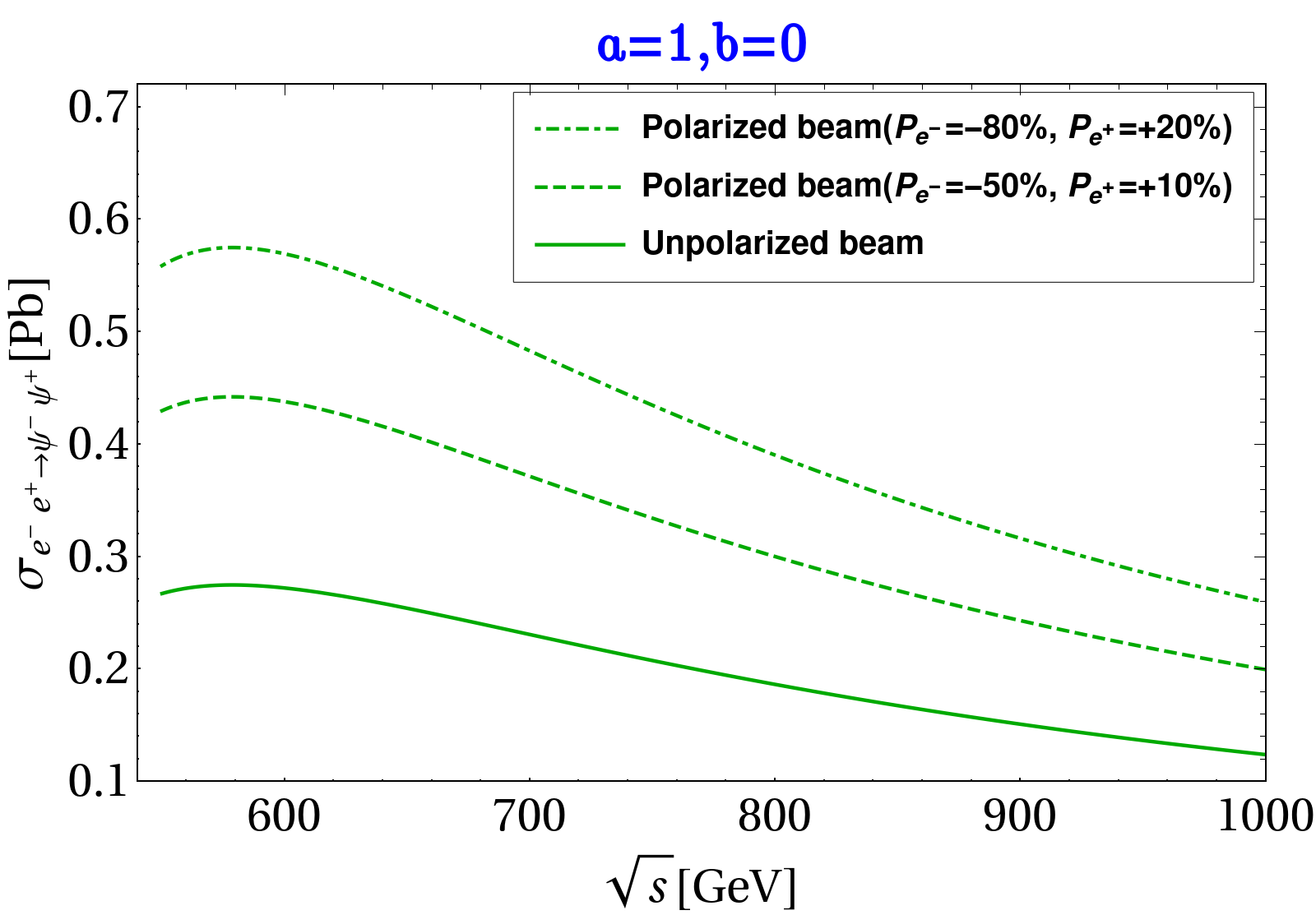}
\includegraphics[scale=0.28]{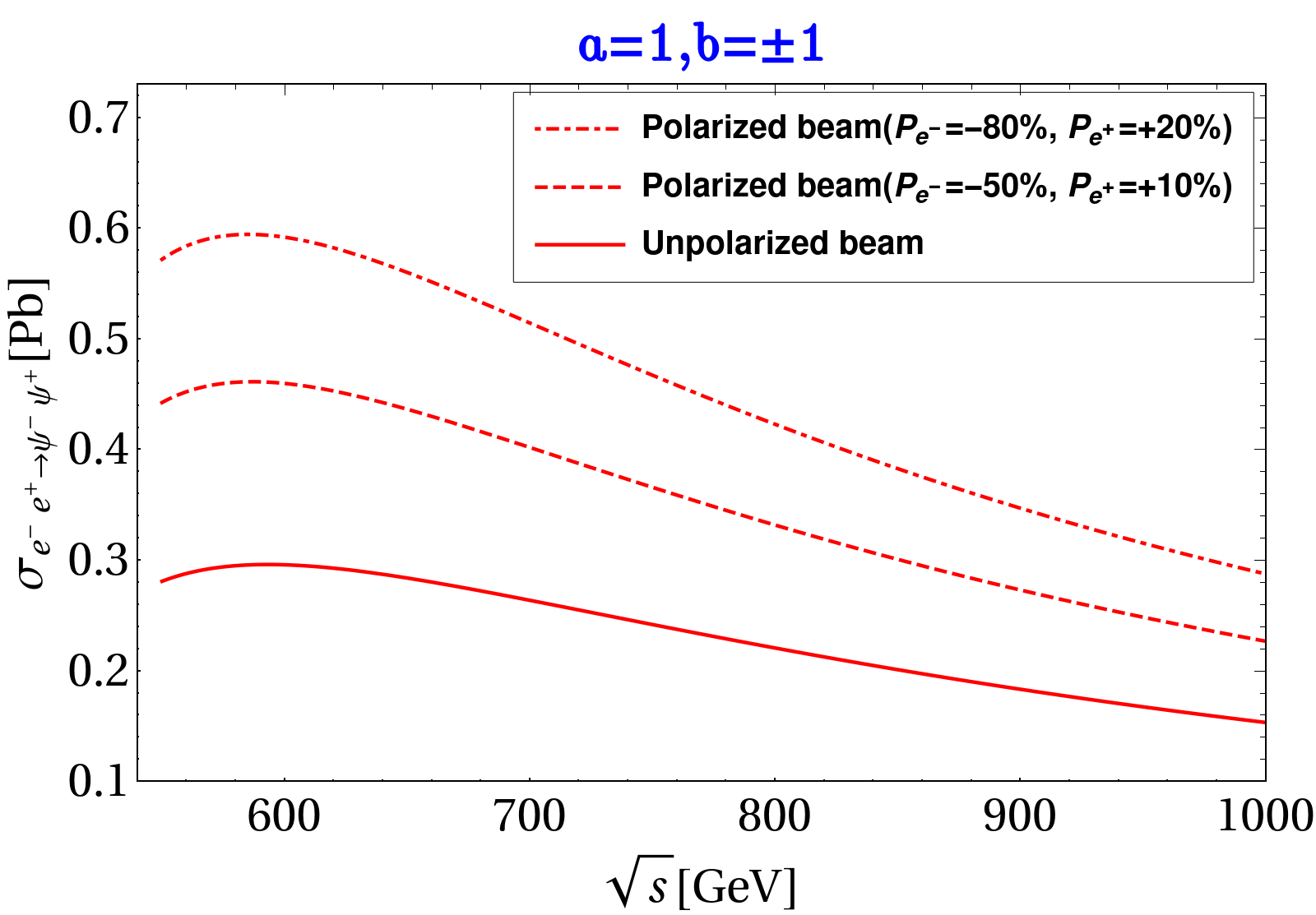}
\includegraphics[scale=0.28]{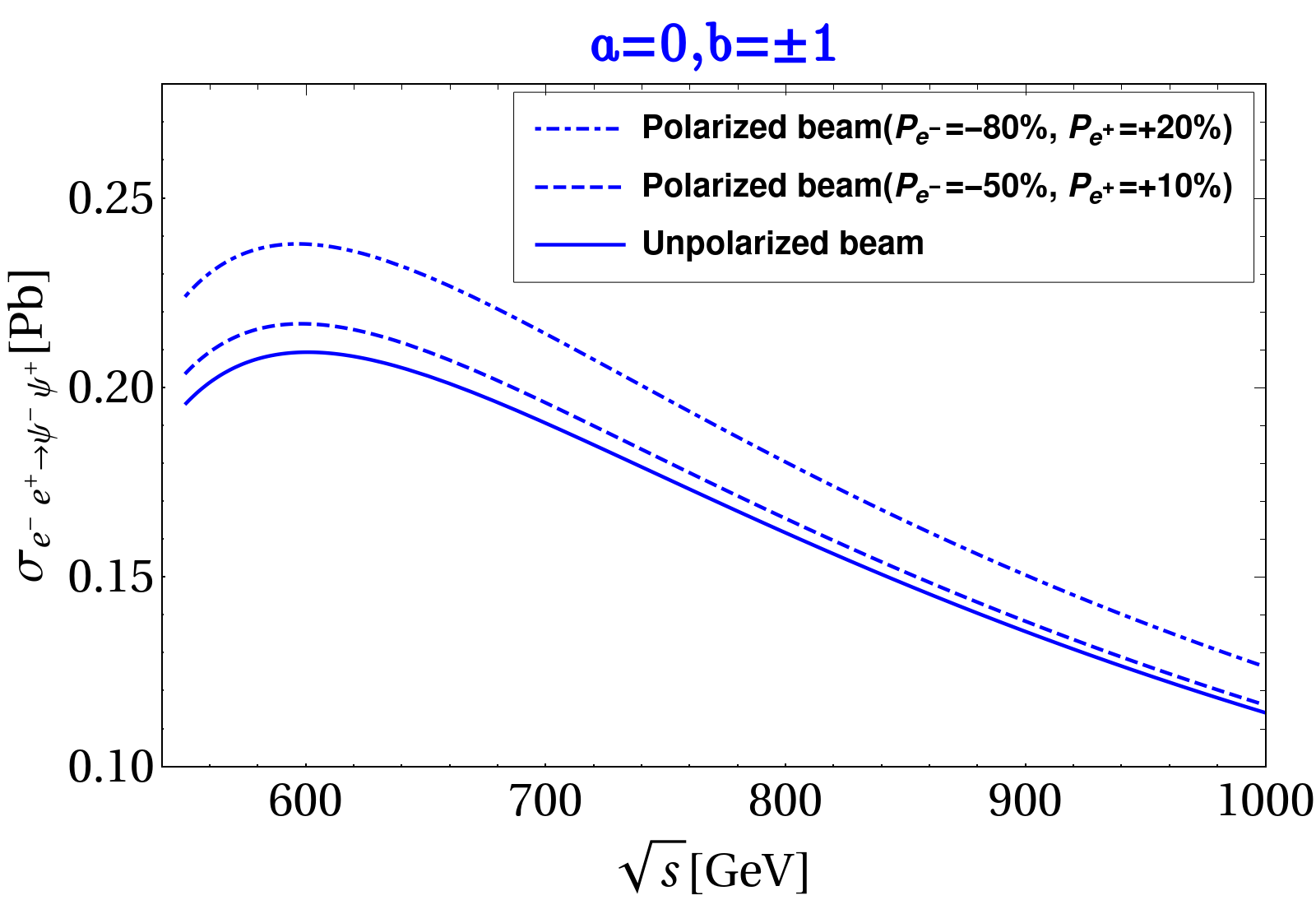}
$$
$$
\includegraphics[scale=0.3]{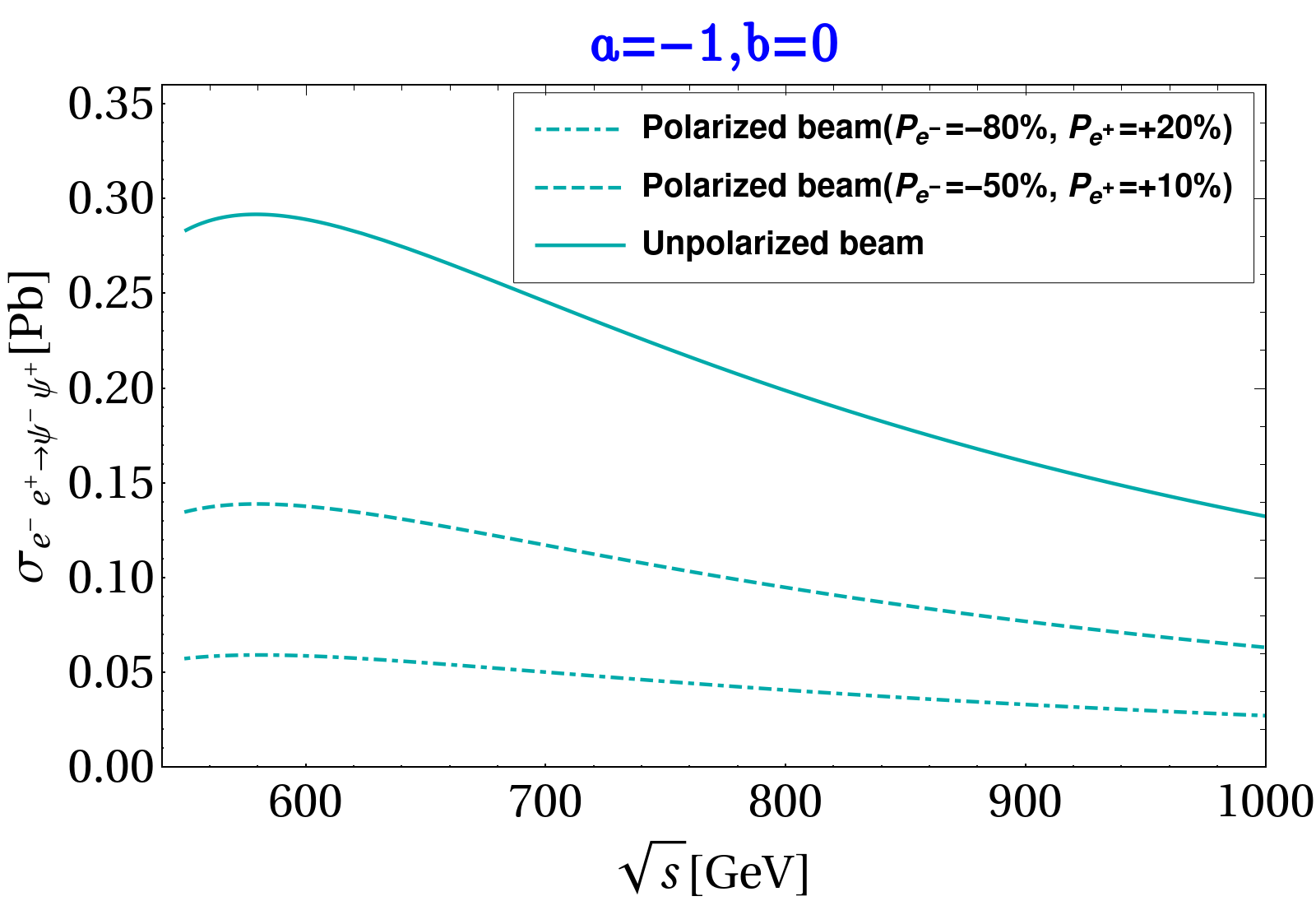}
\includegraphics[scale=0.3]{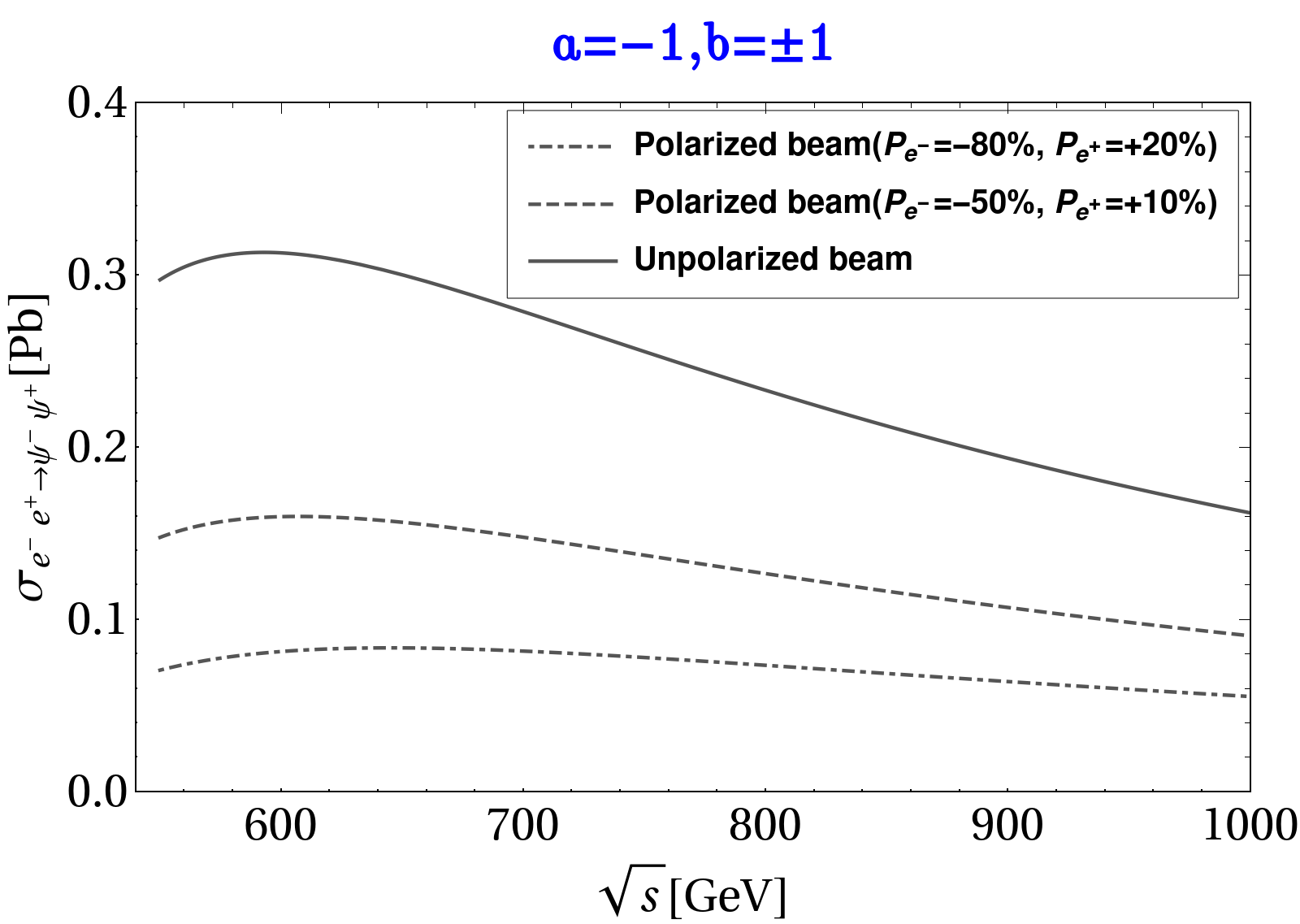}
$$
\caption{Total cross section as a function of the CM energy for unpolarized ($P_{e^\pm}=0$) beams (solid line); $P_{e^\mp}=^{-50\%}_{+10\%}$ (dashed line), and  $P_{e^\mp}=^{-80\%}_{+20\%}$ (dashed-dot line) for various seed values of $a$ and $b$.}
\label{fig:cscom1}
\end{figure}

According to the design report \cite{Behnke:2013xla}, the ILC will produce highly polarized electron beam and moderately polarized positron beam; we will choose $ P_{e^-}=-0.8,\,P_{e^+}=+0.2 $ when considering this option. Also, for polarized beams we have
\bal
a>0: & \quad \sigma\left(P_{e^-},\,P_{e^+} \right) \ge \sigma\left( P_{e^\pm} =0 \right)\,,  \mcr
a<0: & \quad \sigma\left(P_{e^-},\,P_{e^+} \right) \le \sigma\left( P_{e^\pm} =0 \right)\,,
\end{align}
whence it follows that polarization will enhance detectability. We illustrate these features in figures \ref{fig:cscom1} where we plot the total cross section for various seed values of $a,b$ and choices of $ P_{e^\pm}$.

\subsection{Optimal statistical analysis at $\sqrt{s}=500~\rm GeV$}
\label{sec:analysis}
    We now apply the optimal observable method described in Sect.~\ref{sec:method} to the case of $ \psi^\pm $ production at the ILC, using the parameters of Eq.~\eqref{eq:params}; the cases of $250\,\gev$ and $2\,\tev$ CM energy collider are briefly discussed in appendix \ref{sec:1tev}. 

\subsubsection{$\chi^2=1$ surfaces in the $a-b$ plane}
As a first step, we use the above expressions to obtain the coefficients $ c_i $ and functions $ f_i $; for example, for unpolarized beams ($P_{e^\pm}=0$) Eqs.~(\ref{fi}) and (\ref{ci}) give:
\beq
\begin{array}{|c|l|l|}
\hline
\multicolumn{3}{|c|}{\sqrt{s} = 500\, \gev\,, \quad P_{e^\pm}=0 } \cr \hline
m_{\psi^\pm}\, (\gev) & \qquad\qquad\qquad c_i/(\alpha\alpha_0)  
& f_i \times 10^{-8}\, \gev^2 \cr
\hline\hline
& i=1:~ \frac{1}{2}(1-0.086 a +0.522 a^2 +0.245b^2)  & i=1:~  217.60             \\ 
150 & i=2:~ \frac{1}{2}(1.442b - 0.124ab)                & i=2:~ 128.00 \cos \theta  \\
& i=3:~ \frac{1}{2}(1- 0.086a +0.552 (a^2 + b^2))    & i=3:~  102.40 \cos^2\theta \\
\hline
& i=1:~ \frac{1}{2}(1-0.086 a +0.522 a^2 +0.011b^2)  & i=1:~  78.02 \\
245 & i=2:~ \frac{1}{2}(1.442b - 0.124ab)                & i=2:~  7.92 \cos \theta \\
& i=3:~ \frac{1}{2}(1- 0.086a +0.552 (a^2 + b^2))    & i=3:~  1.58 \cos^2 \theta .\\
\hline
\end{array}
\label{eq:c-a-b}
\eeq

\begin{figure}[htb!]
\begin{align*}
\includegraphics[scale=0.254]{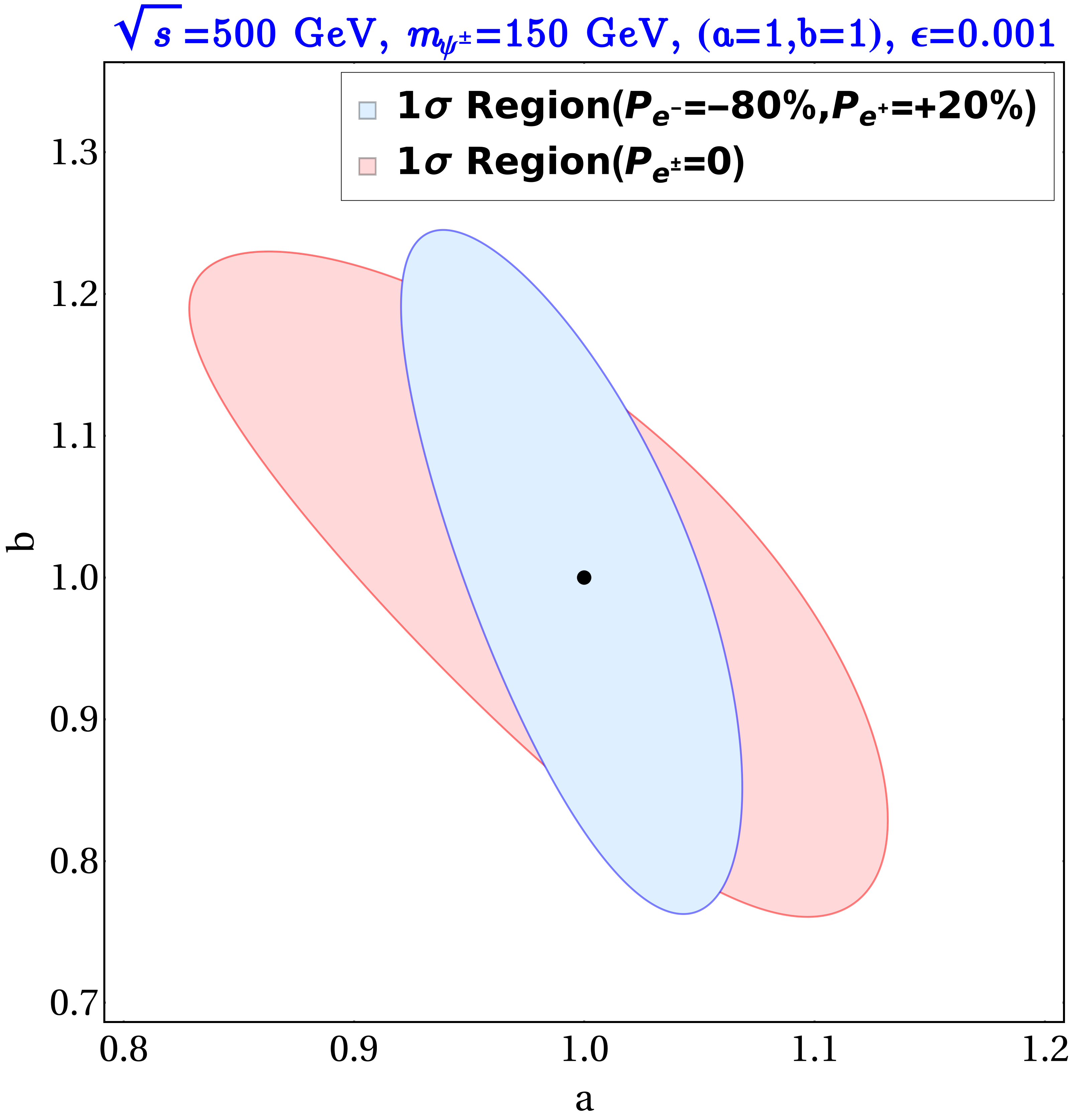} \quad&\quad \includegraphics[scale=0.25]{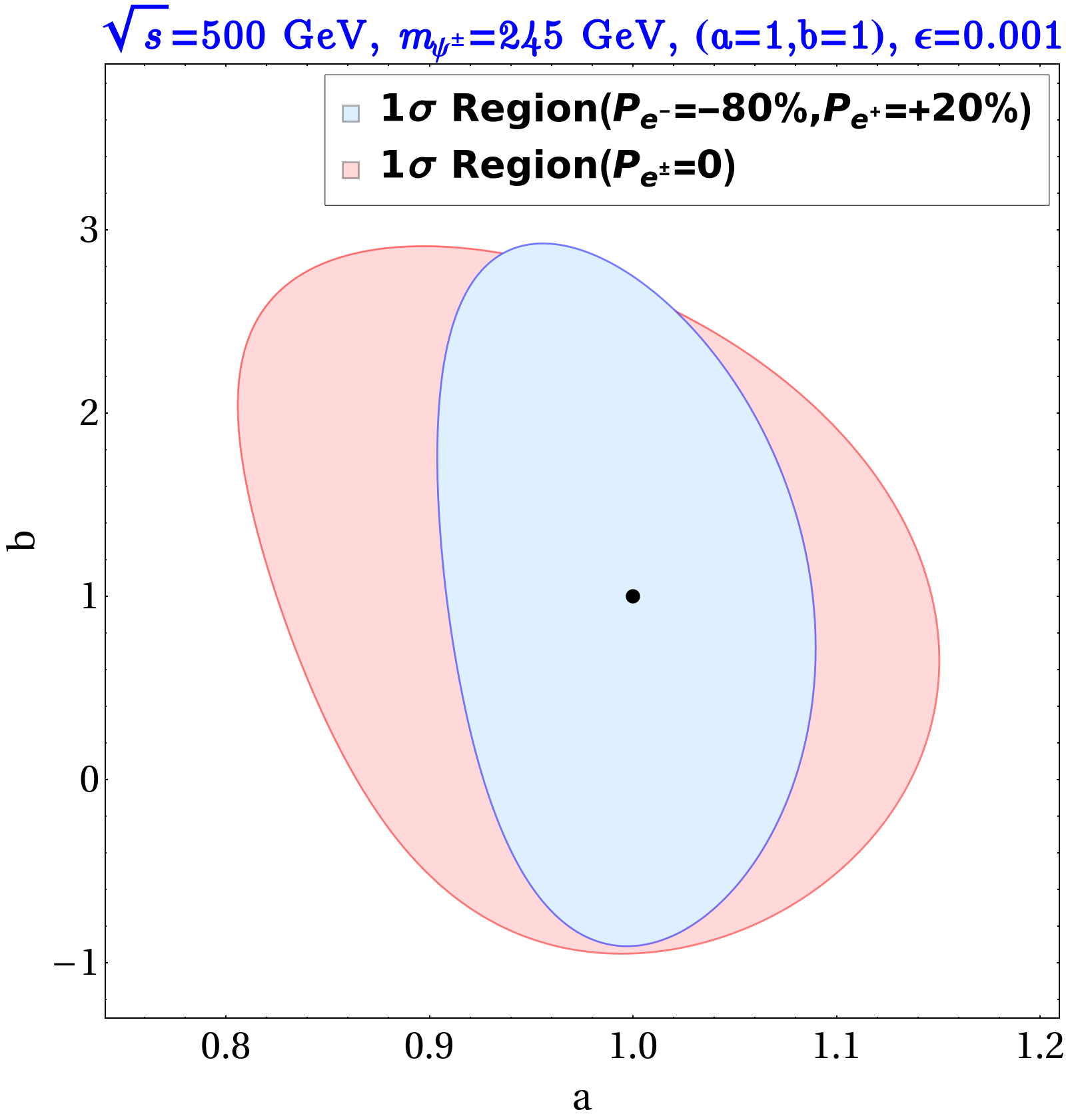} \cr
\includegraphics[scale=0.264]{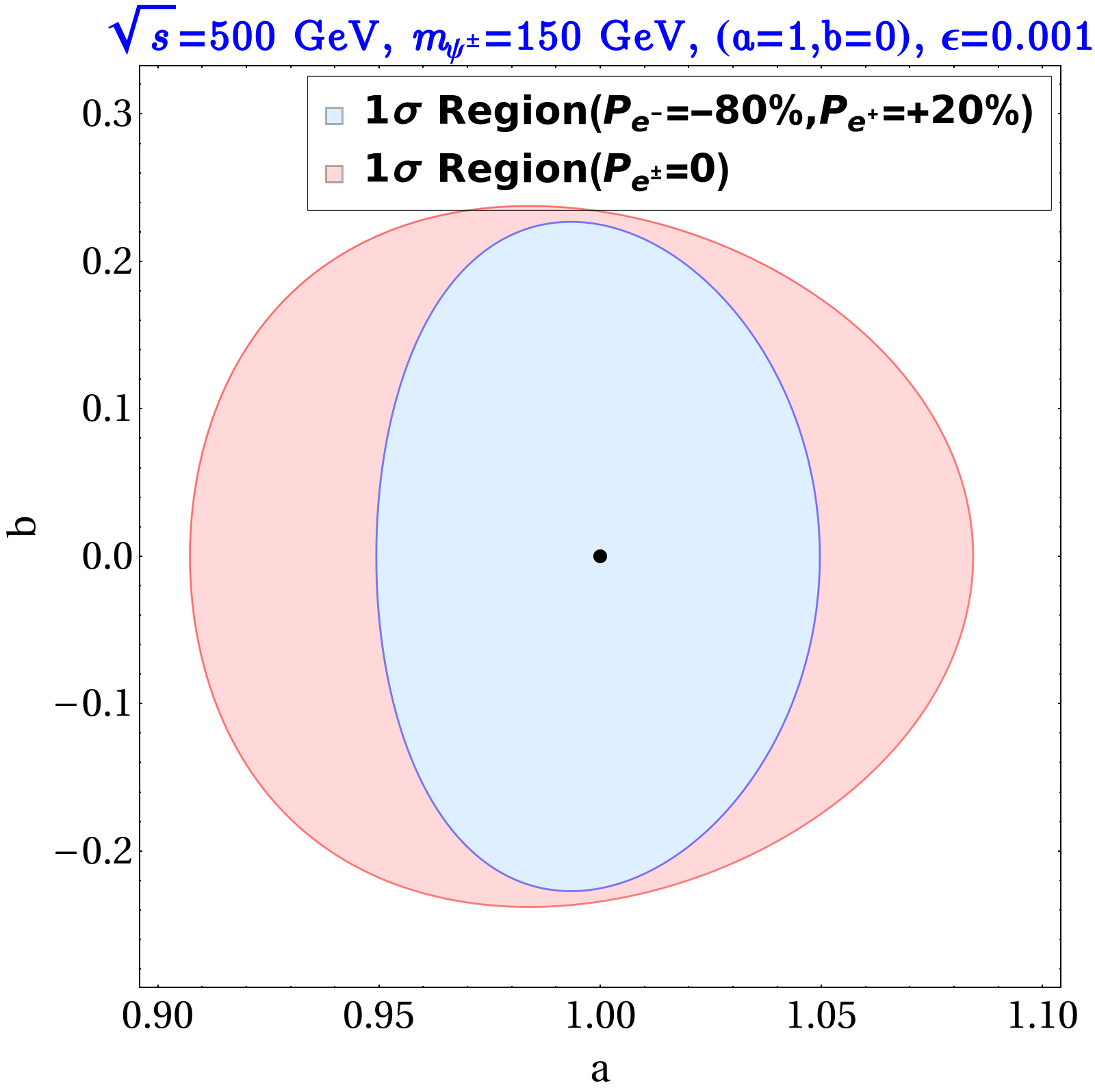} \quad&\quad \includegraphics[scale=0.25]{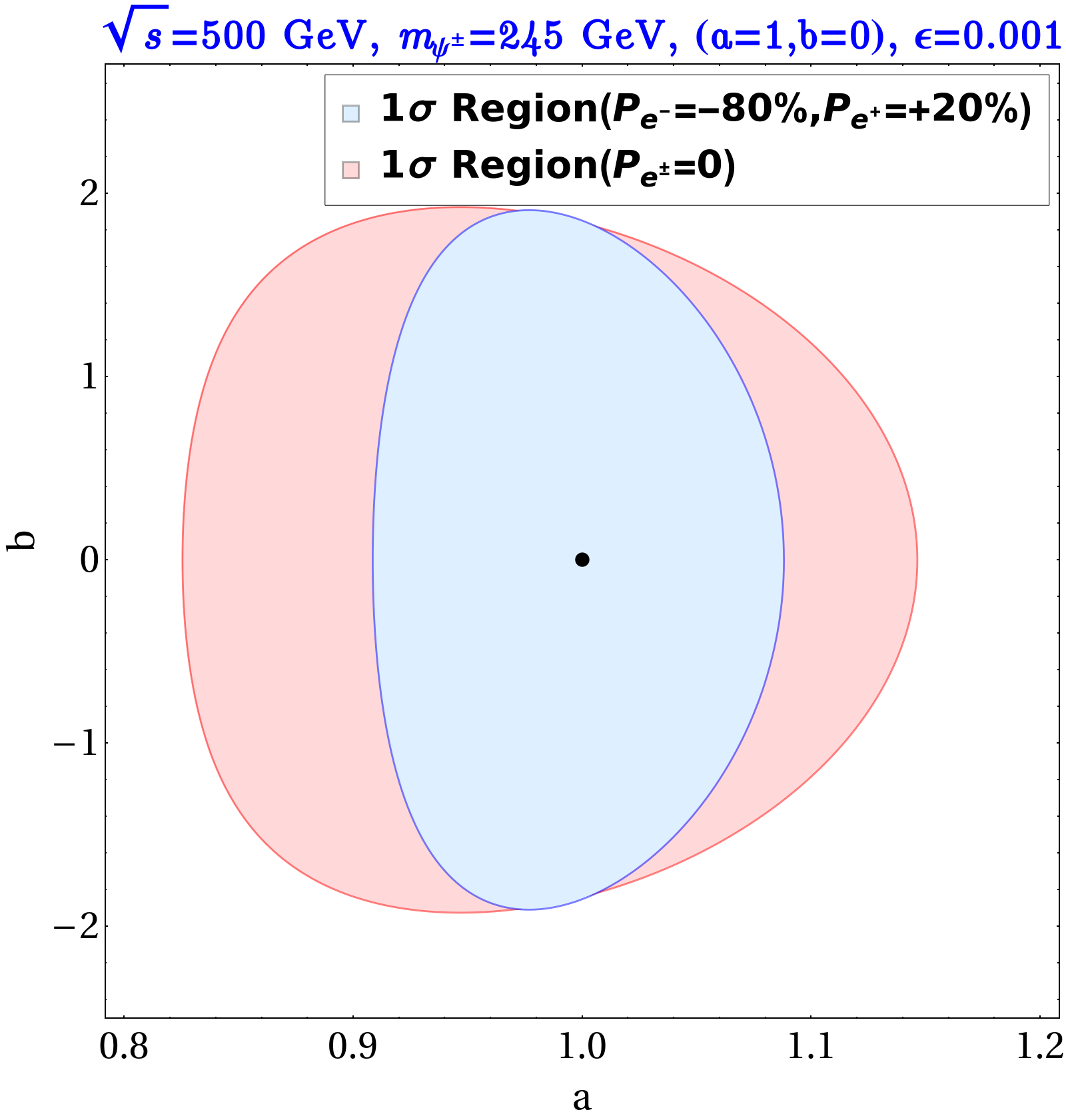} \cr
\includegraphics[scale=0.253]{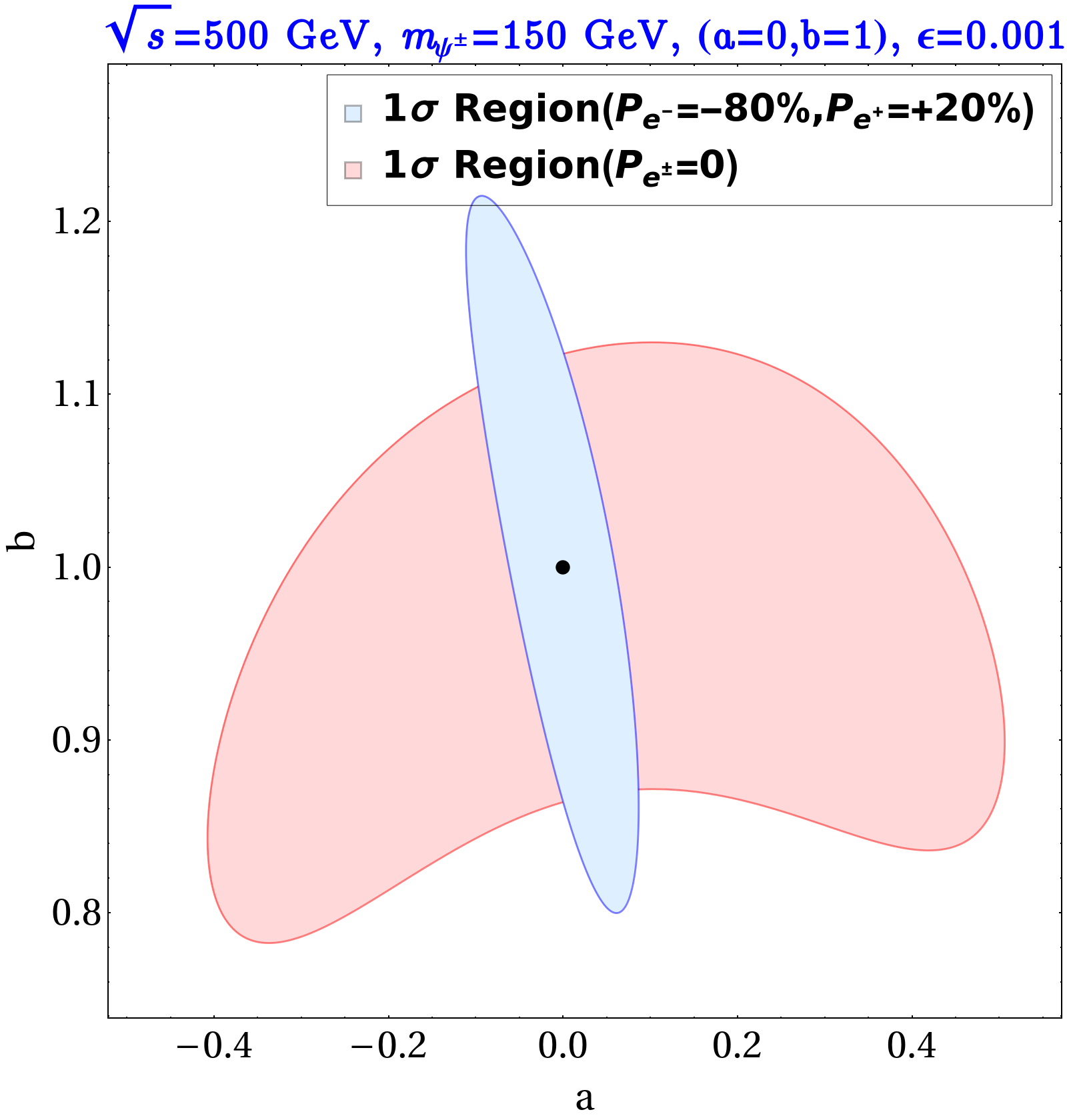} \quad&\quad \includegraphics[scale=0.25]{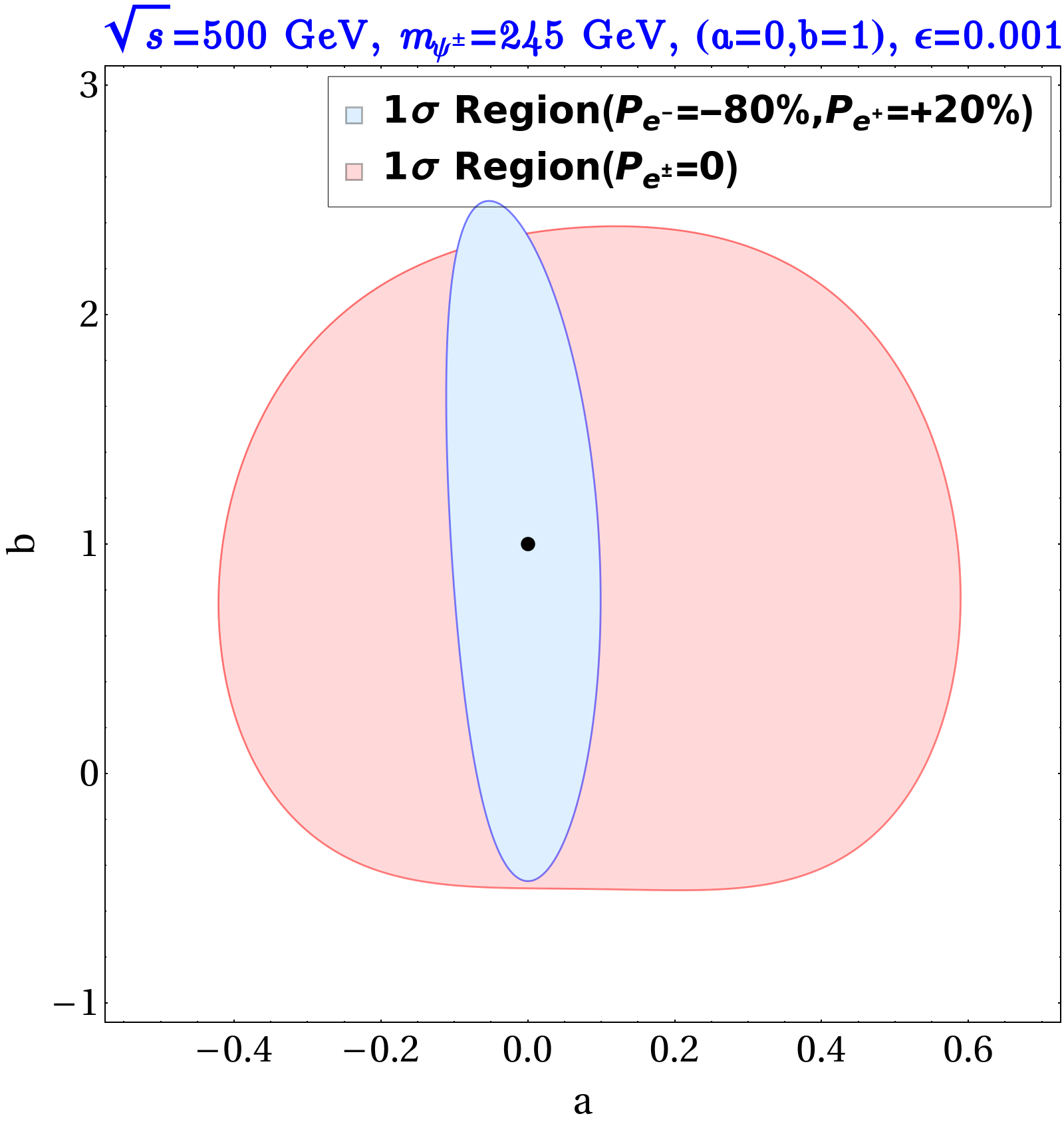} \cr
\end{align*}
\caption{ $\chi^2=1$ surfaces for hypotheses with $ a^0 \ge 0 $, $P_{e^\pm} =0 $ and $ P_{e^\pm} = ^{+20\%}_{-80\%} $, and $ \epsilon = 0.001$. Left (right) column:  $ m_{\psi^\pm} = 150\, (245)\, \gev$. Note: the scales in the graphs are not all equal.}	
\label{fig:poldiff1}
\end{figure}

\begin{figure}[htb!]
\begin{align*}
\includegraphics[scale=0.265]{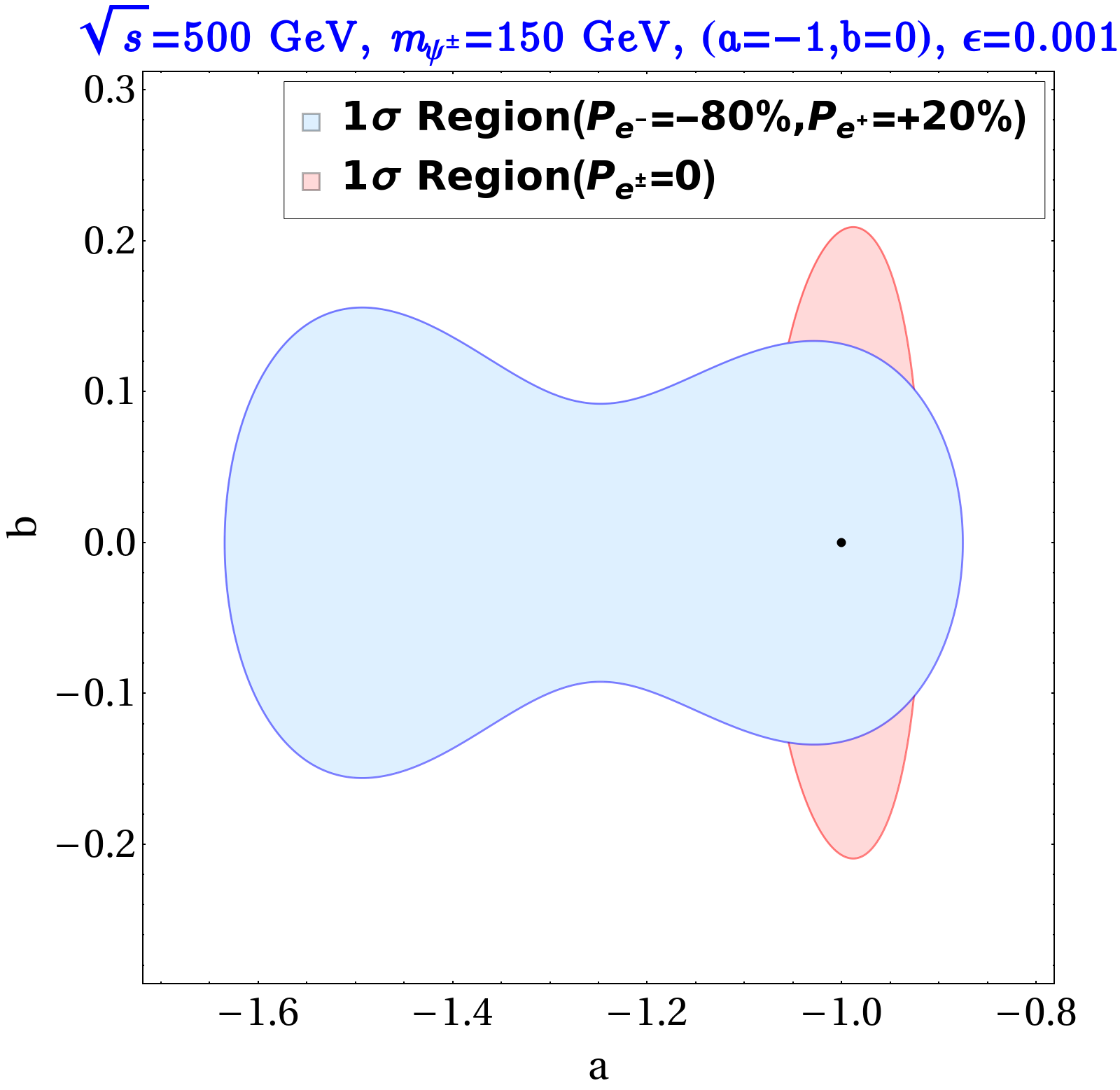}\quad&\quad \includegraphics[scale=0.25]{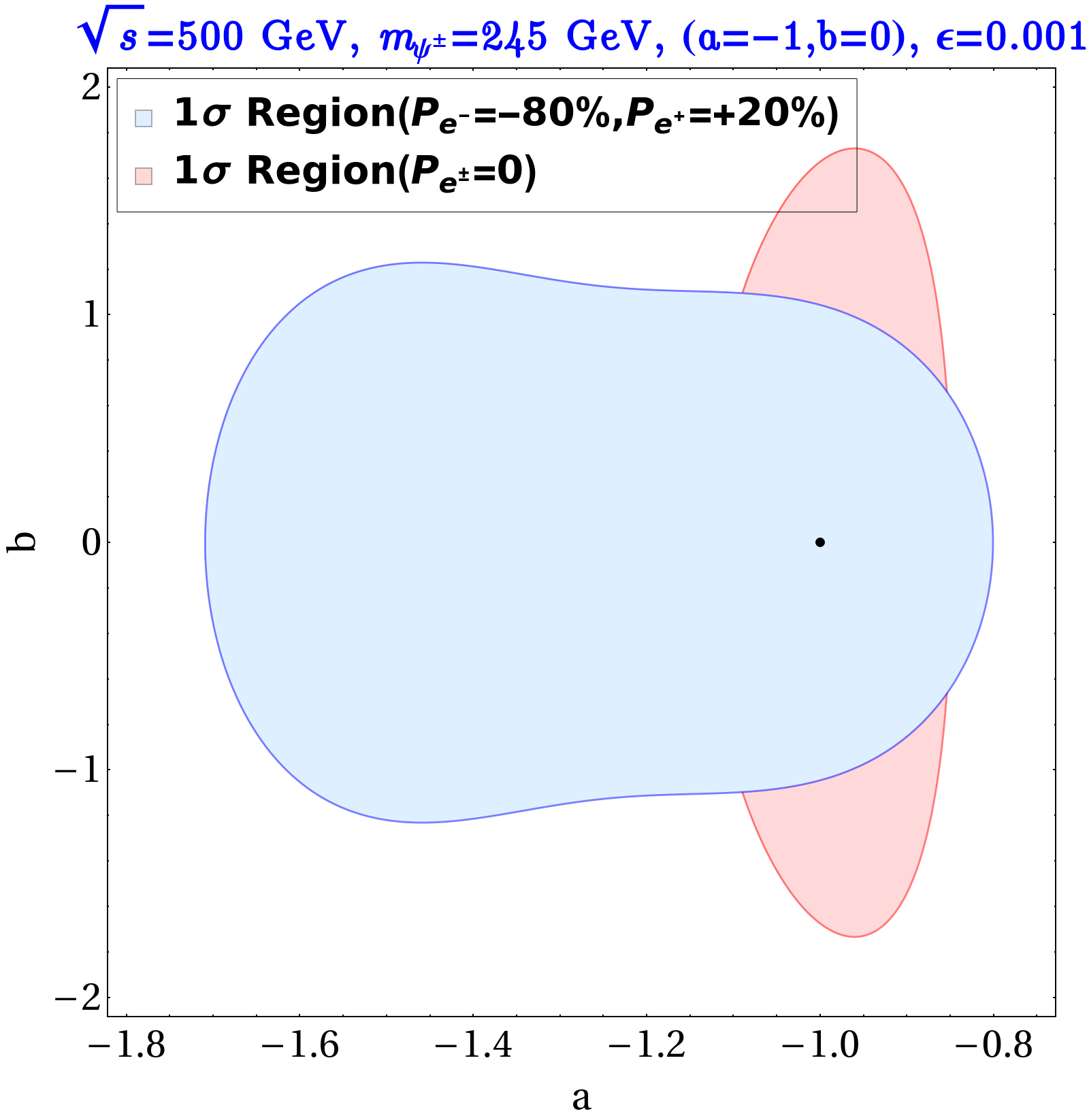} \cr
\includegraphics[scale=0.25]{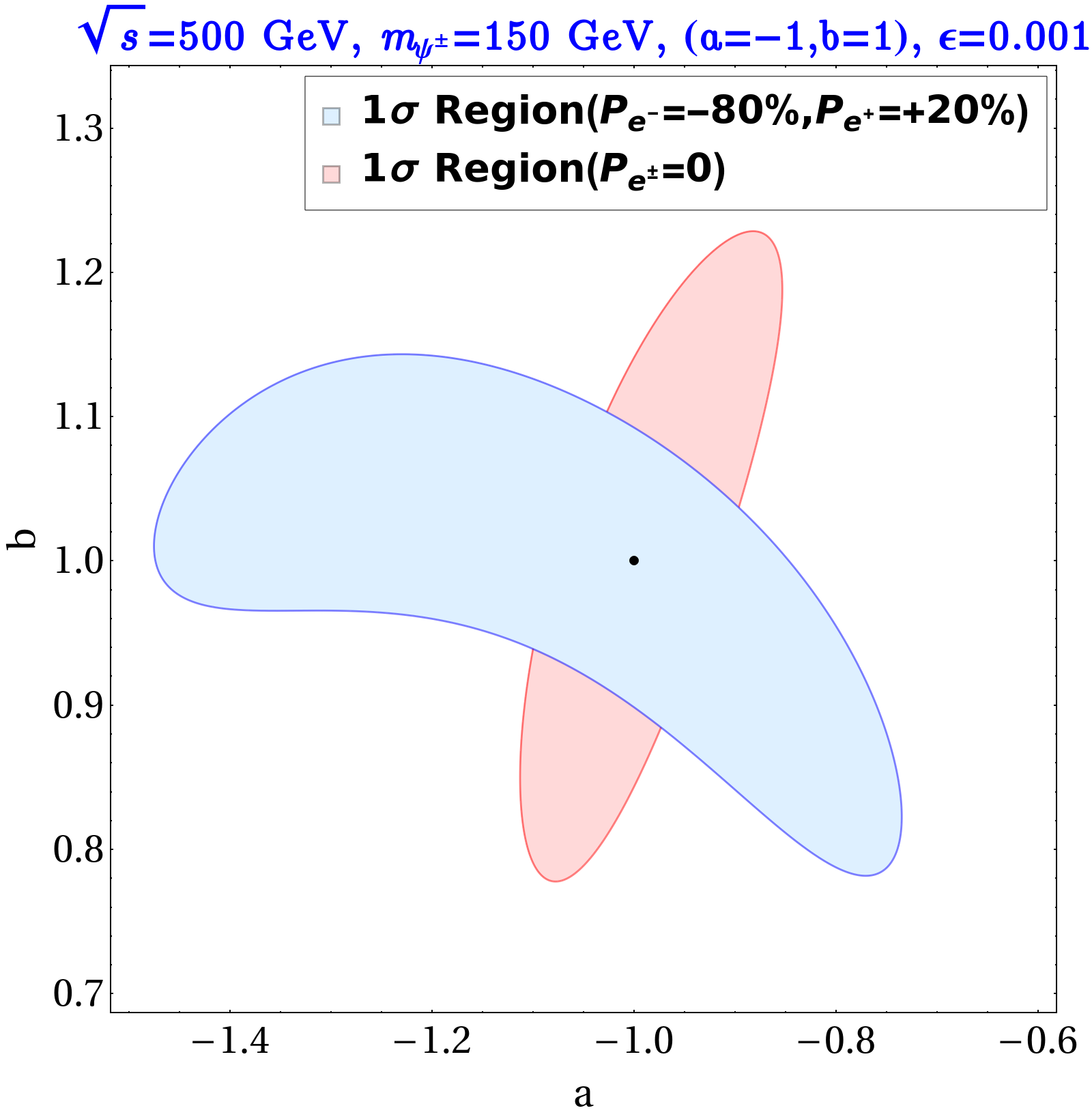}\quad&\quad \includegraphics[scale=0.25]{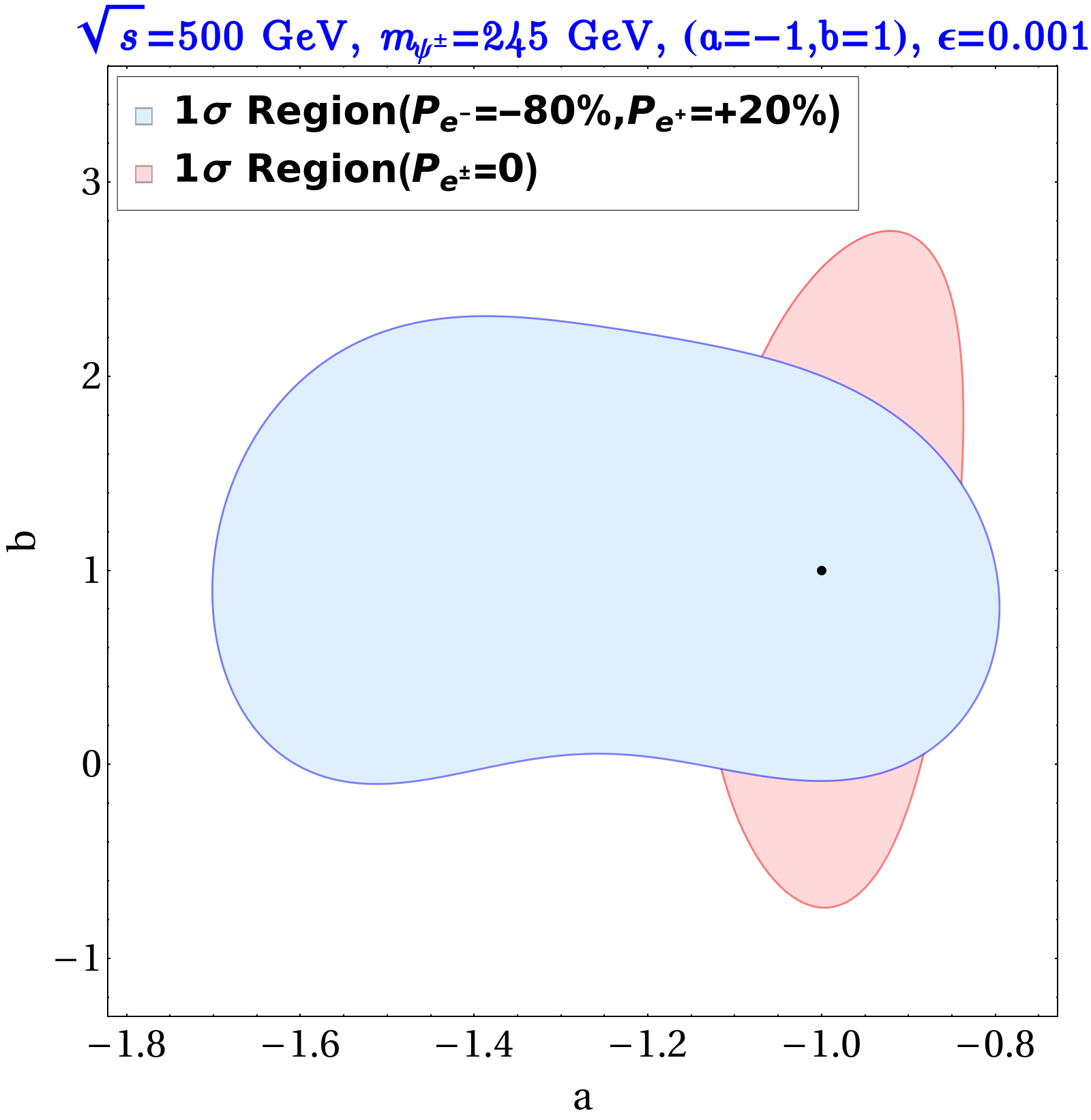} \cr
\end{align*}
\caption{ Same as Fig. \ref{fig:poldiff1} for $a^0<0 $.}	
\label{fig:poldiff2}
\end{figure}

Next, the optimal 1-$\sigma$ statistical uncertainties in the NP parameters $a,\,b$ are obtained from the $ \chi^2 \le1 $ regions using  Eqs.~\eqref{eq:covmat2} and \eqref{chi2} for the parameters in Eq.~\eqref{eq:params}. As illustrative examples we manifest the seed values listed at the beginning of this section, and take~\footnote{It is clear from Eq.~\eqref{chi2} that if $ \chi^2$ is held fixed, the $ \delta c $ will scale as $ 1/\sqrt{\epsilon} $, but the dependence of $ \delta a,\,\delta b $ on $ \epsilon$ is more complicated, see Eq.~\eqref{eq:c-a-b}; in the calculations below we use  $ \delta c = c(a^0 + \delta a,\,b^0 + \delta b) - c^0 $ in Eq.~\eqref{chi2}; see also section \ref{sec:method}.}  $\epsilon=0.001,\,0.005$ as reasonable estimates of the efficiency of signal identification (see Sect. \ref{sec:collider} for a discussion); we consider both unpolarized ($P_{e^\pm}=0$) and polarized ($P_{e^\pm} = ^{+20\%}_{-80\%}$) beams. The results are presented in figure \ref{fig:poldiff1} (for $a^0\ge 0$), figure \ref{fig:poldiff2} (for $a^0=-1$) and Table \ref{table:abe1}.  These results illustrate the advantages that polarization provides in the determination of the couplings of these new particles. 

From these results we can see that of the cases considered, the $a^0=0,\, b^0=\pm1$ (pure axial coupling) hypothesis has the largest statistical errors and is therefore the most challenging. Also worth noting is that, while the magnitude of the total cross section strongly affects the statistical uncertainties, it is not the only factor; this is illustrated by considering the $ a^0=-1$ case where the unpolarized cross section is larger than the polarized one (Fig. \ref{fig:cscom1}) but the uncertainties are larger (Fig. \ref{fig:poldiff2}). We also note that 1$\sigma$ regions for the lower mass ($150$ GeV) are smaller than those for higher mass ($245$ GeV), making the determination of the NP couplings for the latter case more difficult.

\begin{center}
\begin{table}
\begin{tabular}{| c | c | c  c | c  c | c  c |  c c | } 
\hline
\multicolumn{2}{|c|}{Seed parameters} &
\multicolumn{4}{|c|}{$P_{e^\pm}=0$} &
\multicolumn{4}{|c|}{$P_{e^\pm}=^{+20\%}_{-80\%}$} \\
\cline{3-10}
\multicolumn{2}{|c|}{} &
\multicolumn{2}{|c|}{$\epsilon=0.005$} &
\multicolumn{2}{|c|}{$\epsilon=0.001$}&
\multicolumn{2}{|c|}{$\epsilon=0.005$} &
\multicolumn{2}{|c|}{$\epsilon=0.001$}\\
\cline{1-10}
model &$m_{\psi^\pm}$(GeV) & $\pm\Delta a$ & $\pm\Delta b$ & $\pm\Delta a$ & $\pm\Delta b$ & $\pm\Delta a$ & $\pm\Delta b$ & $\pm \Delta a$ & $\pm\Delta b$\\ 
\hline
\multirow{4}*{\minitab[l]{$a=1$ \\ $b=0$}}
& \multirow{2}*{150}$ $ & $+0.04$ & $+0.11$ &$+0.08$ & $+0.24$ &$+0.02$&$+0.10$&$+0.05$&$+0.23$ \\ 
& &$-0.04$ & $-0.11$ & $-0.09$ & $-0.24$&$+0.02$&$+0.10$&$-0.05$ & $-0.23$  \\ 
\cline{2-10}
&$\multirow{2}*{245}$ & $+0.07$ & $+0.87$ & $+0.15$ & $+1.95$ &$+0.04$&$+0.05$ &$+0.09$ & $+0.91$ \\
&  &$-0.07$ & $-0.87$  &$-0.17$ & $-1.95$&$+0.04$&$+0.05$&$-0.09$ & $-0.91$ \\
\hline
\multirow{4}*{\minitab[l]{$a=1$\\ $b=\pm1$}}
& $ \multirow{2}*{150} $ &$+0.06$ & $+0.10$ & $+0.13$ & $+0.23$&$+0.03$&$+0.11$ & $+0.07$ & $+0.25$\\
& &$-0.06$ & $-0.10$ & $-0.18$ & $-0.24$&$-0.03$&$-0.11$& $-0.08$ & $-0.24$ \\
\cline{2-10}
& $\multirow{2}*{245}$ &$+0.07$ & $+0.87$ & $+0.15$ & $+1.96$&$+0.04$&$+0.86$&$+0.09$ & $+1.93$  \\
& &$-0.07$ & $-0.87$ & $-0.21$ & $-1.96$&$-0.04$&$-0.86$&$-0.10$ & $-1.92$ \\
\hline
\multirow{4}*{\minitab[l]{$a=0$\\$b=\pm1$}}
& $ \multirow{2}*{150} $  &$+0.35$ & $+0.06$ & $+0.51$ & $+0.13$&$+0.04$&$+0.09$&$+0.07$ & $+0.22$ \\
&  &$-0.27$ & $-0.11$ & $-0.41$ & $-0.22$&$-0.05$&$-0.09$&$-0.11$ & $-0.22$  \\
\cline{2-10}
&$\multirow{2}*{245}$  &$+0.43$ & $+0.66$ & $+0.59$ & $+1.39$&$+0.04$&$+0.66$&$+0.10$ & $+1.50$\\
& & $-0.26$ & $-0.67$ & $-0.13$ & $-1.51$&$-0.04$&$-0.66$&$-0.11$ & $-1.47$ \\   
\hline
\multirow{4}*{\minitab[l]{$a=-1$\\$b=0$}}
& \multirow{2}*{150}$ $ & $+0.04$ & $+0.09$& $+0.07$ & $+0.21$&$+0.06$&$+0.06$& $+0.12$ & $-0.16$\\
& &$-0.04$ & $-0.09$ & $-0.07$ & $-0.21$ &$-0.08$&$-0.06$&$-0.63$ & $-0.16$ \\ 
\cline{2-10}
&$\multirow{2}*{245}$ & $+0.06$ & $+0.76$ & $+0.15$ & $+1.75$&$+0.10$&$+0.48$& $+0.20$ & $+1.23$  \\
&  &$-0.06$ & $-0.76$  &$-0.15$ & $-1.75$&$-0.21$&$-0.48$&$-0.71$ & $-1.23$ \\
\hline
\multirow{4}*{\minitab[l]{$a=-1$\\$b=\pm1$}}
& $ \multirow{2}*{150} $ &$+0.06$ & $+0.10$ & $+0.15$ & $+0.22$&$+0.17$&$+0.11$&$+0.26$ & $+0.14$ \\
& &$-0.05$ & $-0.10$ & $-0.11$ & $-0.22$&$-0.34$&$-0.11$& $-0.48$ & $-0.22$  \\
\cline{2-10}
& $\multirow{2}*{245}$ &$+0.07$ & $+0.78$ & $+0.17$ & $+1.79$&$+0.11$&$+0.71$&$+0.20$ & $+1.31$\\
& &$-0.06$ & $-0.78$  &$-0.14$ & $-1.74$&$-0.60$&$-0.49$& $-0.70$ & $-1.11$ \\
\hline
\end{tabular}
\caption{Optimal 1$\sigma$ statistical uncertainty in the $a,b$ couplings for both unpolarized and polarized  ($P_{e^\pm} = ^{+20\%}_{-80\%}$) beams and two values of $ \epsilon$; we used the parameters in Eq.~(\ref{eq:params}).}
\label{table:abe1}
\end{table}
\end{center}

\subsubsection{Differentiation of models}
One of the most important uses of the OOT is the ability to estimate the extent to which different hypotheses can be distinguished. Specifically, we consider a ``base'' hypothesis $ a = a^0,\,b=b^0$ and, using Eq.~(\ref{eq:covmat2}), define
\beq
\left[\Delta\sigma(a^0,b^0;\,\bar a,\bar b)\right]^2 = \epsilon\sum_{i,j} \left(c^0_i - \bar c_i \right) \left( c^0_j - \bar c_j \right) \left( V^{-1}_0 \right)_{ij}\,, \quad V_0 = V(c=c^0)\,;
\label{eq:signif-def}
\eeq
(where $ c^0_i= c_i(a^0,\,b^0),\,\bar c_i= c_i(\bar a,\,\bar b)$) which we take as a measure of the degree to which the $ a = \bar a,\,b = \bar b $ hypothesis can be distinguished from the base hypothesis; we refer to $ \Delta \sigma $ 
as the statistical significance of the $ \bar a,\,\bar b$ hypothesis (which depends on the base model chosen).

We will use $ \Delta\sigma $ as a measure of the separation of an alternate model from the base one. The distribution of $ \Delta\sigma $ can also be used to determine the probability that $ \Delta\sigma \le \ell $ occurs; in general this distribution is not simple, but for the cases where $c_0$ and $ \bar c$ have normal distributions with averages ${\mathbbm c}_0,\,\bar{\mathbbm c}$, and covariance matrices $V_0 $ and $ \bar V$ that are approximately proportional to the unit matrix, $ \sigma_0^2 \mati,\, \bar\sigma^2\mati$ respectively, then $ \sqrt{\Delta\sigma} $ is approximately normally distributed with average $ |{\mathbbm c}_0-\bar{\mathbbm c}|/\sigma_0$ and variance $ 1 + (\bar\sigma/\sigma_0 )^2 $; in practice this means that the values quoted for $\Delta\sigma $ will have errors $ \sim \pm \sqrt{1 + (\bar\sigma/\sigma_0 )^2}$. Similar results hold when $ \Delta\sigma $ is written in terms of $a$ and $b$ provided they also are normally distributed. 

\begin{figure}[htb!]
$$
\includegraphics[scale=0.25]{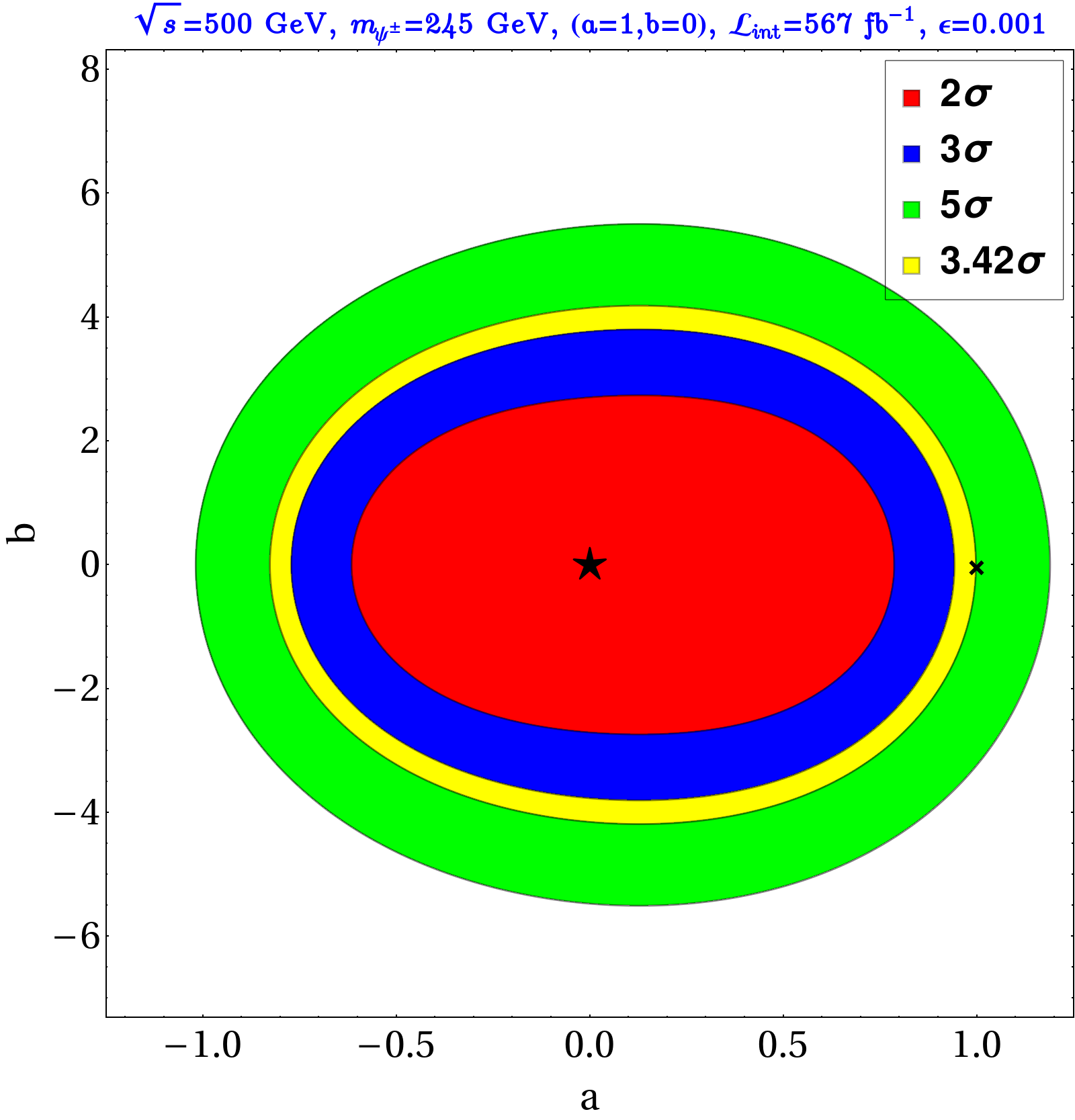}
~~\includegraphics[scale=0.25]{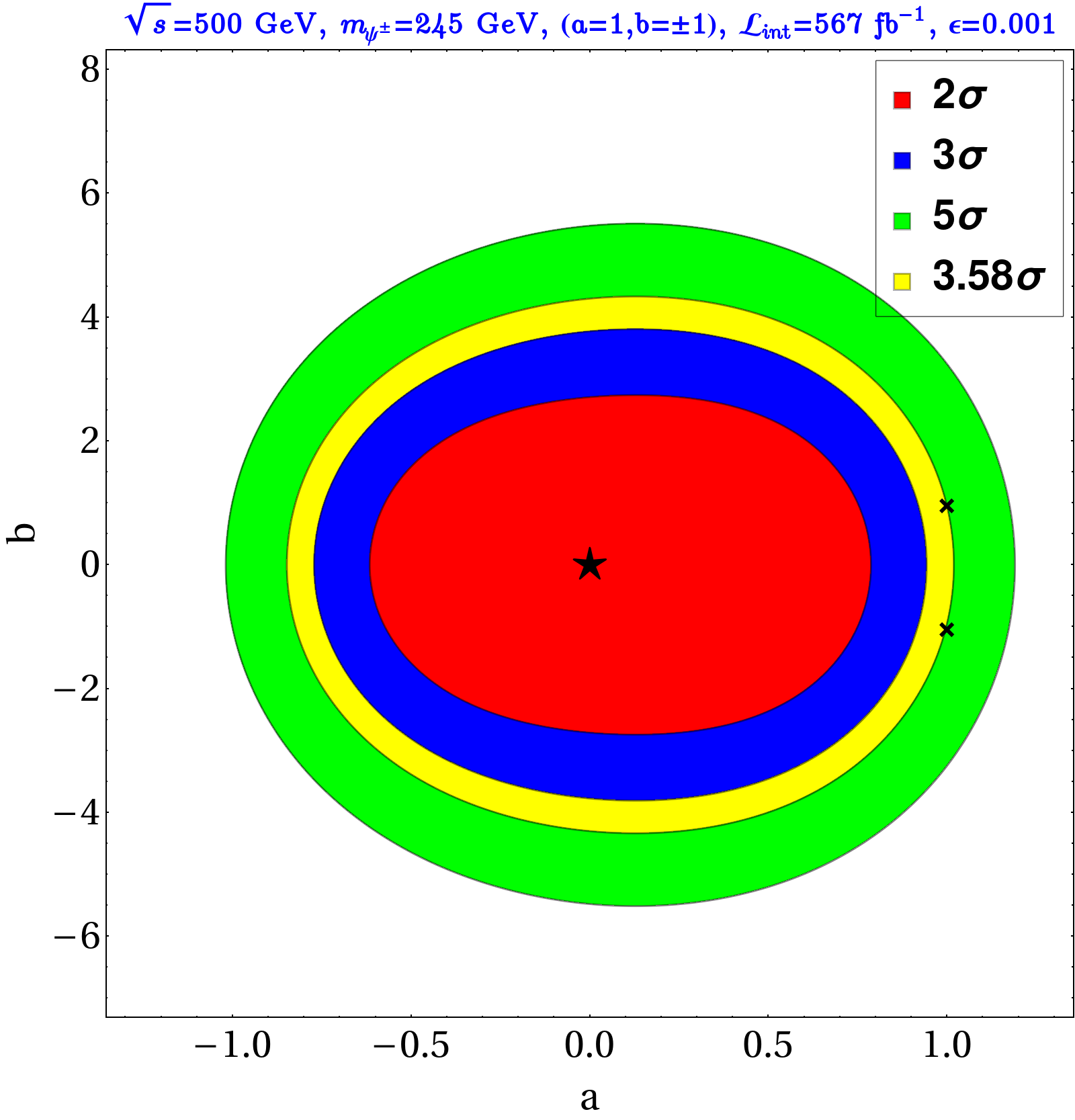}
~~\includegraphics[scale=0.25]{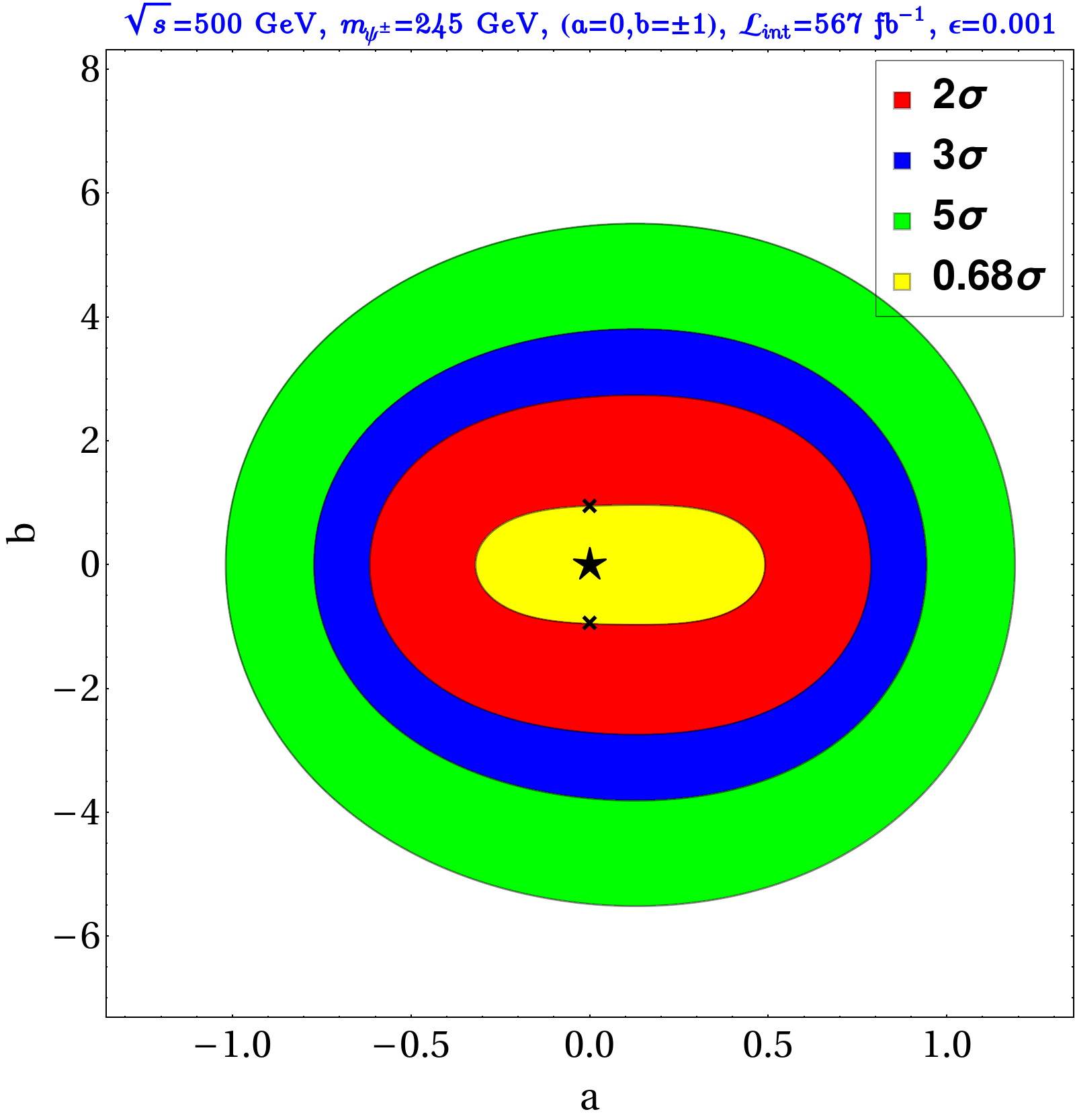}
$$
$$
\includegraphics[scale=0.25]{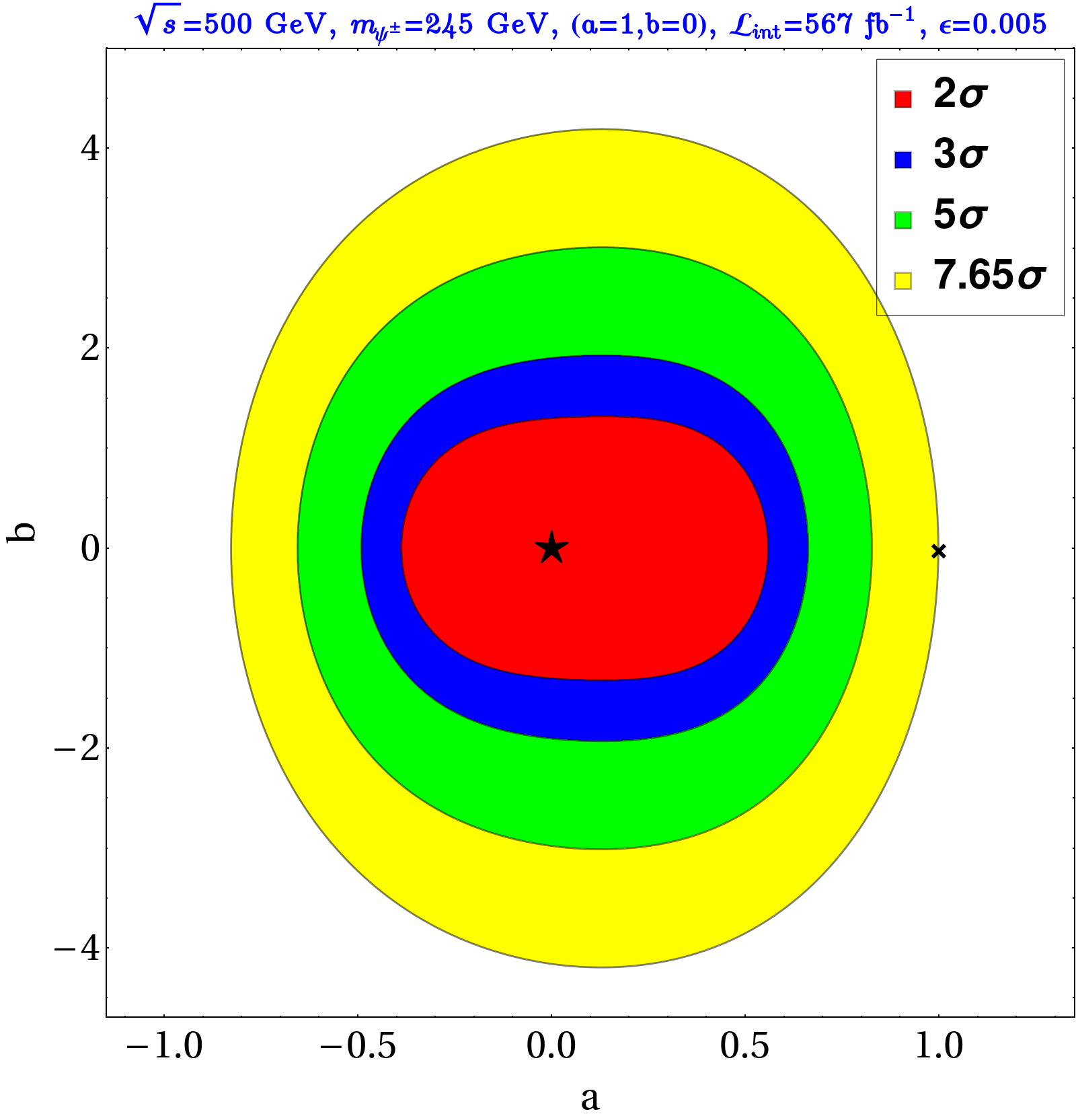}
~~\includegraphics[scale=0.25]{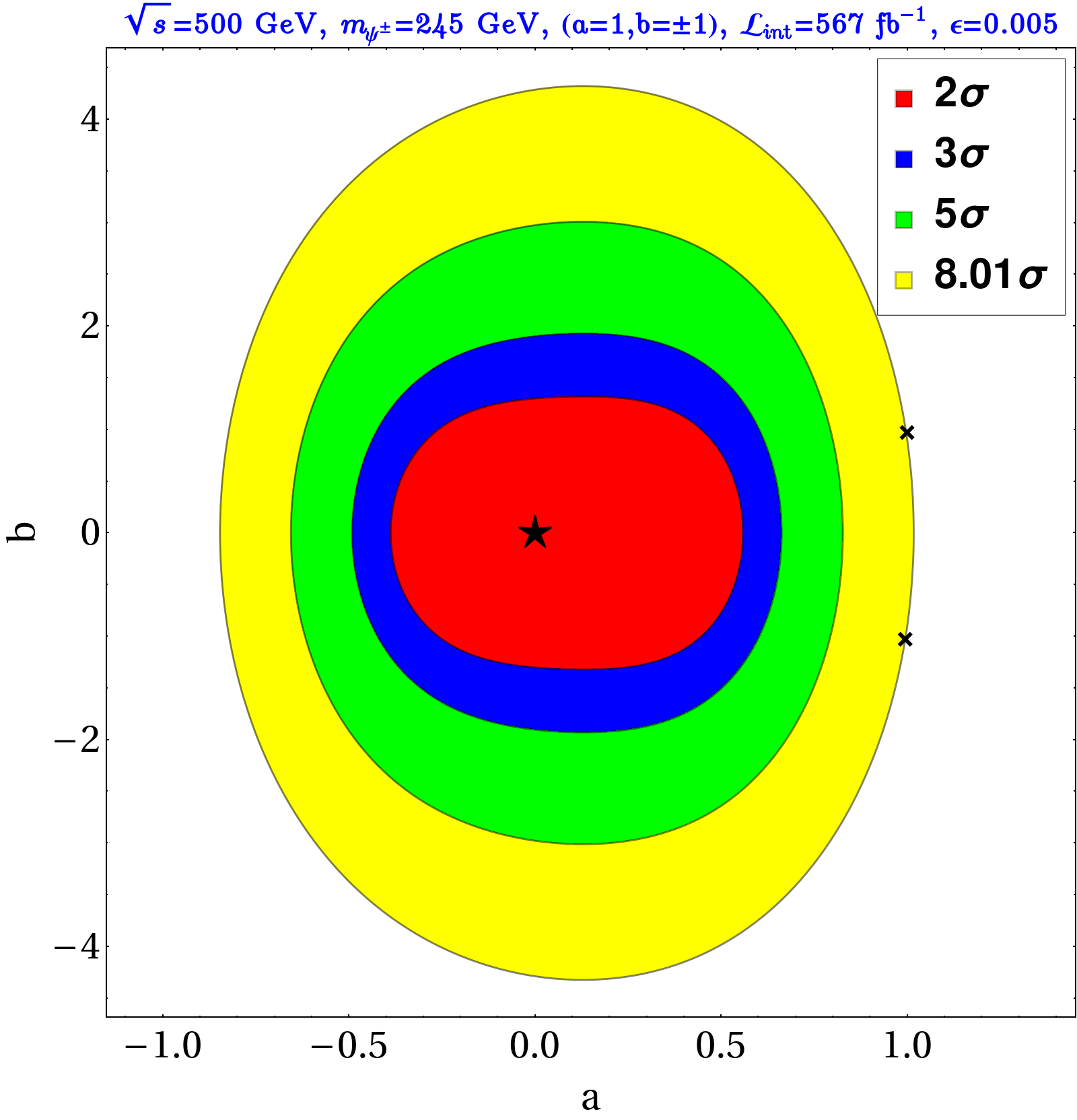}
~~\includegraphics[scale=0.25]{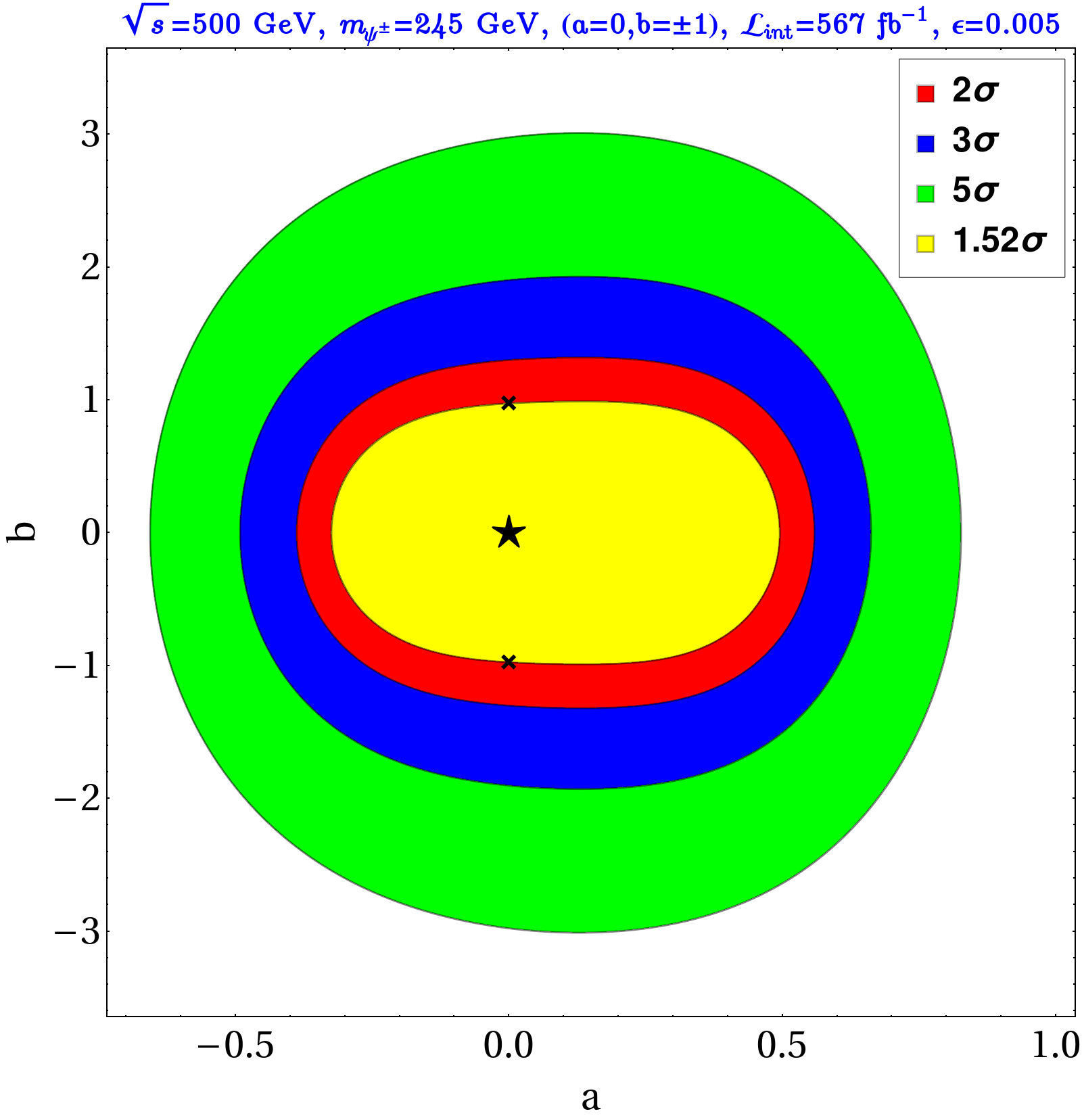}
$$
$$
\includegraphics[scale=0.25]{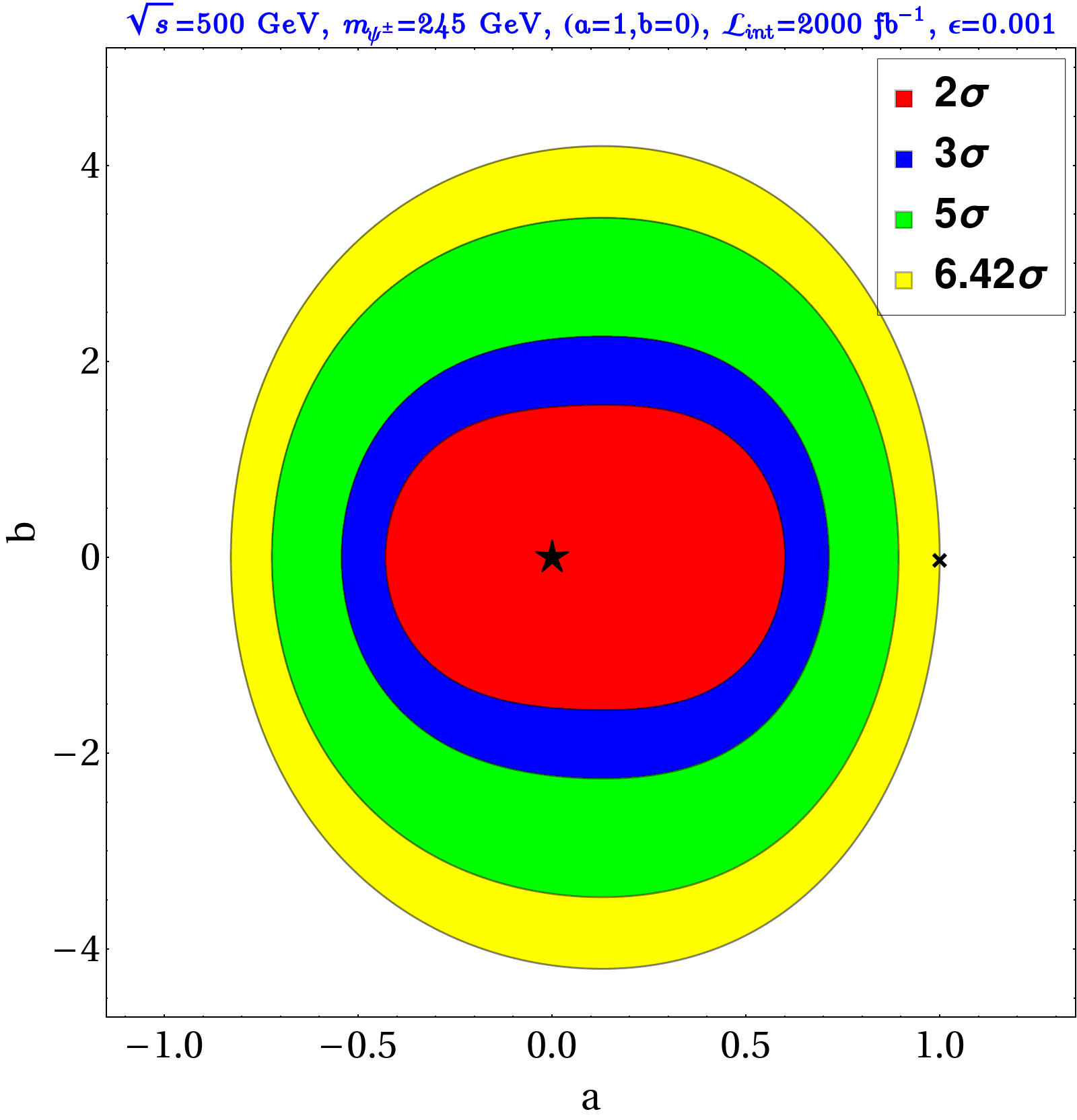}
~~\includegraphics[scale=0.25]{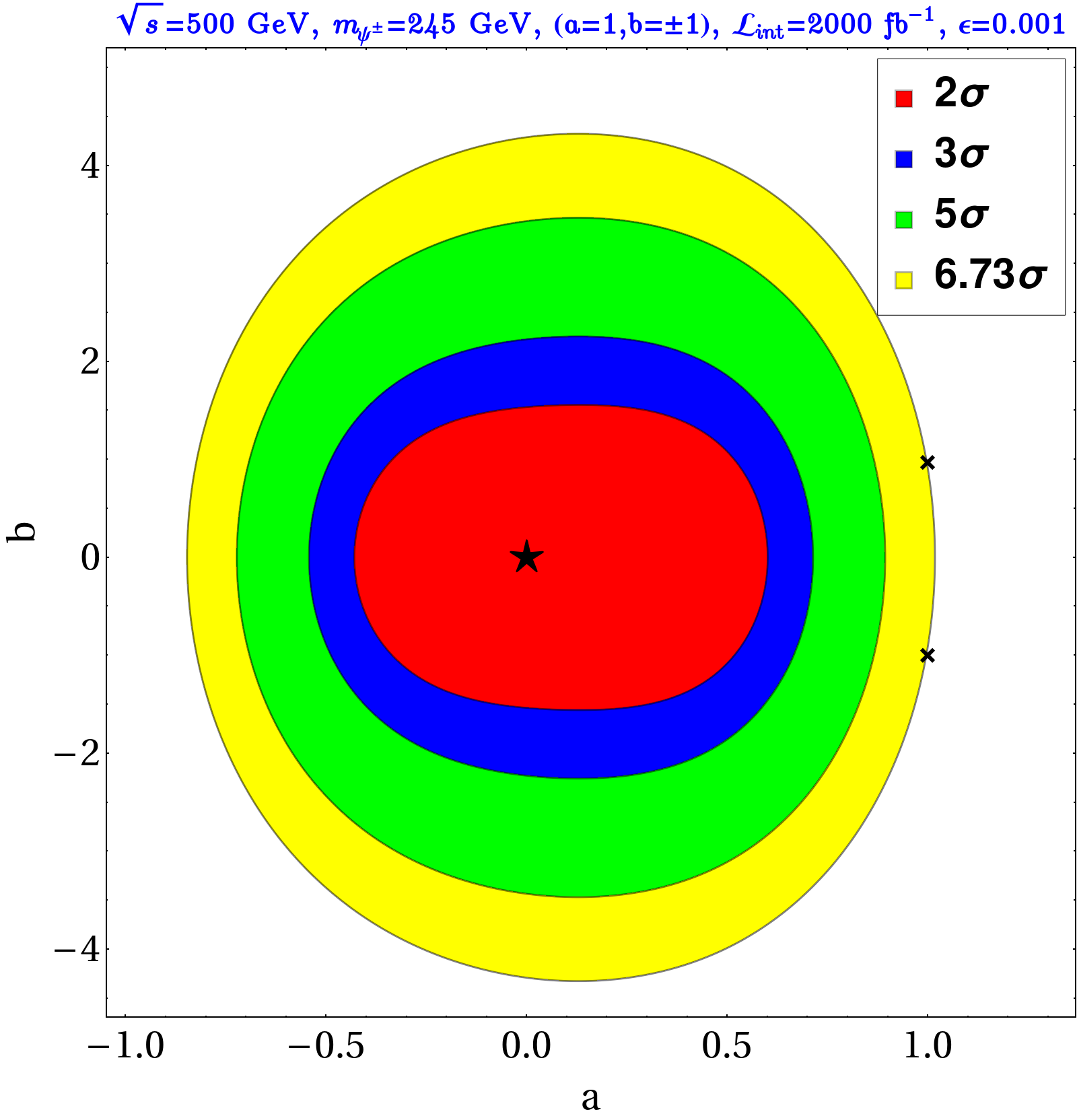}
~~\includegraphics[scale=0.25]{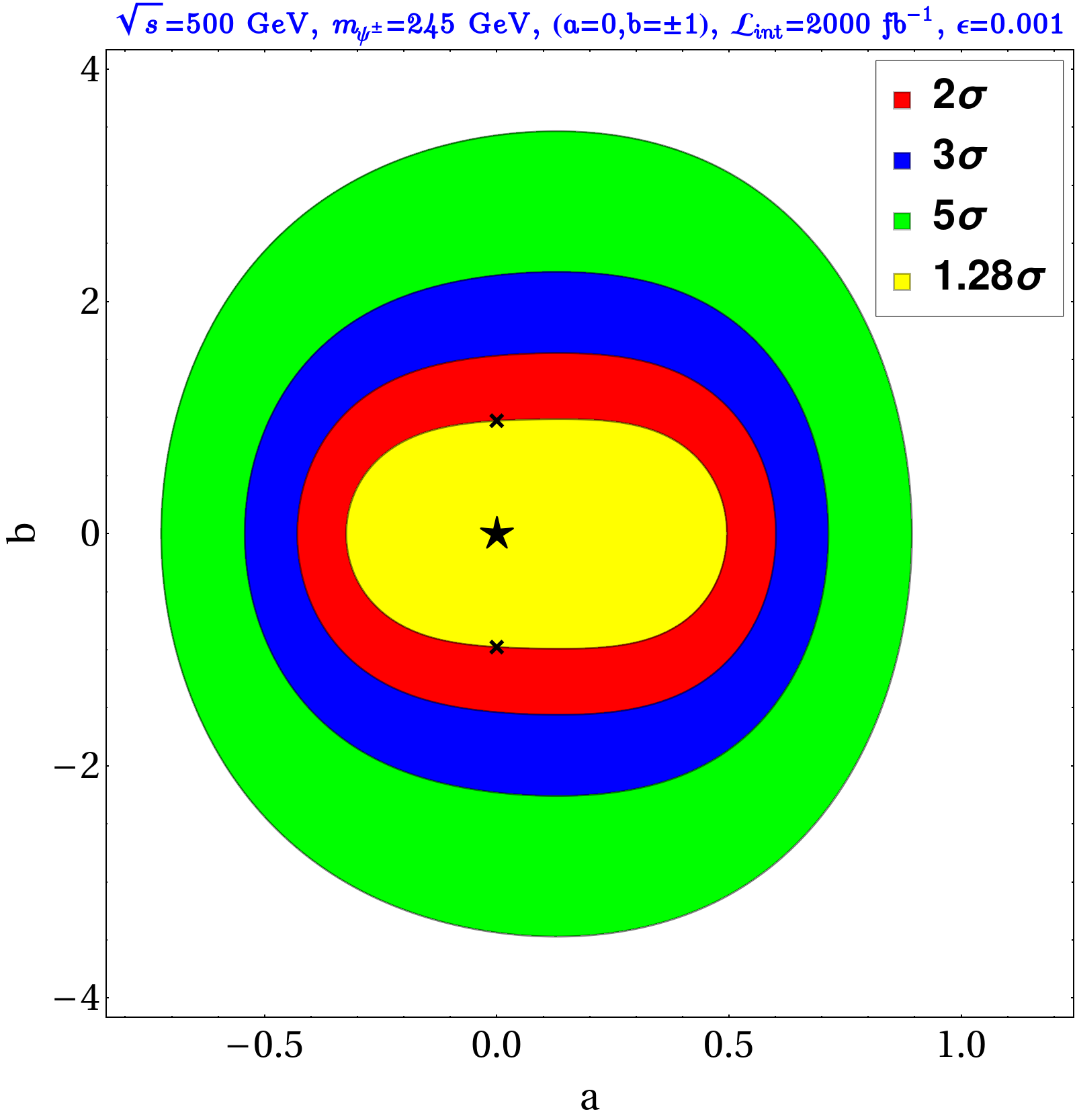}
$$
\caption{$ 2\sigma,\,3\sigma$ and $5\sigma$ regions (red, blue and green, respectively) when $ a^0=b^0=0$, indicated by a star. The yellow area denotes the significance, Eq.~(\ref{eq:signif-def}), of alternate hypotheses $ a=\bar a,\,b=\bar b$ (indicated by crosses) for various choices of luminosity ($ \lcal$) and efficiency ($\epsilon$): $\bar a=1,\, \bar b=0$ (left column), $\bar a=1, \,\bar b=\pm 1$ (middle column) and $\bar a=0,\,\bar b=\pm 1$ (right column). We assumed $ m_{\psi^\pm} = 245\,\gev$ and  unpolarized beams.}
\label{fig:dscvry1}
\end{figure}

\begin{figure}[htb!]
$$
\includegraphics[scale=0.27]{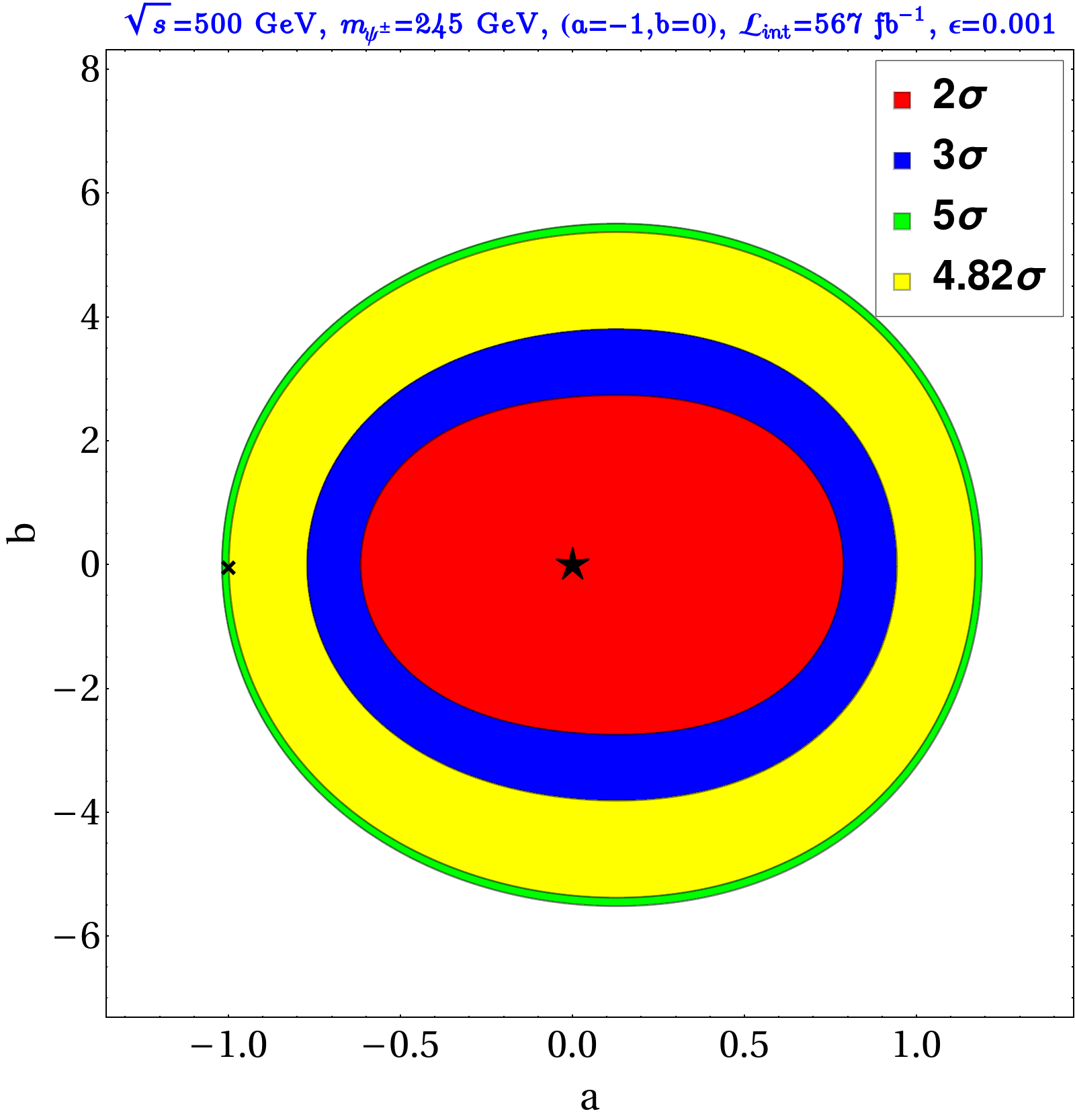}
\includegraphics[scale=0.27]{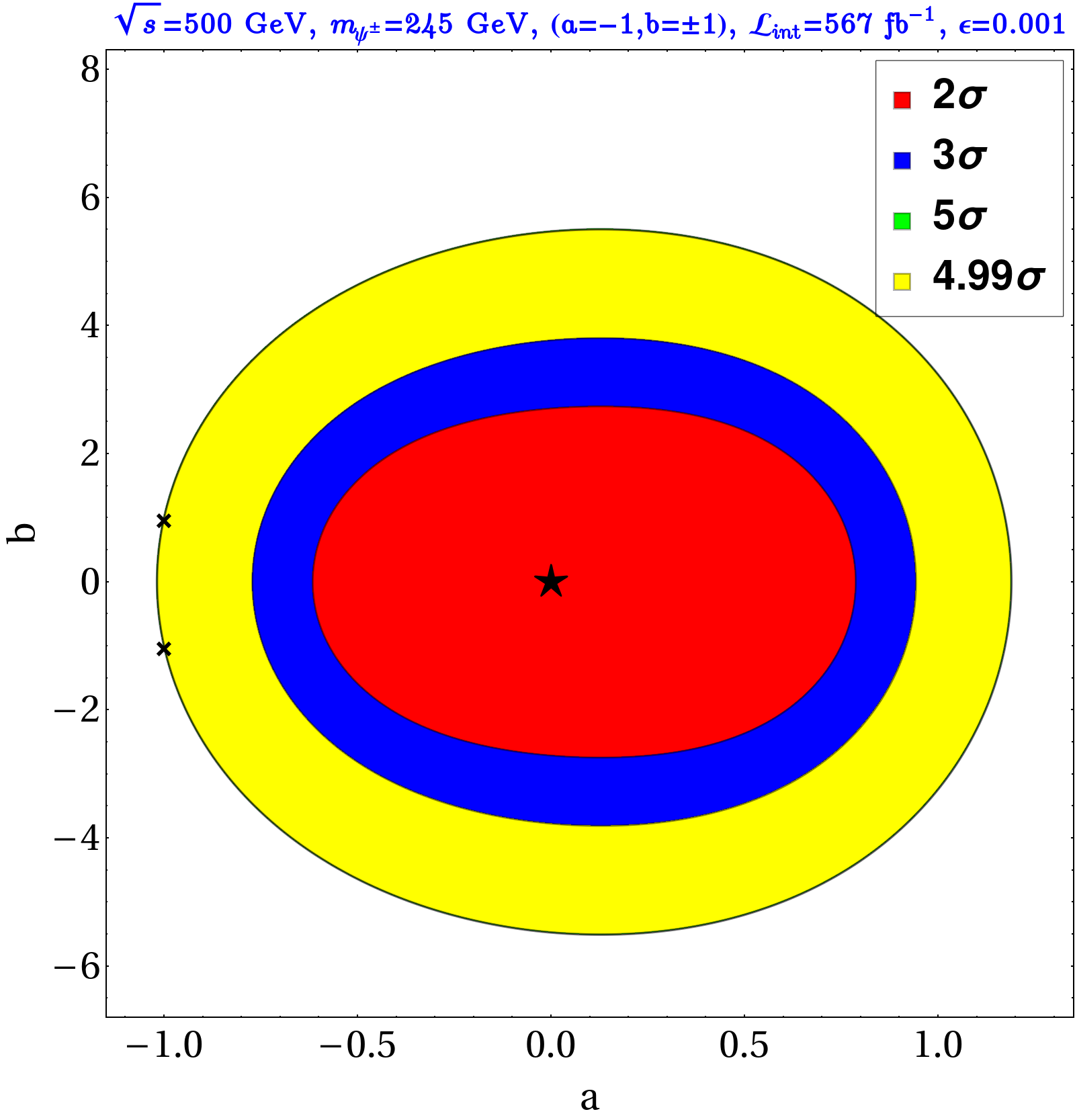}
$$
$$
\includegraphics[scale=0.27]{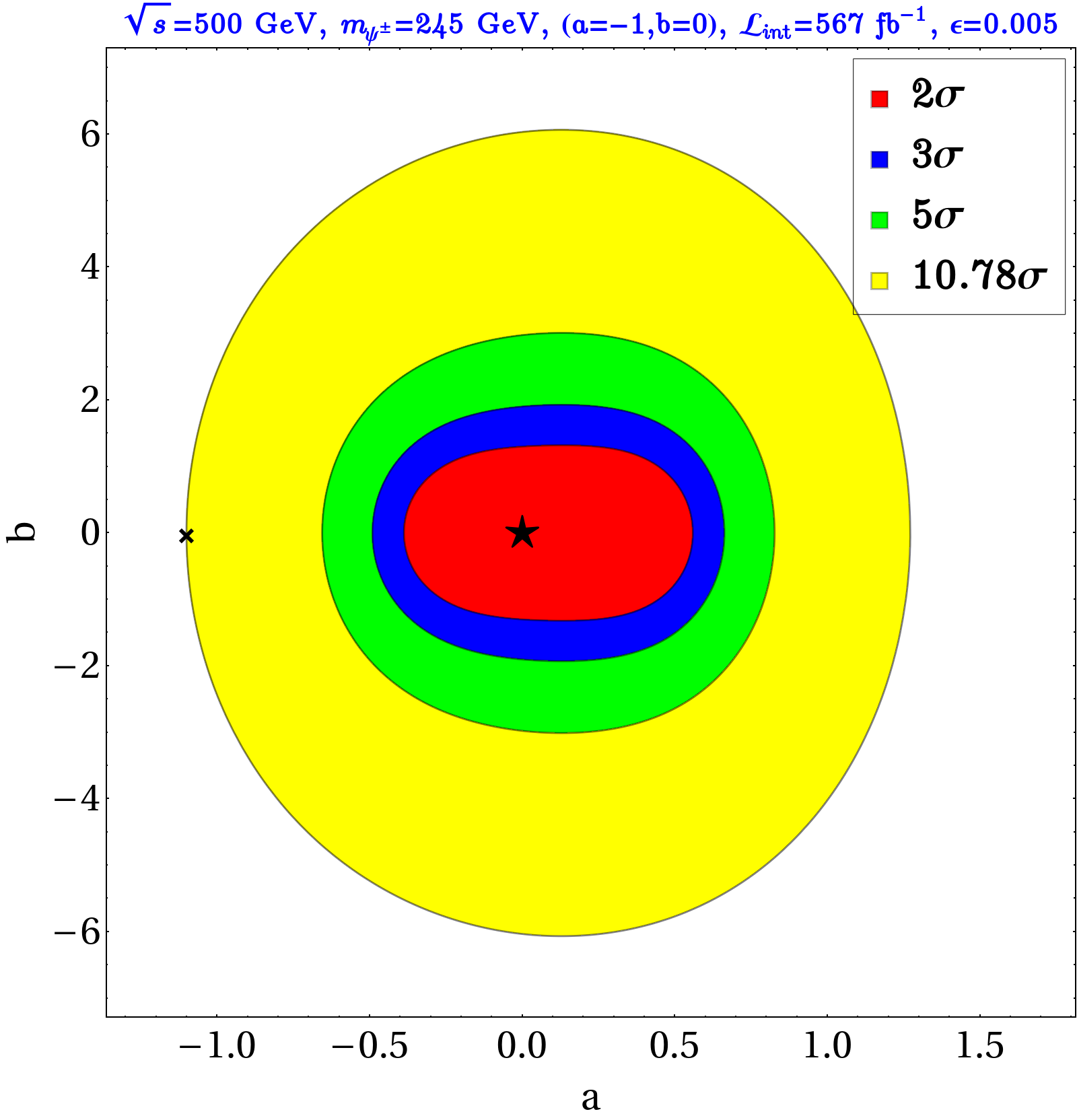}	\includegraphics[scale=0.27]{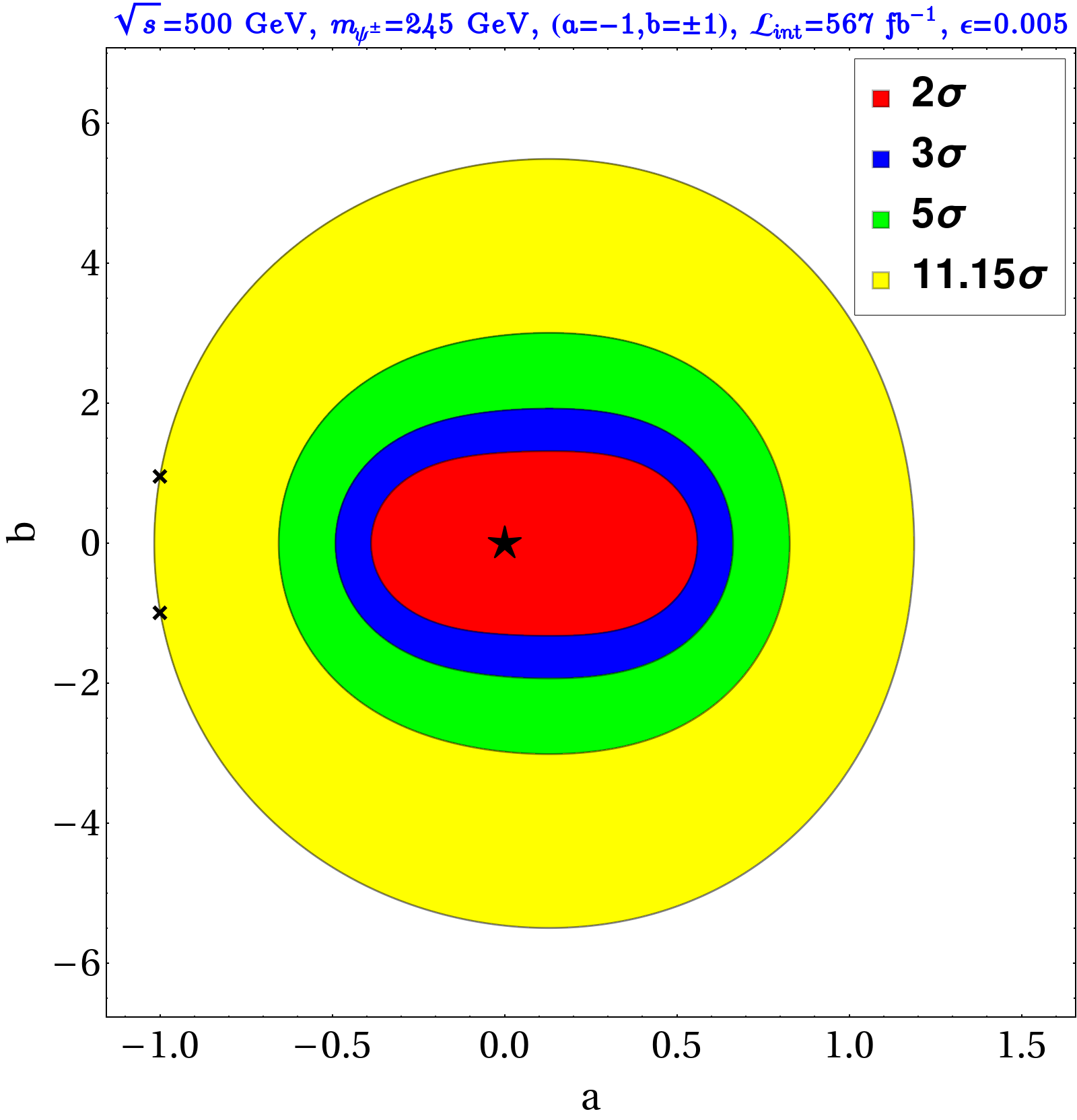}
$$
$$
\includegraphics[scale=0.27]{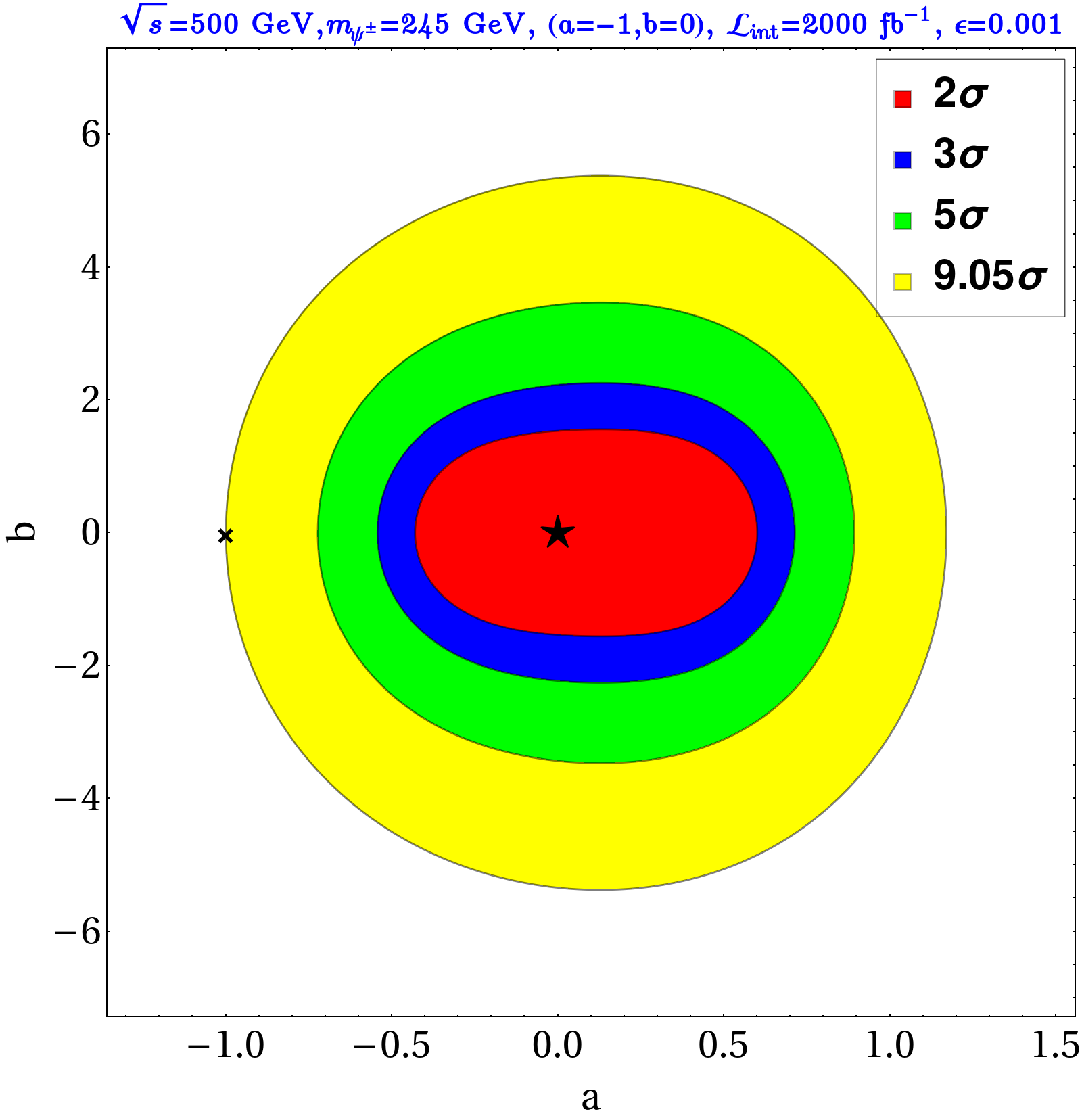}
\includegraphics[scale=0.27]{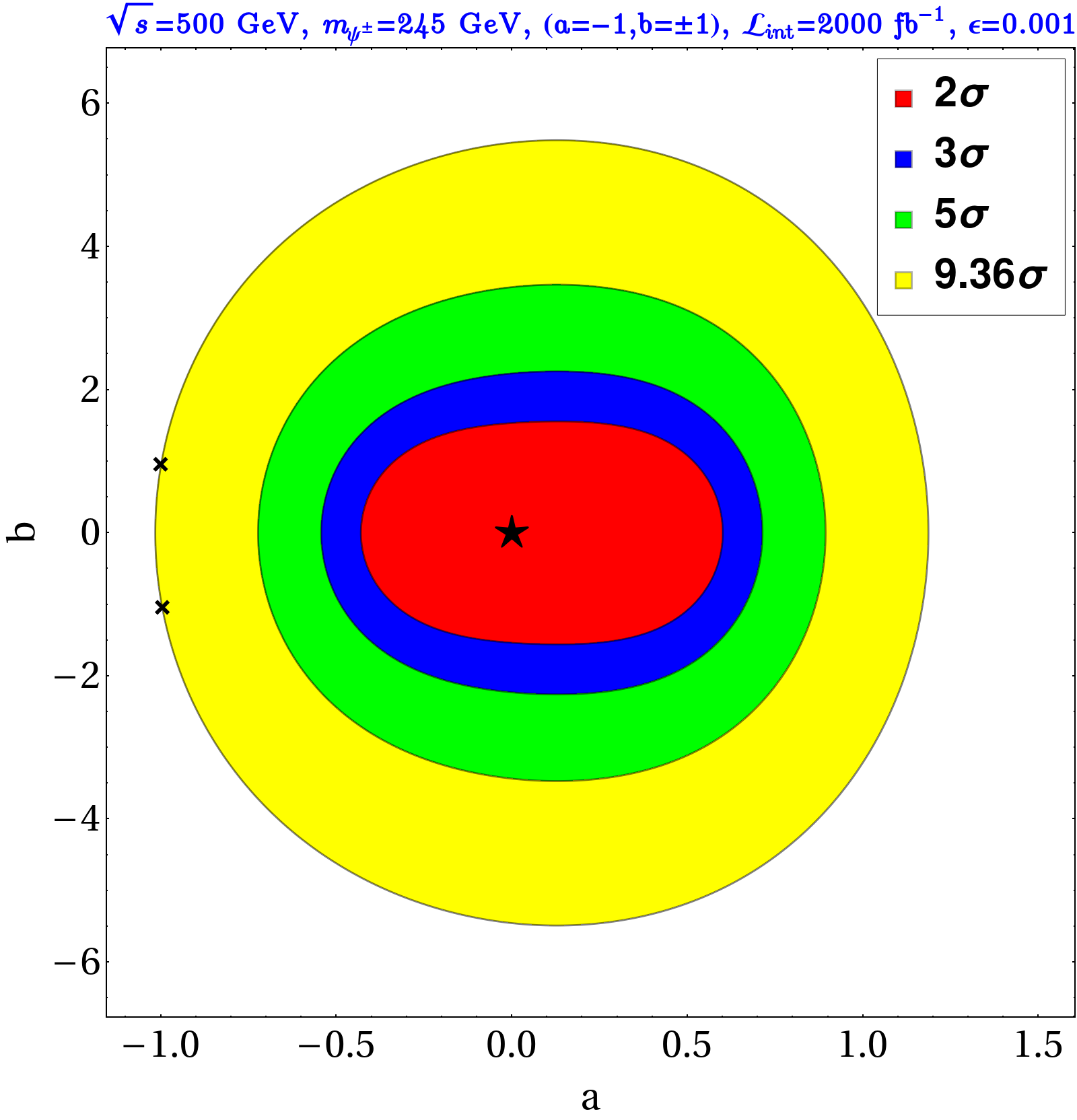}
$$
\caption{Same as figure \ref{fig:dscvry1} for $\bar a = - 1,\,\bar b=0$ (left column) and $ \bar a = -1,\,\bar b=\pm1$ (right column).}
\label{fig:dscvry2}
\end{figure}

\begin{figure}[htb!]
$$
\includegraphics[scale=0.27]{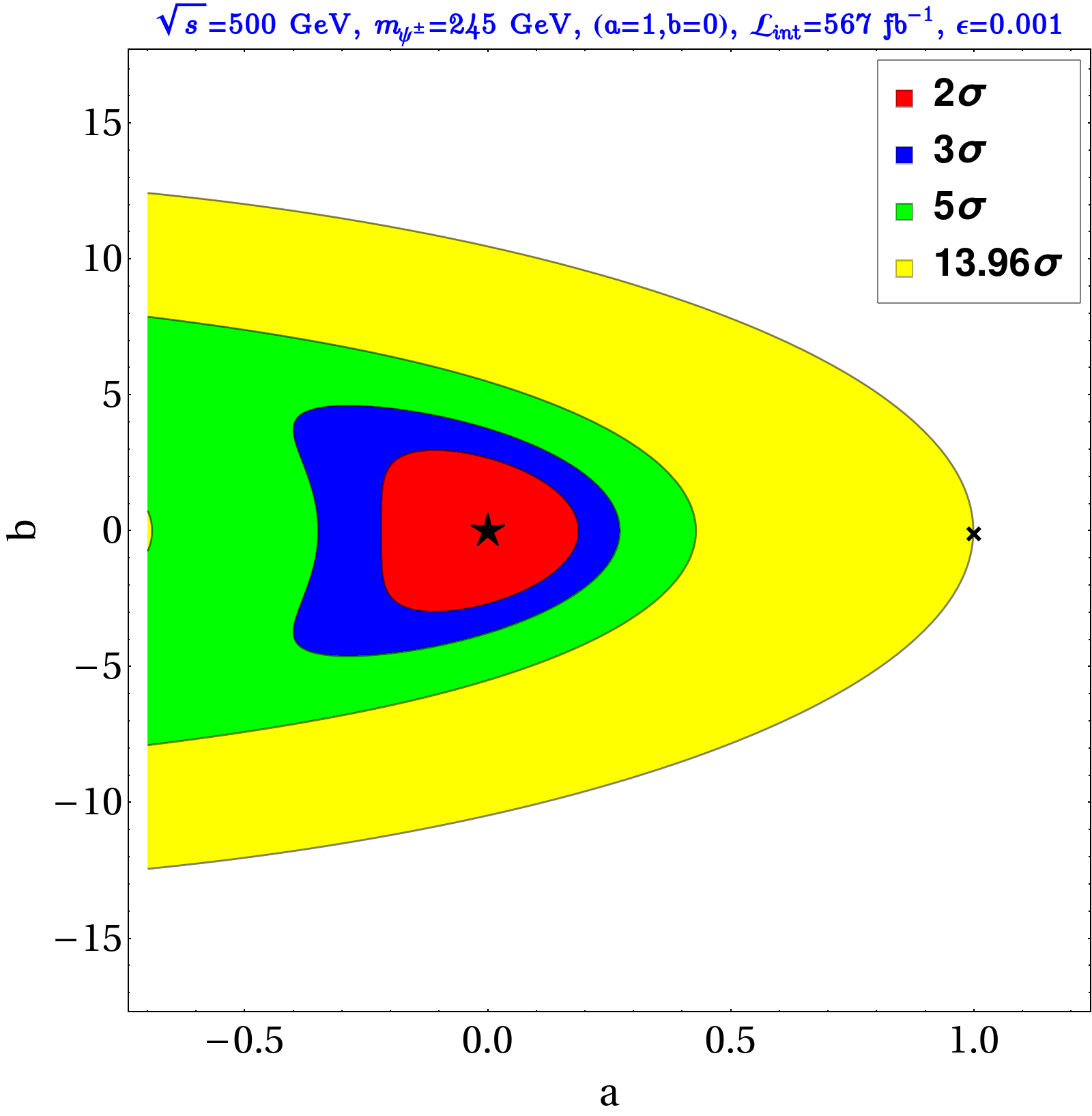}
\includegraphics[scale=0.27]{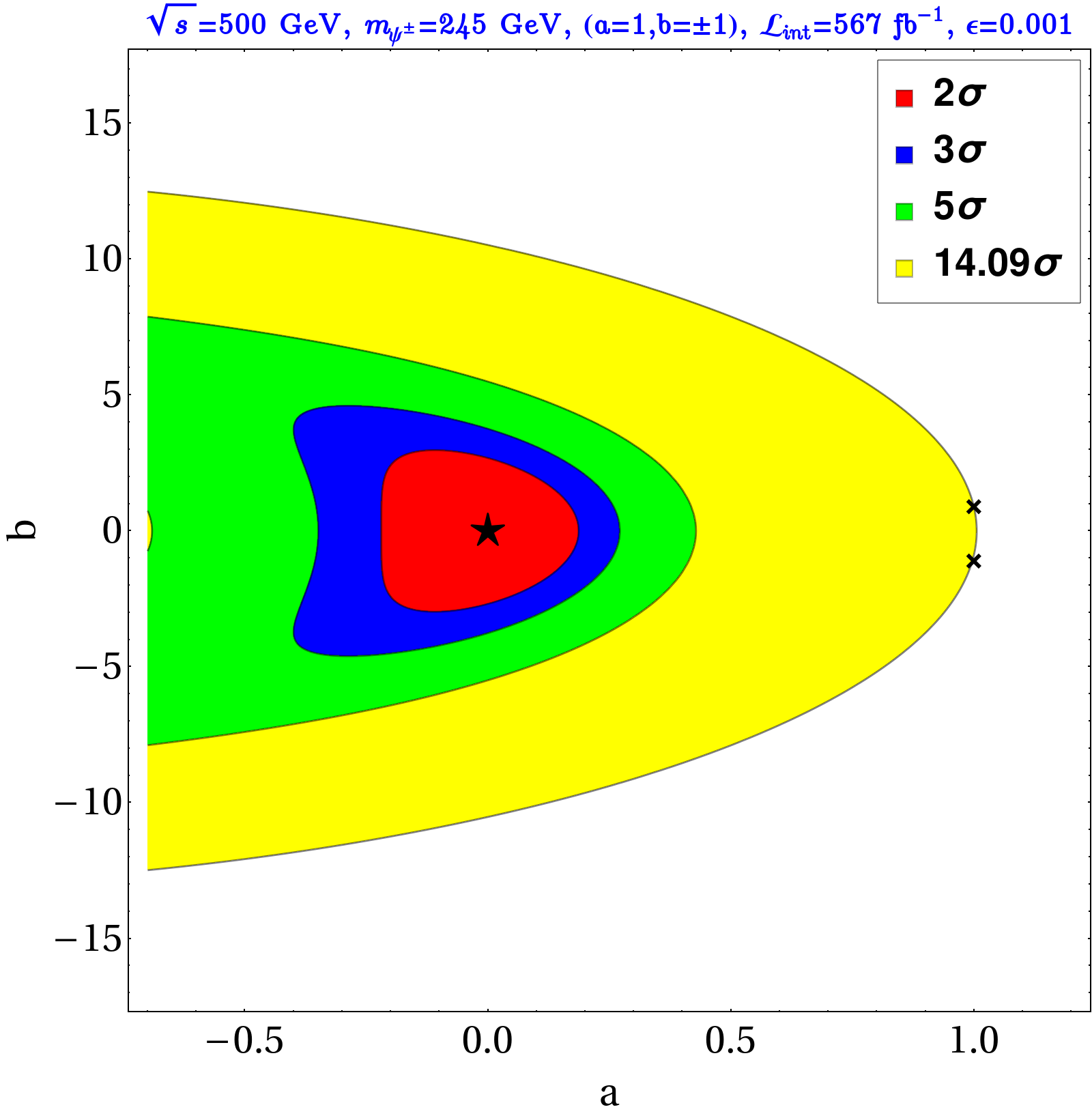}
\includegraphics[scale=0.27]{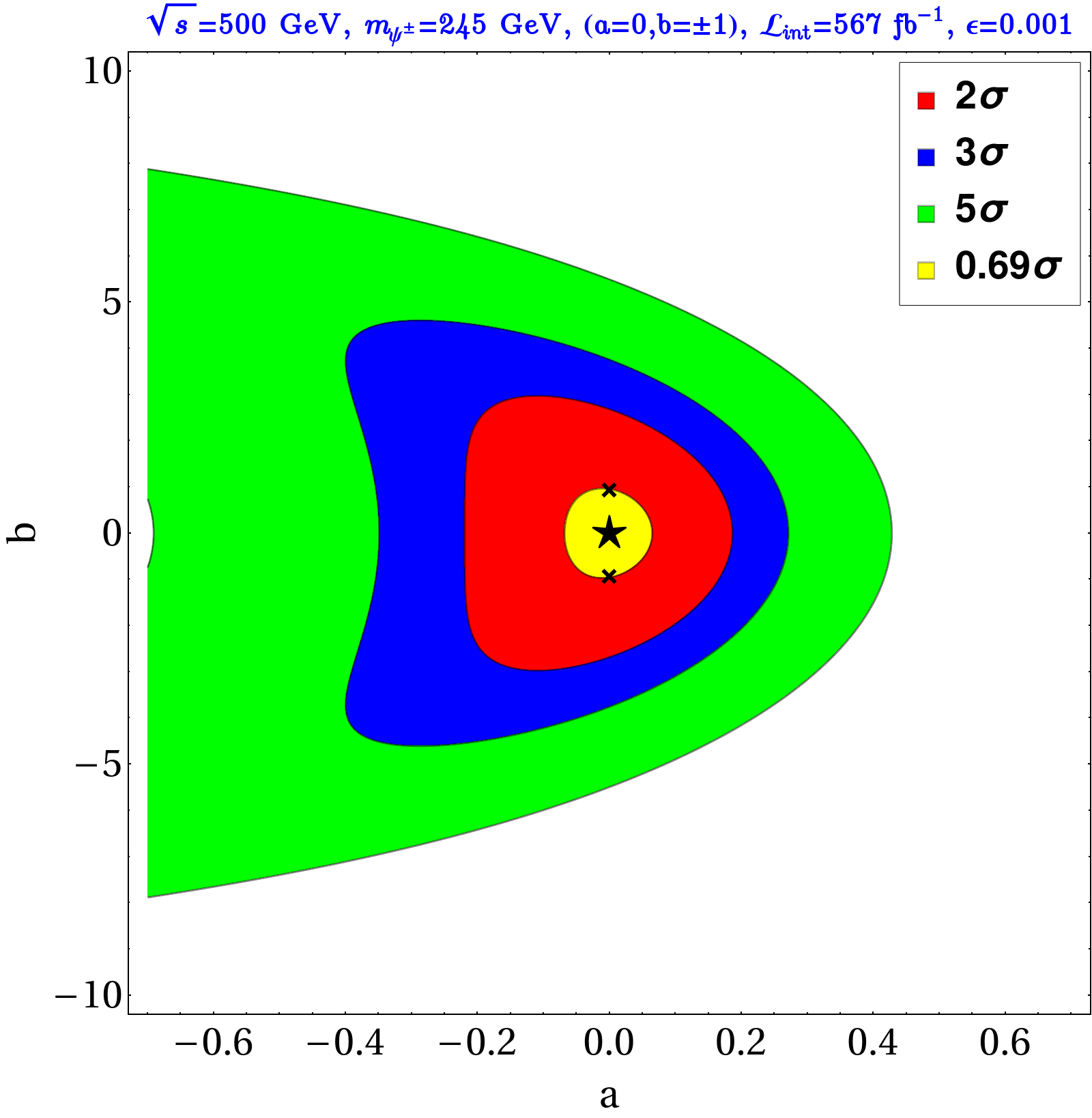}
$$
$$
\includegraphics[scale=0.27]{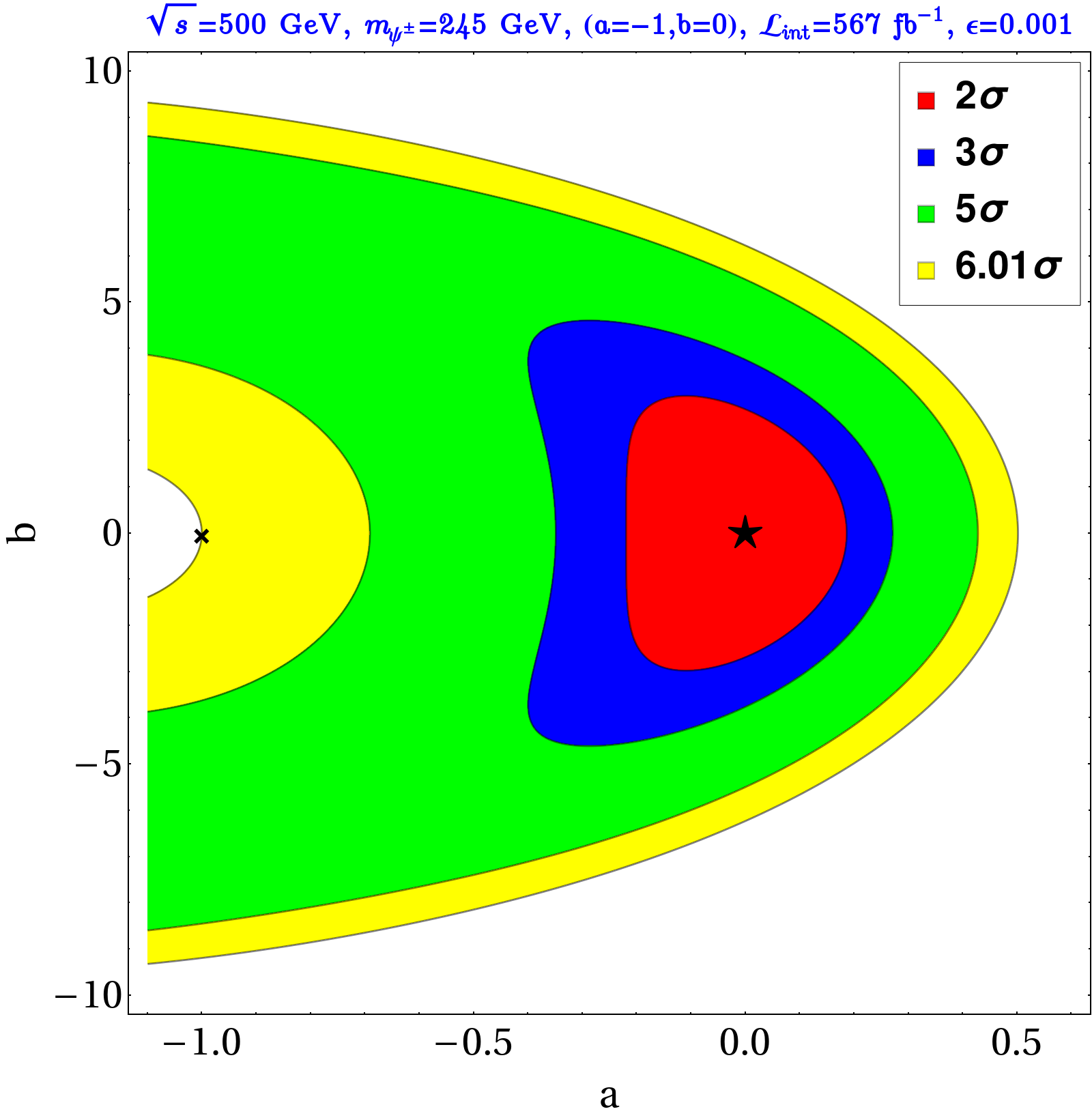}	
\includegraphics[scale=0.27]{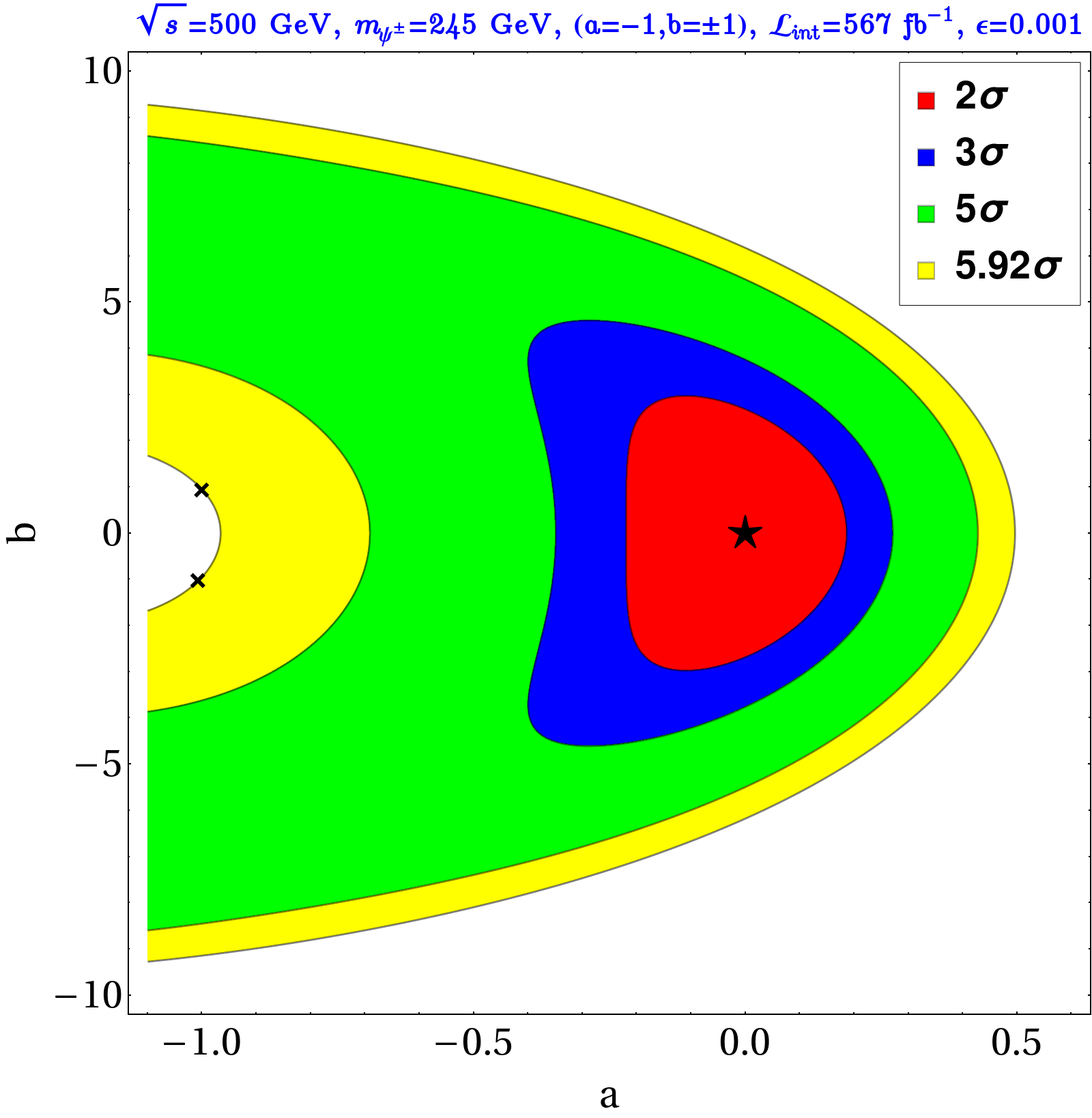}
$$
\caption{Same as figure \ref{fig:dscvry1} for polarized beams ($P_{e^\pm} = ^{+20\%}_{-80\%}$).}
\label{fig:absig1pol2}
\end{figure}

We now consider a few examples\footnote{For these choices $V_0,\,\bar V$ are approximately proportional to $\mati$.} corresponding to some of the cases presented in Table~\ref{table:discovery1} or in Figs.~\ref{fig:dscvry1}, \ref{fig:dscvry2}, \ref{fig:absig1pol2}. If $a^0=1,\,b^0=1$ and $\bar a=-1,\,\bar b=0$, and we choose $m_{\psi} = 150\,\gev$, unpolarized beams, $ \epsilon = 0.005 $, and $\lum=567 \, \text{fb}^{-1}$, we find $ \Delta\sigma \simeq 9 $ with a  $\pm1.43$ uncertainty. Assuming now $m_{\psi} = 245\,\gev$, polarized beam ($P_{e^\pm} = ^{+20\%}_{-80\%}$), $ \epsilon = 0.001 $, and $\lum=567 \,  \text{fb}^{-1}$ and taking $a^0=0,\,b^0=0$ as the base model, we find that when $\bar a=1,\,\bar b=0$ (the purely vector-like case) $ \Delta\sigma = 13.96 $ with $\pm1.05$ uncertainty; w if  $\bar a=1,\,\bar b=1$ we find $ \Delta\sigma = 14.09$ with an uncertainty of $\pm1.07$. We do not consider the  $a^0=0, b^0=\pm 1$ cases since the  $ \Delta\sigma$ distribution is not normal, and a full analysis statistical analysis of the $\Delta\sigma$ statistics lies beyond the scope of this paper; however, we expect that the uncertainties in these cases will continue to be $O(\lesssim10\%)$.

It is worth noting that, as expected, larger efficiency $ \epsilon$ and luminosity $ \lum $ increases the significance, while larger masses reduce it. It is also important to note that though the significance $ \Delta\sigma $ depends on the magnitude of the cross section of the base model, this is not the only factor. For the example considered, the unpolarized cross section of the base model $a^0=b^0=0$ is smaller than the one for  polarized beams by about 30\%, yet the significance of the $ \bar a=0,\,|\bar b|=1 $ models is the same, while that of $ \bar a=1,\,\bar b=0 $ is larger than expected from the cross-section alone.

\begin{center}
\begin{table}
$$		\begin{array}{| c | c | c | c | c | c | c | c | c | c| c|} 
\hline
\multicolumn{1}{|c|}{}&
\multicolumn{1}{|c|}{}&
\multicolumn{1}{|c|}{} &
\multicolumn{4}{|c|}{\text{significance($\Delta\sigma$)\,}} \\
\cline{4-7}
\multicolumn{1}{|c|}{\text{model}}&
\multicolumn{1}{|c|}{\epsilon}&
\multicolumn{1}{|c|}{\lcal_{\tt int}~[\text{fb}^{-1}]}&
\multicolumn{2}{|c|}{\text{$m_{\psi^\pm}=150$ GeV}}&
\multicolumn{2}{|c|}{\text{$m_{\psi^\pm}=245$ GeV}}\\
\cline{4-7}
&& & P_{e^\pm}=0  &P_{e^\pm}=^{+20\%}_{-80\%} & P_{e^\pm}=0  &P_{e^\pm}=^{+20\%}_{-80\%}\\ 
\hline
\multirow{3}*{\minitab[l]{$\bar a=1$ \\ $\bar b=0$}}&0.001&567 &6.13&25.03 &3.42&13.96\\
&0.001&2000&11.51&47.03 &6.62&26.22\\
&0.005&567 &13.72&55.98 &7.65&31.21\\
\hline
\multirow{3}*{\minitab[l]{$\bar a=1$ \\ $\bar b=\pm1$}}&0.001&567 &11.46&29.81 &3.58&14.09\\
&0.001&2000&21.52&56.00 &6.73&26.46\\
&0.005&567 &25.62&66.67 &8.01&31.50\\
\hline
\multirow{3}*{\minitab[l]{$\bar a=0$ \\ $\bar b=\pm1$}}&0.001&567 &7.07&7.09  &0.68&0.69\\
&0.001&2000&13.28&13.32  &1.28&1.29\\
&0.005&567 &15.81&15.86  &1.52&1.54\\
\hline
\multirow{3}*{\minitab[l]{$\bar a=-1$ \\ $\bar b=0$}}&0.001&567 &8.65&10.81  &4.82&6.01\\
&0.001&2000&16.24&20.30 &9.05&11.28\\
&0.005&567 &19.33&24.17&10.78&13.44\\
\hline
\multirow{3}*{\minitab[l]{$\bar a=-1$ \\ $\bar b=\pm1$}}&0.001&567 &14.15&14.92  &4.99&5.92\\
&0.001&2000&26.57&28.02 &9.36&11.12\\
&0.005&567 &31.63&33.36&11.15&13.24\\
\hline
\end{array} $$
\caption{Statistical significance $\Delta \sigma $ (see Eq. (\ref{eq:signif-def})) of hypotheses $\bar a,\,\bar b$ with respect to the base hypothesis $a^0=b^0=0$.}
\label{table:discovery1}
\end{table}
\end{center}

\section{Model example}
\label{sec:DMmodel}

The above analysis focused on the application of the OOT to the study and detectability of the properties of a hypothetical new heavy lepton. 
In this section we turn to a possible underlying economical and UV complete model that contains such a particle. This model provides a viable theoretical underpinning of the previous discussion, a framework for 
studying other aspects of its detectability at the ILC, and can be used to obtain an estimate of the efficiency $ \epsilon $ ({\it cf.} Eq. (\ref{chi2})). 
In addition, the study of this model using event-level simulation allows for a comparison of the expected ILC sensitivity to the optimal statistical 
uncertainties derived above. Finally, we will see that the model proposed contains a viable dark matter candidate, satisfying the relic-density, 
direct-search and electroweak constraints in a large region of parameter space.

The model consists of an extension of the SM by two vector-like leptons: a weak iso-doublet, $\psi=(\psi^0 ,\, \psi^-\, )$ of hypercharge $-1$, and an iso-singlet $\chi$ of zero hypercharge; both $ \psi$ and $ \chi $ are  odd under an exact $\zBB_2$ symmetry under which all the SM fields are even \cite{Bhattacharya:2015qpa, Bhattacharya:2018fus}. Upon electroweak symmetry breaking (EWSB) the $\psi\chi H$ Yukawa coupling (see Eq.~(\ref{lag:lagVF}) below) generates a mixing between the neutral component $\psi^0$ and  $\chi$, the resulting lighter mass eigenstate will be  odd under $\zBB_2$ and therefore stable, and serves as a DM candidate. The quantum numbers under the SM$\times \zBB_2$ symmetry are summarized in Table \ref{tab:tab5}.

\begin{table}[htb]
	\begin{center}
\begin{tabular}{|c||c|c|c|c|}
\hline
field & $\su3_{\tt C}$ & $\su2_{\tt L}$ & $\ui_{\tt Y}$ & $\zBB_2$ \cr \hline
$\psi$ & $1$ & $2$ & $-1$ & odd \cr
$\chi$ & $1$ & $1$ & $0 $ & odd \cr \hline
\end{tabular}
\caption{Quantum numbers of the additional dark-sector fermions under $ \su3_{\tt C} \times \su2_{\tt L} \times \ui_{\tt Y}  \times \zBB_2 $. SM fields have the usual gauge quantum numbers and are even under $ \zBB_2$.}
\label{tab:tab5} 
\end{center}
\end{table}

The Lagrangian of the model is
\beq
\label{lag:lagVF}
\mathcal{L}^{\tt VF} = \bar\psi \left[ i \left(\slashed\partial - i \frac g2 \sigbf\cdot\slashed\WW - i \frac{g'}2 \slashed B\right)-m_{\psi^\pm} \right]\psi + \bar\chi \left(i \slashed\partial-m_{\chi} \right)\chi - \left(Y_1\bar\psi \widetilde{H}\chi + \text{H.c} \right)\,,
\eeq
(plus the usual SM terms); $H$ denotes the SM Higgs isodoublet, $ \WW_\mu$ and $ B$ the $\su2_L$ and $\ui_Y$ gauge fields, respectively, and $g,\,g'$ the corresponding gauge couplings.

After electroweak symmetry breaking  $H$ acquires a \vev\ $v/\sqrt{2}$:
\beq
H \to \frac{ v + h }{\sqrt{2}} \bpm1 \cr 0 \epm,
\label{eq:hssb}
\eeq 
and, as noted above, the $ \chi$ and $ \psi_0 $ will mix through the Yukawa interaction $ \propto Y_1 $. The mass Lagrangian then becomes
\beq
-\mathcal{L}_{\tt mass} =  \left( \bar\chi , \bar\psi^0 \right) \bpm m_\chi & \mu \cr \mu & m_{\psi^\pm} \epm \bpm \chi \cr \psi^0 \epm +m_{\psi^\pm}{\psi^+}\psi^- ; \qquad \mu = \frac{ Y_1 v}{\sqrt{2}}.
\eeq
The mass eigenstates $ \psi_{1,2} $ are then given by
\beq
\bpm \chi \cr \psi^0 \epm = \bpm \cg & - \sg \cr \sg & \cg \epm \bpm \psi_1 \cr \psi_2 \epm \,; \qquad \ttg = \frac{2\mu}{m_\chi - m_{\psi^\pm}}.
\label{ref:mixang}
\eeq
We will assume~\footnote{The case $ |\mu| < m_{\chi} \ll m_{\psi^\pm} $ is excluded by DM direct-detection and relic abundance constraints.} $ |\mu| \ll m_{\chi} < m_{\psi^\pm} $ so that $2\mu \ll |m_{\chi} - m_{\psi^\pm}|$; in this case $ \gamma $ is small and
\beq
m_{\psi_1}  \simeq  m_\chi - \frac{\mu^2}{m_{\psi^\pm} - m_\chi}\,, \qquad m_{\psi_2}  \simeq  m_{\psi^\pm} + \frac{\mu^2}{m_{\psi^\pm} - m_\chi};
\eeq
so that $ m_{\psi_2} > m_{\psi^\pm}> m_{\psi_1} $ and $ \psi_1 $ is the DM candidate. Note that we also have
\beq
Y_1 = - \sin(2\gamma) \frac{\Delta m}{\sqrt{2}\, v}\,, \qquad \Delta m = m_{\psi_2} - m_{\psi_1} > 0. 
\label{ref:reltn}
\eeq

In the mass-eigenstate basis the interaction Lagrangian becomes
\bal
\mathcal{L}^{\tt VF}_{int} =&  \frac{e_0}{\stw}\left[ s^2_\gamma (\bar {\psi_1} \gamma^\mu  \psi_1) + c^2_\gamma (\bar {\psi_2}\gamma^\mu  \psi_2) + s_\gamma c_\gamma (\bar {\psi_1} \gamma^\mu \psi_2 + \bar {\psi_2} \gamma^\mu \psi_1) - \ctw ({\psi^+}\gamma^{\mu}\psi^-)\right] Z_{\mu}  \cr
& - e_0 ({\psi^+}\gamma^{\mu}\psi^-)A_{\mu} +\frac{e_0}{\sqrt2\sw}\left\{ \left[ s_\gamma (\bar {\psi_1}\gamma^\mu \psi^-) + c_\gamma (\bar {\psi_2}\gamma^\mu  \psi^-) \right]W_\mu^+ + \text{H.c.} \right\}  \cr
& -\frac{Y_1}{\sqrt2}h \left[s_{2\gamma} (\bar {\psi_1}{\psi_1}-\bar {\psi_2}\psi_2)+ c_{2\gamma}(\bar {\psi_1}\psi_2+\bar {\psi_2}\psi_1)\right],
\label{eq:interaction}
\end{align}
where $ s_\gamma = \sin\gamma$, etc., and $h$ is defined in Eq. (\ref{eq:hssb}).
We see that the charged heavy fermions (${\psi^\pm}$) have vector-like interactions with $Z$ boson (corresponding $a^0=\ctw\sim1/2,\,b^0=0$ in Eq.~\eqref{eq:vertex1}); the $W$ couplings are also vector-like. Comparing with Table \ref{table:abe1} we that for $ \epsilon =0.005$ and polarized beams we expect~\footnote{The case at hand is similar to $ a^0=1,\,b^0=0$.} the ILC to be able to measure $a^0,\,b^0 $ to within $ \lesssim 10\% $ at $ 1 \sigma $ (ignoring systematic uncertainties).

The strongest limits on the model parameters come from dark matter constraints. The interactions in Eq.~\eqref{eq:interaction} show that the DM relic density is determined by the $h$ and $Z$-mediated annihilation and co-annihilation channels, while nuclear scattering, probed by direct-search experiments, is dominated by the $Z$ exchange process alone. The experimental constraint on the spin-independent cross section $ \sigma^{\tt SI}_{\tt dir.\,det.}  \lesssim 10^{-47} \rm{cm}^2 $  (XENON1T collaboration, \cite{Aprile:2018dbl}), and the fact that this cross section is $ \propto \sin^4\gamma $ gives
\beq
\sin\gamma \lesssim 0.05 ~;
\eeq
with a weak dependence on the DM mass. This limit on $\sin\gamma$ sharply reduces DM annihilation cross-section via $Z$ mediation, and also via Higgs portal interactions since $Y_1 \propto \sin(\gamma) $. Though the SM$\to$DM annihilation channels are suppressed, the relic-abundance restriction~\footnote{Here {\sf h} denotes the Hubble parameter in units of  $100\,$km\,s$^{-1}$\,Mpc$^{-1}$.} $\Omega_{\tt DM} {\sf h}^2=0.11933\pm 0.00091$ (PLANCK collaboration \cite{Ade:2013zuv, Aghanim:2018eyx}) can still be met through  co-annihilation channels involving $ \psi^\pm $, provided $ |\Delta m| \ll m_{\psi^\pm} $ \cite{Bhattacharya:2018cgx}.

\begin{figure}[htb!]
	$$ 
	\includegraphics[height=5.1cm]{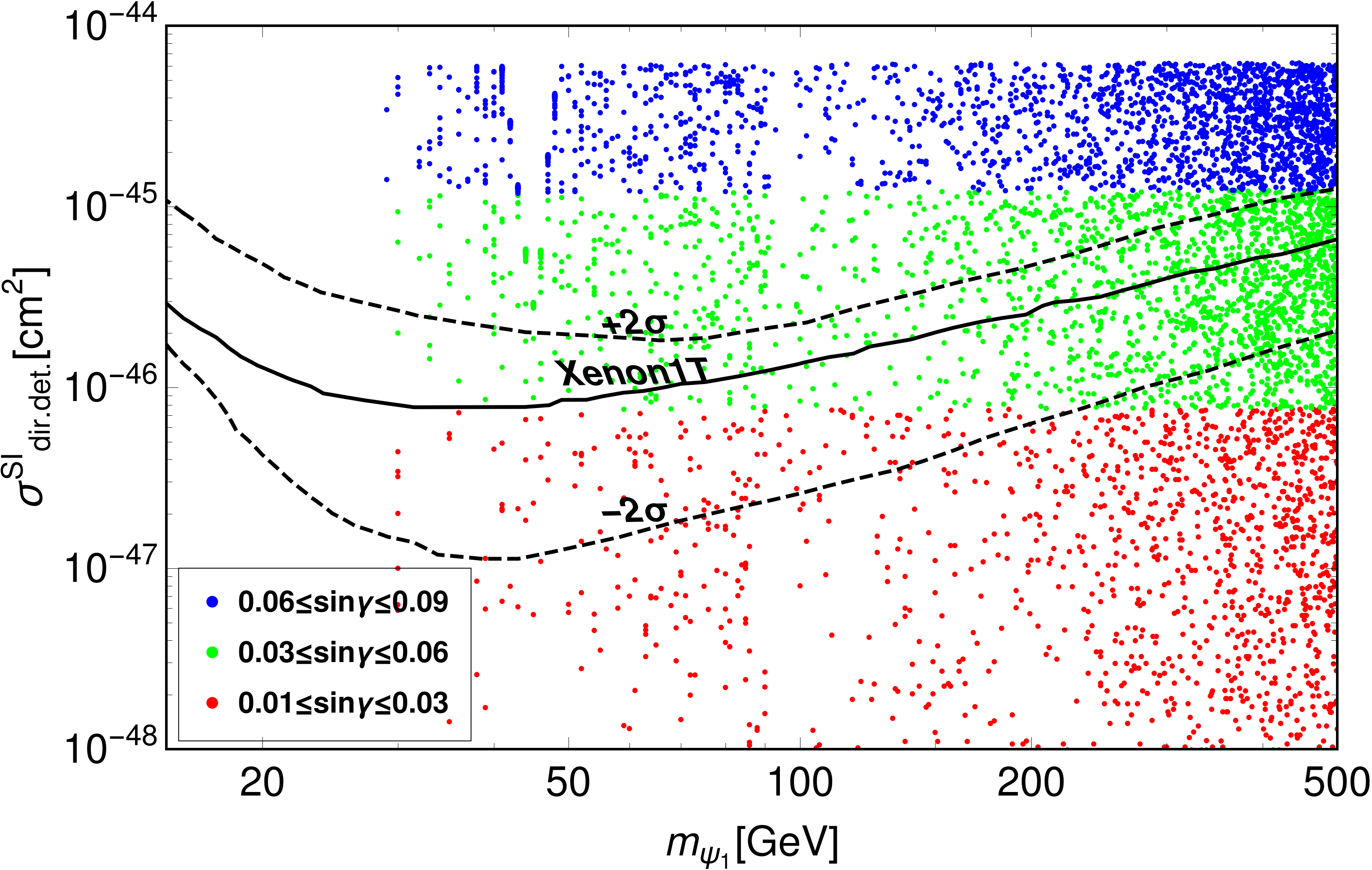} 
	$$
	$$
	\includegraphics[height=5.1cm]{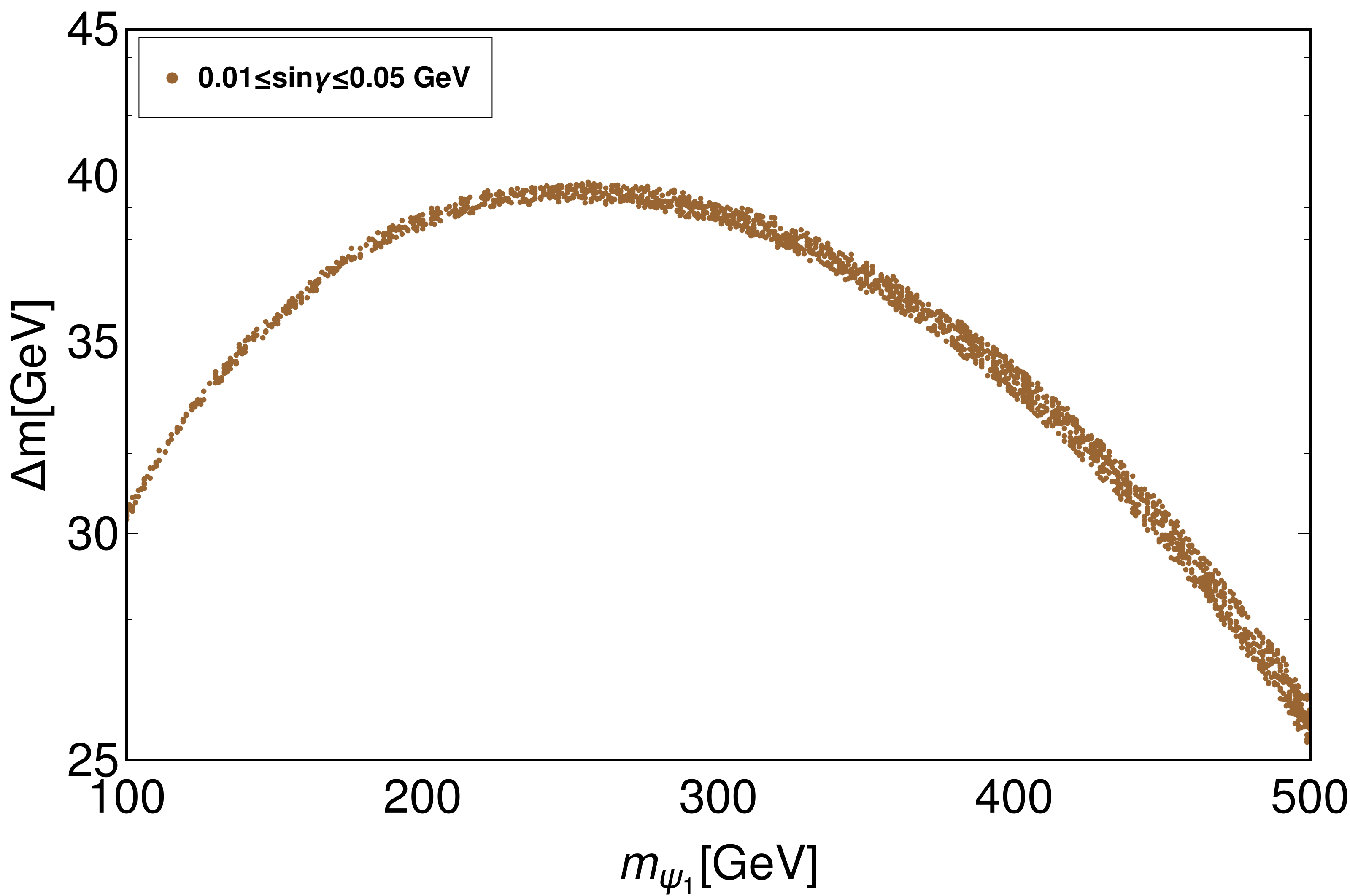} \quad
	\includegraphics[height=5.1cm]{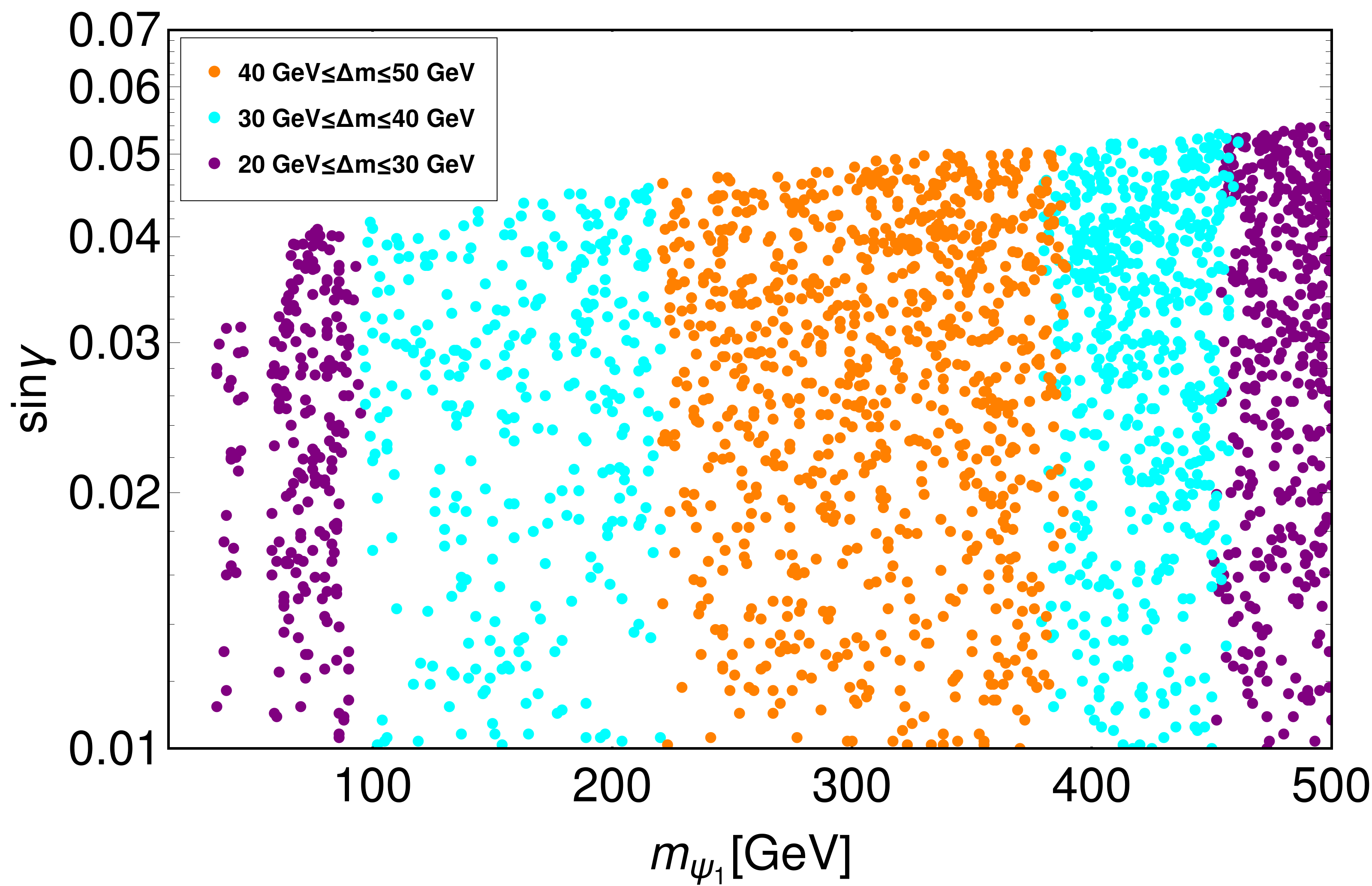}
	$$
	\caption{Regions allowed by the direct-detection and relic density constraints. Top: in the SI direct-detection cross-section ($\sigma^{\tt SI}_{\tt DD}$) vs DM mass ($m_{\psi_1}$) plane for various ranges of $ \sin \gamma$. Bottom left: in the $\Delta m -m_{\psi_1}$ plane for $0.01\le \sin\gamma \le0.05$; bottom right: in the $\sin \gamma - m_{\psi_1}$ plane for different ranges of $\Delta m$.}
	\label{fig:relicdd}
\end{figure}

Fig.~\ref{fig:relicdd} displays various regions allowed by the direct-detection and relic-density constraints. The top panel displays the spin-independent direct-detection cross-section $ \sigma^{\tt SI}_{\tt DD}$ a function of DM mass ($m_{\psi_1}$) for various ranges of $\sin\gamma$. The allowed region in the $ m_{\psi_1} - \Delta m $ plane is displayed in the bottom left panel, while the allowed region in the $ m_{\psi_1} - \sin\gamma $ plane for several ranges of $ \Delta m$ is displayed on the right bottom panel of that figure. The parabola-like region in the bottom left panel is responsible for having two allowed values of $ m_{\psi_1} $ for each choice of $ \Delta m $ range on the right bottom panel; this paraboloid shape can be traced to the contribution from co-annihilation channels $\psi^\pm \psi_1\to$SM to the DM annihilation cross section:
\beq
\vevof{ \sigma v}_{\tt tot} \simeq \vevof{ \sigma v}_{\psi_1\bar{\psi_1}\to \rm SM} + \vevof{ \sigma v}_{\psi_1\psi^\pm\to \rm SM} \left(1+\frac{\Delta m}{m_{\psi_1}} \right)^{3/2}e^{-\Delta m/T} +\cdots
\label{eq:coann}
\eeq
where $T$ denotes the temperature of the bath, and the ellipses indicate other co-annihilation channels ({\it e.g.} $ \psi^+ \psi^-\to$SM, $\psi_1\bar{\psi_2} \to$SM) with a stronger exponential suppression; for details, see \cite{Bhattacharya:2018fus}. The relic density is then
\beq
\Omega_{\tt DM} {\sf h}^2 =\left. \frac{1.09 \times 10^9 ~{\rm GeV}^{-1} (m_{\psi_1}/T)}{{g_*}^{1/2}M_{\tt Pl}}\frac{1}{\vevof{ \sigma v}_{\tt tot}} \right|_{T=T_f};
\eeq
where $x = T/m_{\psi_1}$, $g_*$ denotes the effective relativistic degrees of freedom, $M_{\tt Pl}$ the Planck mass, and $ T_f $ the value of $T$ at freeze-out. From these expressions it follows that for small $ \Delta m $ the allowed values increase with  $m_{\psi_1}$, but only up to a point beyond which $ \Delta m $ must drop to balance the  the exponential suppression in Eq.~\eqref{eq:coann}.

Collider data also impose constraints on this model, with the strongest limits from those on production of chargino pairs \cite{Aad:2019vnb}, or chargino and second neutralino production \cite{Aad:2019qnd}, in supersymmetric theories. Chargino pair production is the exact parallel of the one we  study below (see Fig. \ref{fig:signal} with  $ \psi^\pm $  replaced by charginos, and $ \psi_1 $ by neutralinos), in the limit where the chargino is wino-dominated and the sneutrinos are heavy~\footnote{The sneutrinos generate a $t$-channel graph not present in our model, and other contributions to the charginos generate chiral couplings to the $W$.}, and assuming the on-shell production of charginos dominates the cross section. These SUSY limits  give 
\beq
m_{\psi_2} + 115 \,\gev \gtrsim m_{\psi^\pm} \quad \text{for}~ 250\,\gev > m_{\psi^\pm} > 150 \,\gev
\eeq
 which, for $ m_{\psi^\pm} = 150 \, \gev \, (245\, \gev) $, requires $ m_{\psi_1} > 35\, \gev \, (111\, \gev) $.  The model is also consistent with electroweak precision observables, and with the invisible decay widths for the Higgs and $Z$ boson  whenever $m_{\psi_1}>m_h/2$, which we assume. 
  
\begin{figure}[htb!]
$$
\includegraphics[scale=0.405]{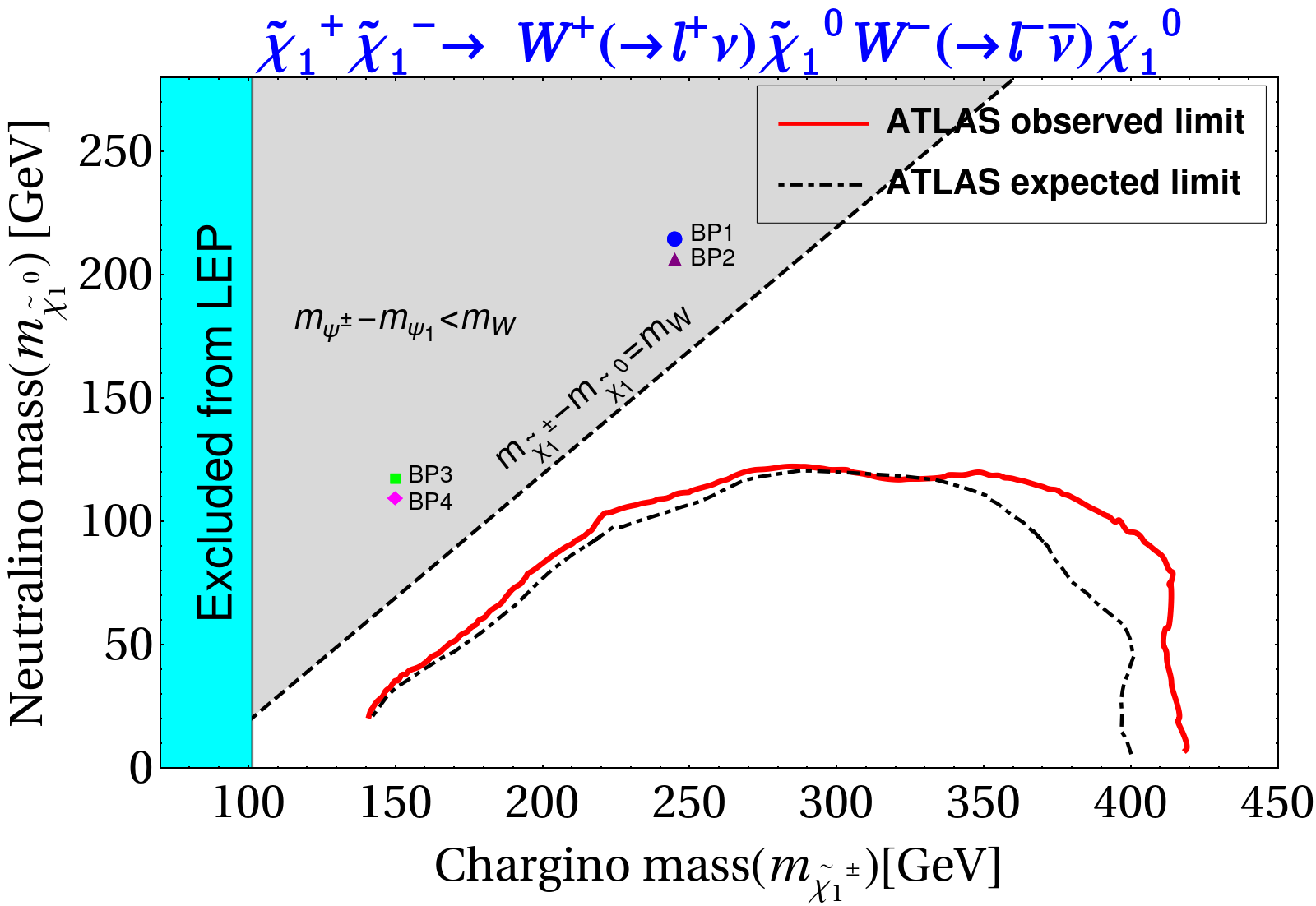}
\includegraphics[scale=0.4]{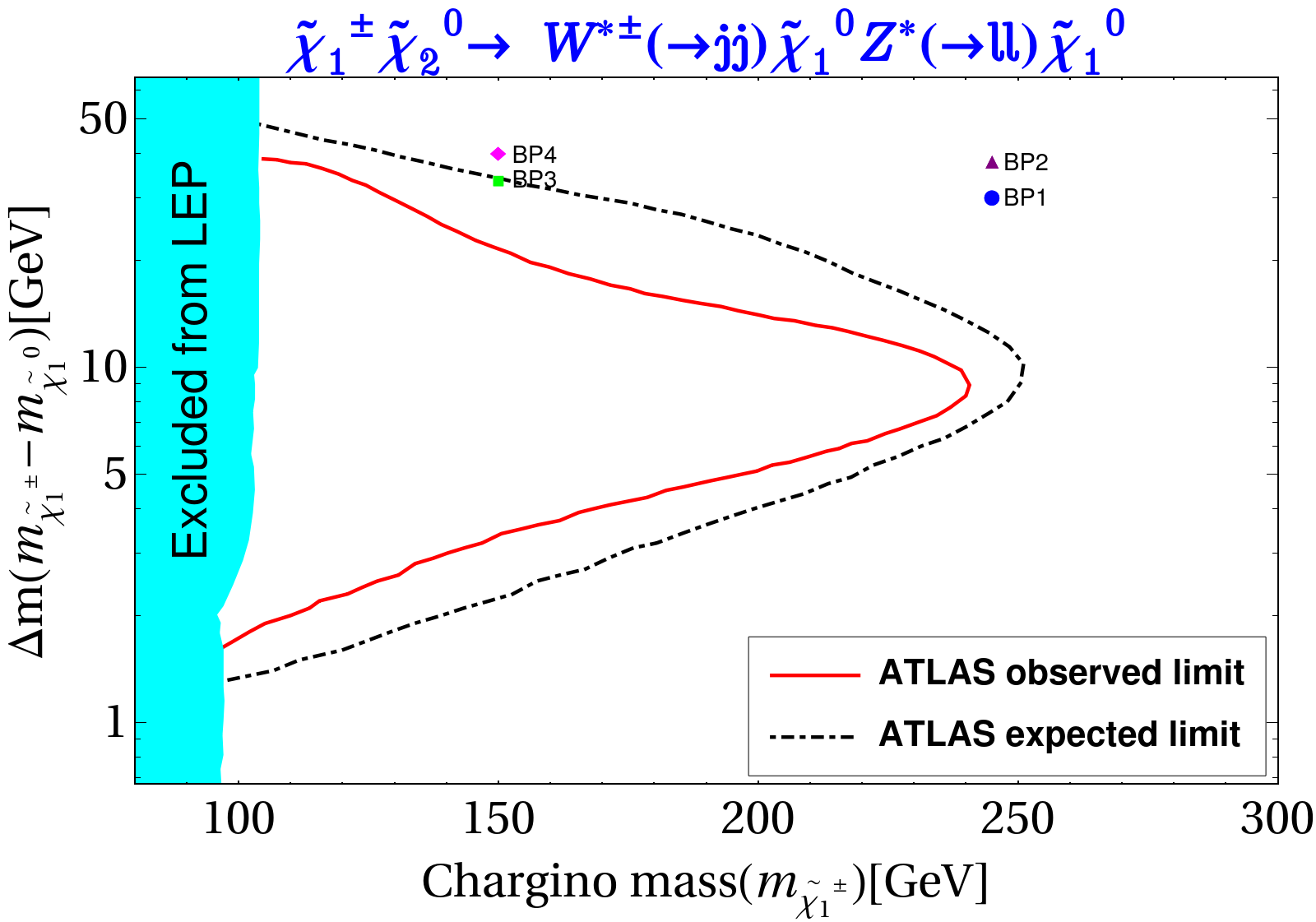}
$$
\caption{Current experimental limits from the LHC for supersymmetric chargino-neutralino production from the dilepton plus missing energy channel \cite{Aad:2019vnb} (left) and dilepton plus dijet plus missing energy channel \cite{Aad:2019qnd} (right). The benchmark points for our model (Table.~\ref{tab:BP}) are included for comparison.}
\label{fig:masslim}
\end{figure}

With these constraints in mind, we select several benchmark points, listed in Table \ref{tab:BP}, where all constraints are obeyed and which we will use in our study of the model at the ILC; for these we also assumed $\Delta m <m_W$, so that the decay of heavy fermion occurs via an off-shell $W$. We will show that for such relatively small mass splitting there is better segregation of the signal from the SM background at the ILC. The benchmark points are compared with limits from ATLAS \cite{Aad:2019vnb,Aad:2019qnd}~\footnote{The limits from CMS \cite{CMS:2020bfa} agree well with Fig.~\ref{fig:masslim} and also allow the chosen benchmark points.} in Fig.~\ref{fig:masslim}. 

\begin{table}[htb]
	\begin{center}
\begin{tabular}{| c | c | c | c | c | } 
\hline
Benchmark Points&$m_{\psi^\pm}$ (GeV)& $m_{\psi_1}$ (GeV) & $\Delta m$ (GeV)  \\ 
\hline
BP1&\multirow{2}*{\minitab[l]{$245$}}&215 & 30  \\
\cline{1-1}\cline{3-4}
BP2&&207 & 38 \\ 
\hline
BP3&\multirow{2}*{\minitab[l]{$150$}}&117 & 33 \\
\cline{1-1}\cline{3-4}
BP4&&110 & 40 \\
\hline
\end{tabular}
\caption{Benchmark points chosen for collider analysis for singlet-doublet fermion model;  in all cases we took $\sin\gamma=0.05$.}
\label{tab:BP}
\end{center}
\end{table}

\subsection{Simulation of collider events}
\label{sec:collider}

\begin{figure}[htb!]
	$$ 
	\includegraphics[height=5.2cm]{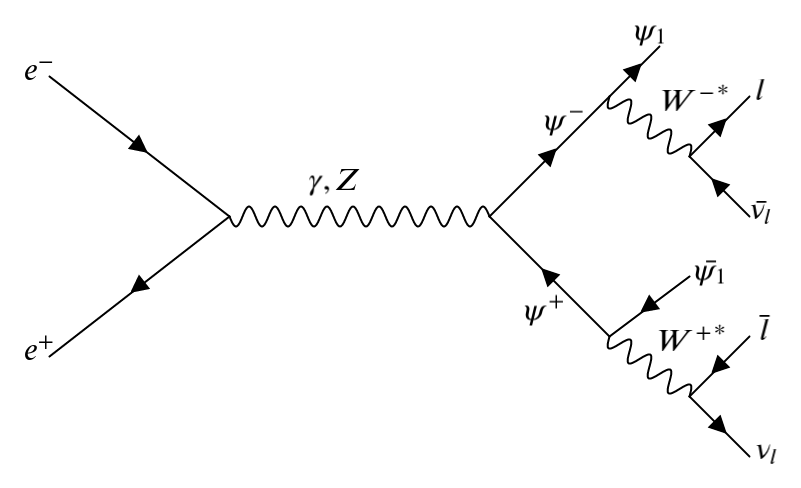} 
	$$
	\caption{Production and decay of the heavy charged fermions at the ILC for the model described in section \ref{sec:DMmodel}.}
	\label{fig:signal}
\end{figure}

We now turn to the ILC collider signatures for this model for the chosen benchmark points (table \ref{tab:BP}) using the simplest signal: $ \psi^\pm$ on-shell pair production with their subsequent decay into DM + opposite-sign leptons ({\tt OSL}) via off-shell $W$ bosons (Fig. \ref{fig:signal});  we adopt the above mass hierarchy, $ m_{\psi_2} > m_{\psi^\pm}> m_{\psi_1} $. We note that $ \psi_1 \, \psi_2 $ production, followed by $ \psi_2 \to Z \psi _1 $ generates a similar final state ({\tt OSL} plus missing energy), but the cross is $ \propto \sin^4\gamma $ and significantly smaller. It is also possible to pair produce $ \psi_1,\,\psi_2 $ but the final state signature is different.

We simulated {\tt OSL} events at the ILC with $\sqrt s=500$ GeV as follows: the model was implemented in {\sf Feynrules} \cite{Alloul:2013bka}, and parton-level signal events were generated using {\sf CalcHEP} \cite{Belyaev:2012qa}, and then showered and analyzed using {\sf Pythia }\cite{Sjostrand:2006za}; SM background events were generated using {\sf MadGraph} \cite{Alwall:2014hca} and showered using  {\sf Pythia}. For event reconstruction, we use the following criteria:
\begin{itemize}
	\item Leptons are required to have at least transverse momentum $p_T> 10$ GeV; we consider only electrons and muons with pseudorapidity $|\eta|< 2.4$ -- we do not consider $ \tau $ signals. 
	Two leptons are assumed isolated if $\Delta R_{\ell\ell}=\sqrt{(\Delta \eta)^2+(\Delta \phi)^2} \ge 0.2$, while a lepton and a 
	jet are assumed isolated if $\Delta R_{\ell j}\ge 0.4$.
	
	\item We impose a zero-jet requirement, where jets are reconstructed using the cone jet algorithm around initiating parton. We further require $p_T>20$ GeV and $|\eta|< 3.0$.

	\item Background signal was minimized by imposing cuts at $20$ and $30$ GeV (see below) on the missing transverse energy, which is defined by
	\beq
	\etm = \left| \pp\up{\tt vis}_\perp \right|^2
	\eeq
	where $\pp\up{\tt vis}_\perp$ is the total visible momentum perpendicular to the beam direction.
\end{itemize}

\begin{figure}[htb!]
	$$
	\includegraphics[scale=0.4]{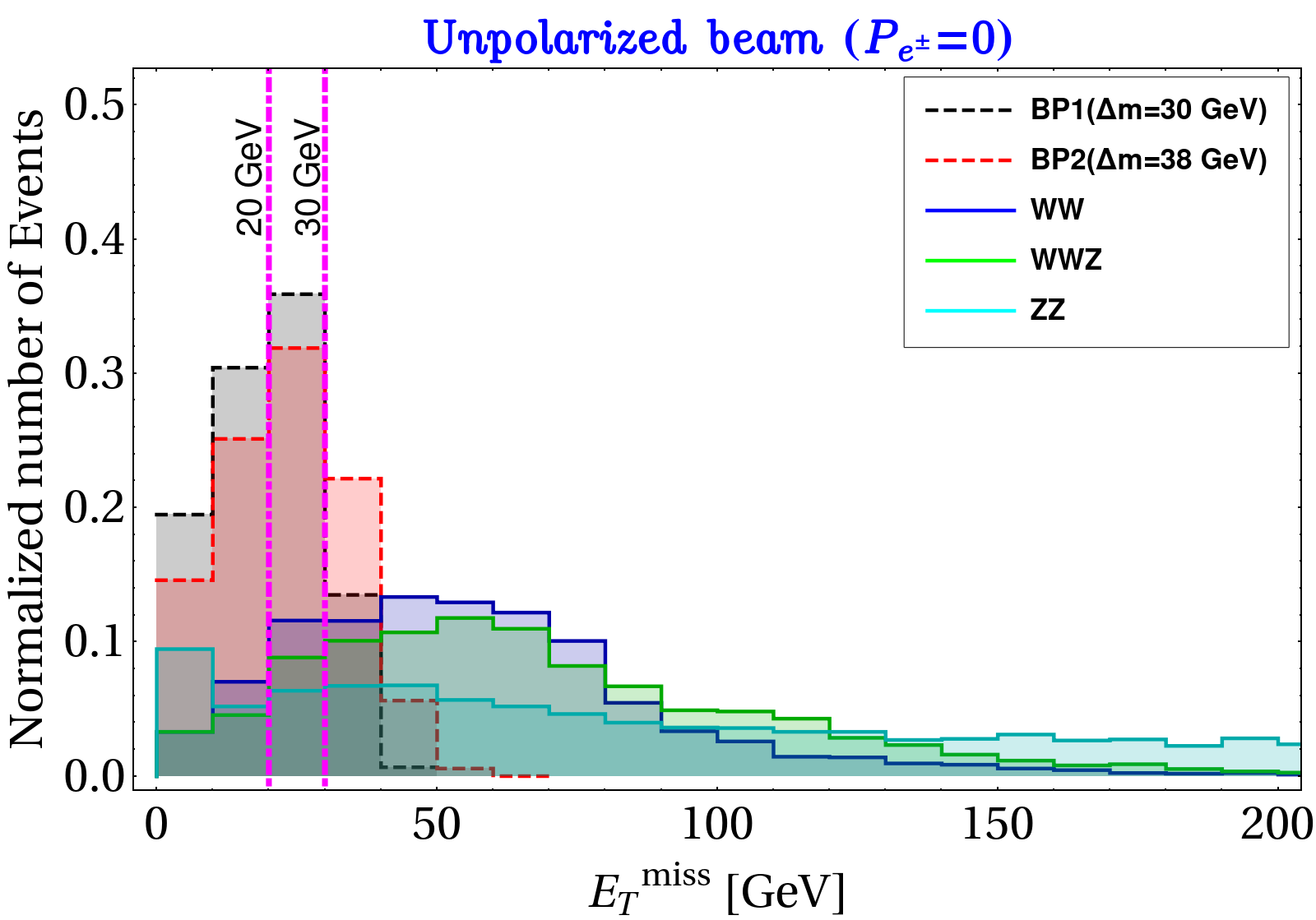}
	\includegraphics[scale=0.4]{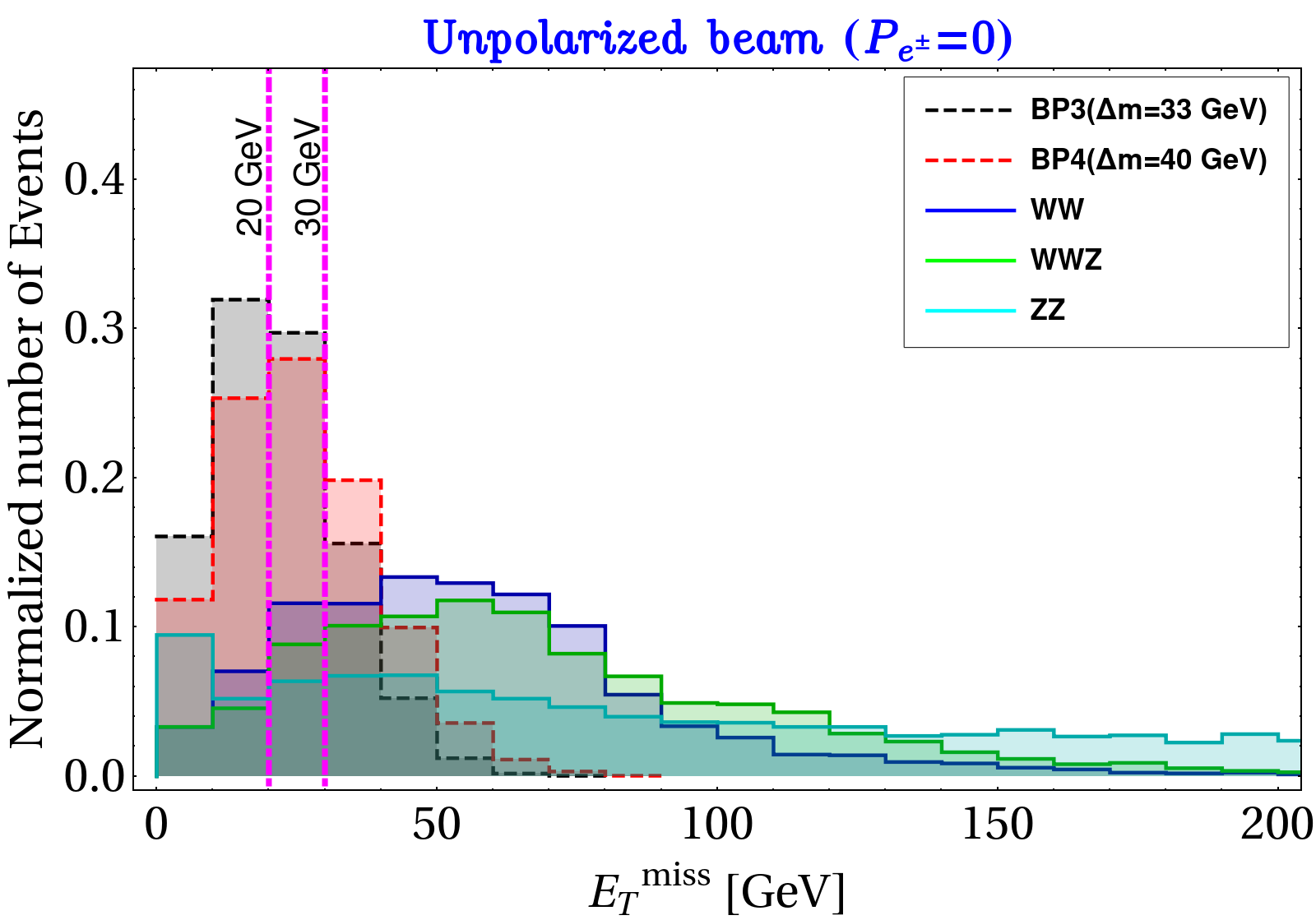}
	$$
	$$
	\includegraphics[scale=0.422]{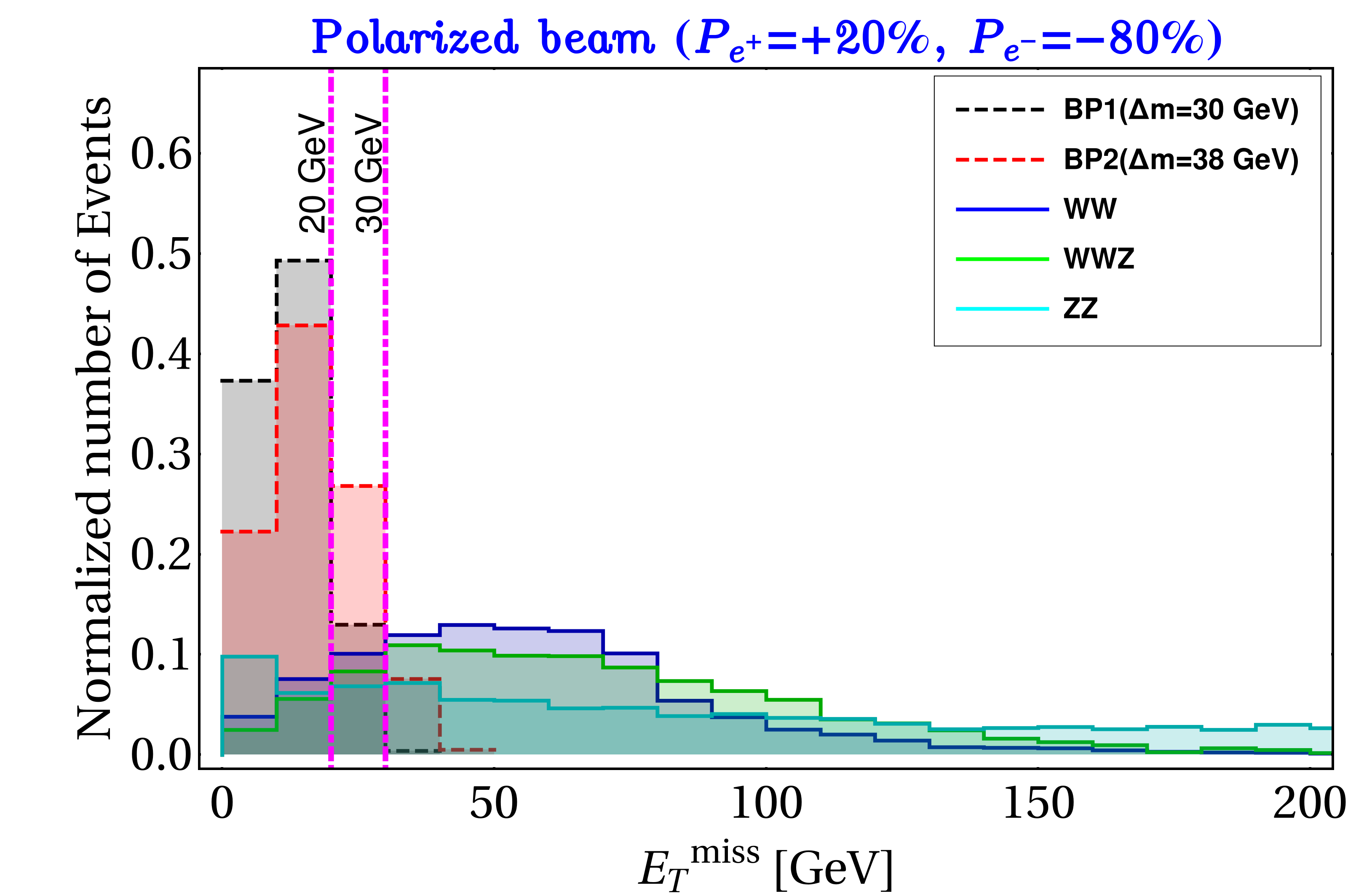}
	\includegraphics[scale=0.4]{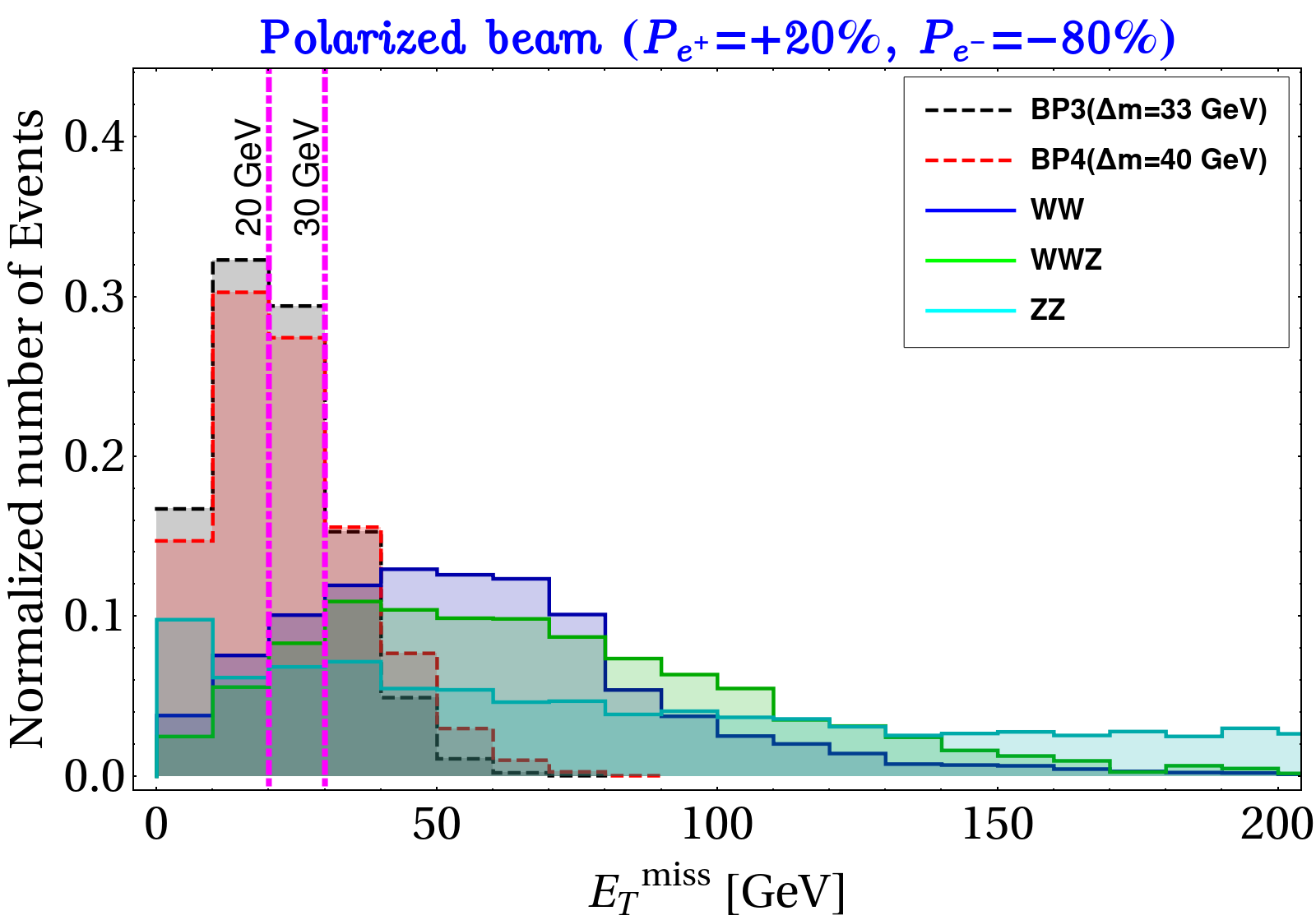}
	$$
	\caption{Normalized $\etm$ distribution for the {\tt OSL} final state events at the ILC with $\sqrt s$= 500 GeV. Top: results for unpolarized beams; left (right), benchmark points BP1, BP2 (BP3, BP4) ({\it cf.} Table \ref{tab:BP}). Bottom: same for polarized beams ($P_{e^\pm} = ^{+20\%}_{-80\%}$). The SM background distributions from $WW,\, WWZ,\, ZZ$ production are also shown. The cuts $\etm<20,\,30$ GeV cut used in the analysis are also indicated.}
	\label{fig:c10c21}
\end{figure}

\begin{table}[htb]
	\begin{center}
		\begin{tabular}{| c | c | c | c | c | c | c | } 
			\hline
			
			\multicolumn{1}{|c|}{\multirow{2}*{\minitab[l]{Background}}}&
			\multicolumn{2}{c|}{$\sigma^{\psi^+ \psi^-}_{\tt SM}$ [pb]} & 
			\multicolumn{1}{c|}{\multirow{2}*{\minitab[l]{$\etm$ [GeV]}} } &
			\multicolumn{2}{c|}{$\sigma^{\tt OSL}_{\tt SM} $ [fb]} \\ 
			\cline{2-3}
			\cline{5-6}
			\multicolumn{1}{|c|}{}&
			\multicolumn{1}{c|}{$ P_{e^\pm}=0$}&
			\multicolumn{1}{c|}{$P_{e^\pm}=^{+20\%}_{-80\%}$}&
			\multicolumn{1}{c|}{}&
			\multicolumn{1}{c|}{$ P_{e^\pm}=0$}&
			\multicolumn{1}{c|}{$P_{e^\pm}=^{+20\%}_{-80\%}$}\\
			\hline
			\multirow{2}*{\minitab[l]{$WW$}}& \multirow{2}*{\minitab[l]{$7.4 $}}&\multirow{2}*{\minitab[l]{$15.5$}} & $<20$ &0.90 &2.73   \\  
			& &   &$<30$ & 0.90&2.73 \\
			\hline
			\multirow{2}*{\minitab[l]{$WWZ$}} & \multirow{2}*{\minitab[l]{$0.04$}}& \multirow{2}*{\minitab[l]{$0.085$}}&$<20$  &$5.3\times10^{-4}$ &$1.5\times10^{-3}$   \\  
			& &  &$<$30 & $1.6\times10^{-3}$ &$4.5\times10^{-4}$  \\
			\hline
			\multirow{2}*{\minitab[l]{$ZZ$}} & \multirow{2}*{\minitab[l]{$0.41$}} & \multirow{2}*{\minitab[l]{$0.66$}}&$<20$ &$3.1\times10^{-3}$ &$9.4\times10^{-3}$  \\  
			& &   &$<30$ & $4.7\times10^{-3}$&$1.4\times10^{-2}$  \\
			\hline
		\end{tabular}
		\caption{SM background cross sections for the $ \psi^\pm$ and {\tt OSL} final states at the ILC with $\sqrt{s}$=500 GeV with the selection cuts adopted (see text), and for  unpolarized and polarized beams ($P_{e^\pm} = ^{+20\%}_{-80\%}$).}
		\label{bg1}
	\end{center}
\end{table}

\begin{table}[htb]
	\begin{center}
		\begin{tabular}{| c | c | c | c|c |c|c|c|c| } 
			\hline
			\multicolumn{2}{|c|}{$\sigma^{\psi^+ \psi^-}$[pb]} &
			\multicolumn{1}{c|}{\multirow{2}*{\minitab[l]{BPs}}} &
			\multicolumn{1}{c|}{\multirow{2}*{\minitab[l]{$\etm$ [GeV]}}}&  
			\multicolumn{2}{c|}{$\sigma^{\tt OSL}$[fb]} & 
			\multicolumn{2}{c|}{Efficiency ($\epsilon$)}\\
			\cline{1-2}
			\cline{5-8}
			\multicolumn{1}{|c|}{$ P_{e^\pm}=0$}&
			\multicolumn{1}{c|}{$P_{e^\pm}=^{+20\%}_{-80\%}$}&
			\multicolumn{1}{c|}{}&
			\multicolumn{1}{c|}{}&
			\multicolumn{1}{c|}{$ P_{e^\pm}=0$}&
			\multicolumn{1}{c|}{$P_{e^\pm}=^{+20\%}_{-80\%}$}&
			\multicolumn{1}{c|}{$ P_{e^\pm}=0$}&
			\multicolumn{1}{c|}{$P_{e^\pm}=^{+20\%}_{-80\%}$}\\
			\hline
			\multirow{4}*{\minitab[l]{$0.14$}} & \multirow{4}*{\minitab[l]{$0.32$}} & \multirow{2}*{\minitab[l]{BP1}} & $<20$&0.93 & 2.35 & $6.43\times10^{-3}$&$6.70\times10^{-3}$  \\
			& &  &  $<30$  & 0.98 &2.50 &$6.98\times10^{-3}$&$7.11\times10^{-3}$\\ 
			\cline{3-8}
			& & \multirow{2}*{\minitab[l]{BP2}} &$<20$ & 0.70 &1.80 & $5.00\times10^{-3}$&$5.62\times10^{-3}$ \\
			& & &  $<30$    &  0.73 & 1.88&$5.21\times10^{-3}$ &$5.87\times10^{-3}$\\
			\hline
			\multirow{4}*{\minitab[l]{$0.45$}} & \multirow{4}*{\minitab[l]{$1.13$}} &\multirow{2}*{\minitab[l]{BP3}} & $<20$  &2.97 & 7.92& $6.60\times10^{-3}$ &$7.00\times10^{-3}$ \\  
			& &    &  $<30$  &3.32 &   8.45      & $7.37\times10^{-3}$ &$7.47\times10^{-3}$ \\
			\cline{3-8}
			&& \multirow{2}*{\minitab[l]{BP4}} &$<20$ & 2.30 & 6.05 & $5.09\times10^{-3}$ &$5.36\times10^{-3}$ \\
			& &    &  $<30$  & 2.40 & 6.32 &$5.30\times10^{-3}$&$5.60\times10^{-3}$ \\
			\hline
		\end{tabular}
		\caption{Signal ({\tt OSL}) and $ \psi^\pm$ pair production cross sections and associated signal efficiency $\epsilon$ Eq.~(\ref{eq:eff.def}) at the ILC for $\sqrt{s}=500$ GeV with two missing-energy selection cuts and other selection cuts (see text) for unpolarized and polarized beams ($P_{e^\pm} = ^{+20\%}_{-80\%}$).}
		\label{tab:efficiency}
	\end{center}
\end{table}

\begin{table}[htb]
	\begin{center}
		\begin{tabular}{| c | c | c | c | c | c | c | c|c | } 
			\hline
			
			\multicolumn{1}{|c|}{\multirow{1}*{\minitab[l]{Benchmark}}}&
			\multicolumn{1}{c|}{\multirow{1}*{\minitab[l]{$\etm$}}}&
			\multicolumn{6}{c|}{$\sigma^{\tt signal}$ [fb]}\\
			\cline{3-8}
			\multicolumn{1}{|c|}{ Points}&
			\multicolumn{1}{|c|}{[GeV]}&
			\multicolumn{2}{c|}{OSL +  0 photon }&
			\multicolumn{2}{c|}{OSL+ $\le$ 1 photon}&
			\multicolumn{2}{c|}{OSL+ $\le$ 2 photon} \\
			\cline{3-8}
			\multicolumn{1}{|c|}{}&
			\multicolumn{1}{|c|}{}&
			\multicolumn{1}{|c|}{$ P_{e^\pm}=0$}&
			\multicolumn{1}{c|}{$P_{e^\pm}=^{+20\%}_{-80\%}$}&
			\multicolumn{1}{|c|}{$ P_{e^\pm}=0$}&
			\multicolumn{1}{c|}{$P_{e^\pm}=^{+20\%}_{-80\%}$}&
			\multicolumn{1}{|c|}{$ P_{e^\pm}=0$}&
			\multicolumn{1}{c|}{$P_{e^\pm}=^{+20\%}_{-80\%}$}\\
			\hline
			 	\multirow{2}*{\minitab[l]{BP1}} & $<20$ & 0.76& 2.17 & 0.84& 2.26 & 0.93 &2.35  \\ 
			      & $<30$ & 0.80& 2.31 & 0.87& 2.39 & 0.98 &2.50  \\ 
			\hline
			\multirow{2}*{\minitab[l]{BP2}} & $<20$ & 0.55& 1.79 & 0.62& 1.76 & 0.70 &1.80  \\ 
			& $<30$ & 0.60& 1.71 & 0.67 & 1.79 & 0.73 &1.88  \\ 
			\hline
			\multirow{2}*{\minitab[l]{BP3}} & $<20$ & 2.82& 7.70 & 2.90& 7.81 & 2.97 &7.92  \\ 
			& $<30$ & 2.15& 8.21 & 2.24 & 8.33 & 3.32 &8.45  \\
			\hline
			\multirow{2}*{\minitab[l]{BP4}} & $<20$ & 2.17& 5.87 & 2.23& 5.95 & 2.30 & 6.05  \\ 
			& $<30$ & 2.26& 6.16 & 2.34 & 6.23 & 2.40 & 6.32 \\
			\hline
		\end{tabular}
		\caption{Signal cross-section including ISR and FSR photon count.}
		\label{tab:sig}
	\end{center}
\end{table}

We present in Fig.~\ref{fig:c10c21} the missing transverse energy ($\etm$) distributions (normalized to one event) at the benchmark points for both the signal and 
the  dominant SM background processes events ($WW,\,WWZ,\,ZZ$); and for both polarized and unpolarized beams.  Since the intermediate $W$ bosons are off 
shell, the peak of the missing energy distribution for the signal is at a much smaller value than those SM background, where $W$ production is on-shell. 
Based on these distributions we choose upper cuts  $\etm < 20,\,30\,\gev$\footnote{ILC projections indicate this collider will be able to measure missing energy very accurately, so that the cut used in our analysis is viable \cite{Behnke:2013xla}.}, which retain a significant part of the signal and eliminate most background events, as illustrated in Table \ref{bg1}.

The values of the $ \psi^\pm$ pair production and signal event cross sections ($\sigma^{\psi^+\psi^-}$ and $\sigma^{\tt OSL}$, respectively) after imposing the above selection criteria and $\etm$ cuts are given in Table~\ref{tab:efficiency} for both unpolarized and polarized ($P_{e^\pm}=^{+20\%}_{-80\%}$) beams. Following the discussion in Sect. \ref{sec:method} we define
\bea
\epsilon=\frac{\sigma^{\tt OSL}}{\sigma^{\psi^+\psi^-}}~, 
\label{eq:eff.def}
\eea
(see Eq. \eqref{eq:eps.def}) whose values are listed in Table~\ref{tab:efficiency}. It is clear that $\epsilon\sim 0.001$ used throughout the previous OOT analysis is a conservative choice ($\epsilon\sim 0.005$ more closely corresponds to the results derived in this section). 

We also note that signal events are often accompanied with initial state radiation (ISR) and final state radiation (FSR) photons simulated with inbuilt functions in {\tt Pythia} event generator. Using the photon selection criteria  $p^{\tt photon}_T>5\,\gev$  and $|\eta_{\tt photon}|<0.24$,  the signal cross-section with zero photons, and the inclusive $\le1,\,2$ photon cross sections are listed in table \ref{tab:sig}. From this we can see that the inclusive diphoton cross sections match quite accurately the signal cross-section without photon tagging listed in Table \ref{tab:efficiency}\footnote{With three photon and four photon events being very rare, inclusive di-photon event counts match quite accurately to signal cross-section without photon tagging as in Table~\ref{tab:efficiency}.}.

\begin{figure}[htb!]
	$$
	\includegraphics[scale=0.4]{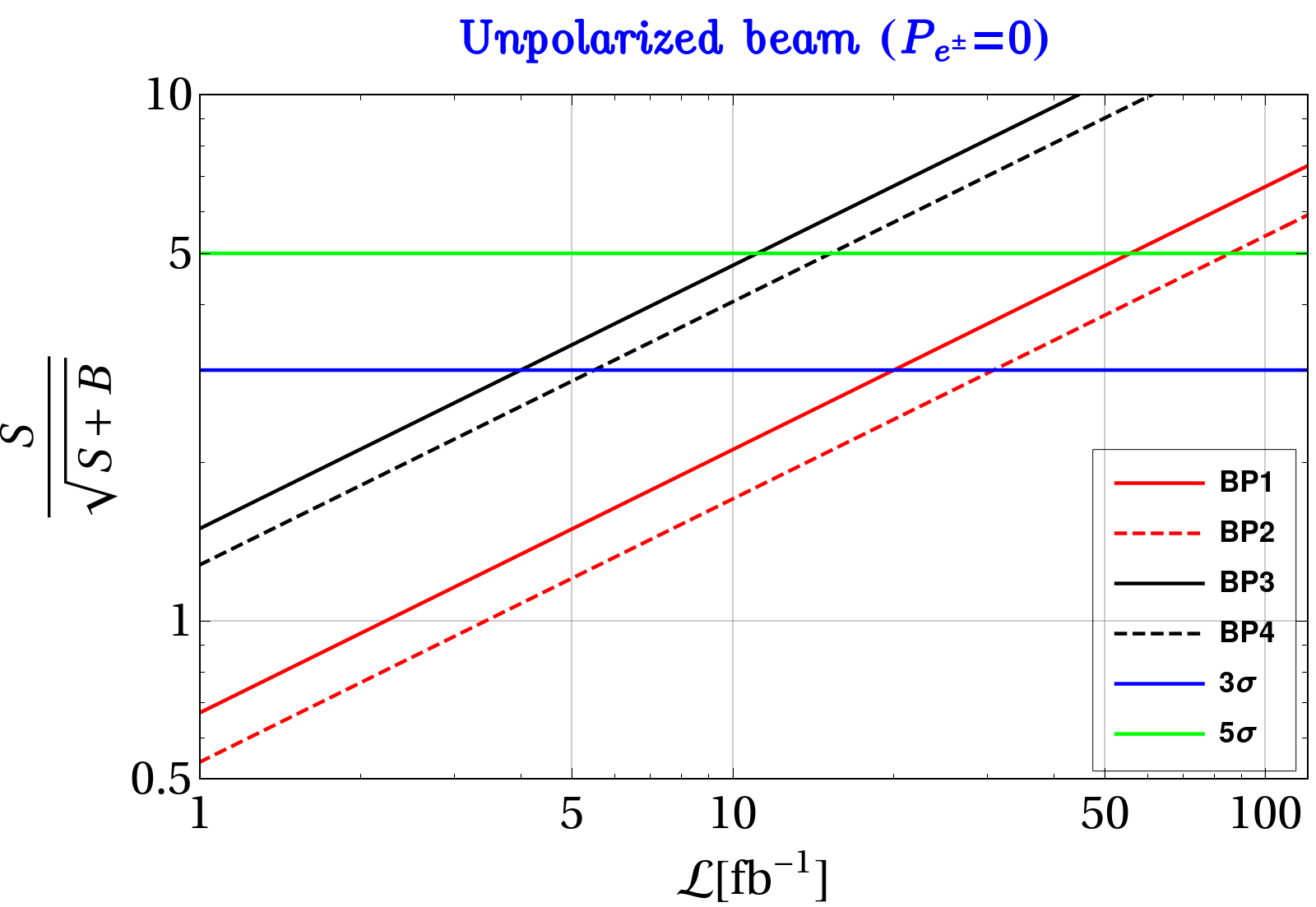}
	\includegraphics[scale=0.4]{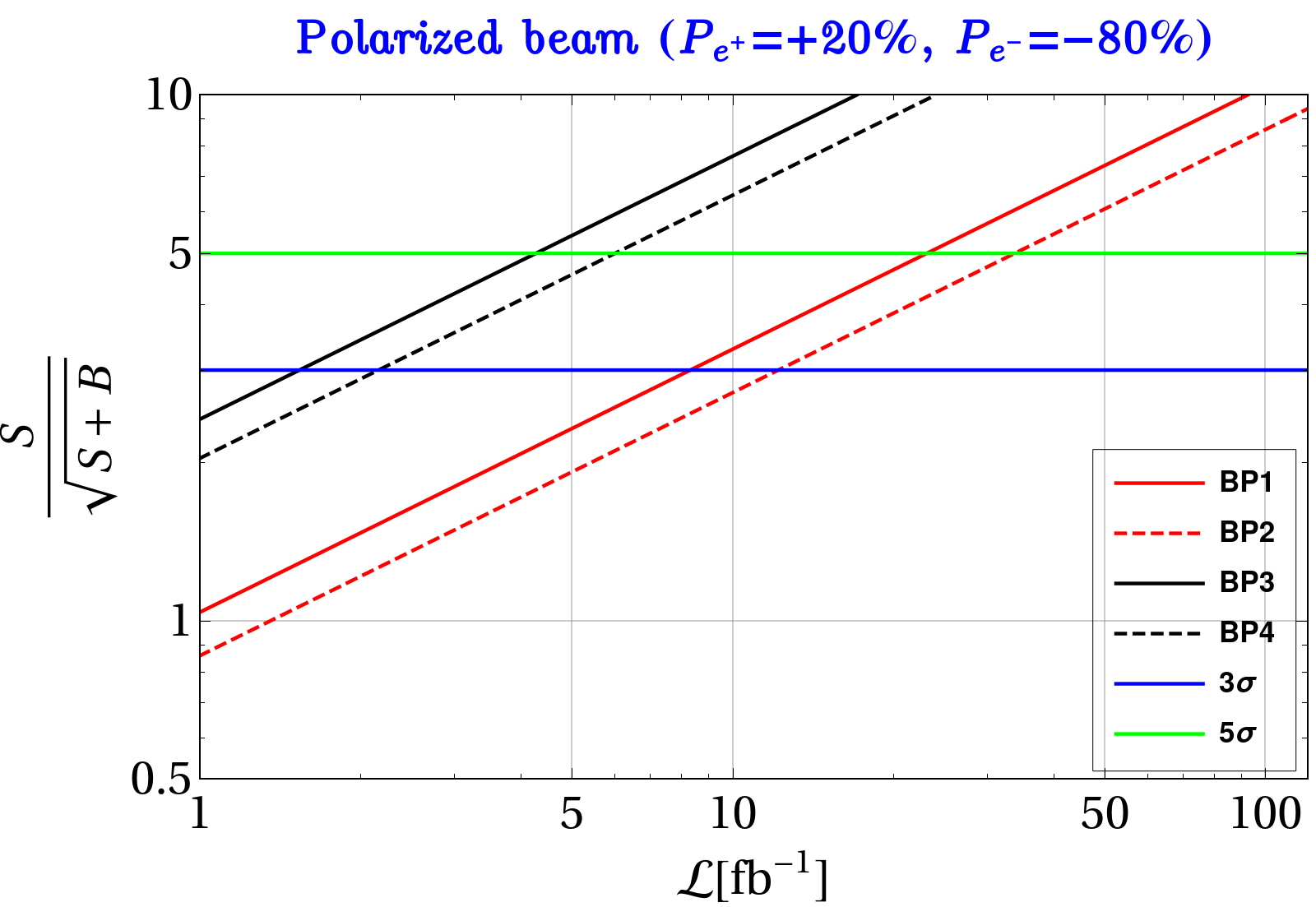}
	$$
	\caption{ Signal significance ($S/\sqrt{S+B}$) for the benchmark points in Table \ref{tab:BP} for unpolarized (left) and polarized  ($P_{e^\pm}=^{+20\%}_{-80\%}$) beams (right). Blue and green lines correspond to $3\sigma$ and $5\sigma$ exclusion and discovery limits, respectively.}
	\label{fig:sig1}
\end{figure}

We determine the discovery potential of the {\tt OSL} signal at the ILC by plotting signal significance ($S/\sqrt{S+B}$, where $S$ and $B$ denote, respectively the number of signal and SM background events), as a function of luminosity $\lcal $. The results are presented in Fig.~\ref{fig:sig1}; of particular interest is the advantage provided by using polarized beams, which require a lower luminosity for either the discovery or exclusion of the selected signal; in either case the design luminosity ({\it cf.} Eq. (\ref{eq:params})) will be sufficient to exclude or detect the NP signal here investigated.

It is intriguing to investigate whether the optimal uncertainty of the NP parameters obtained in the preceding section can be realized in a collider environment given the SM background contribution as analyzed in model specific scenario here. The experimental determination of model parameters will depend on the choice of signatures, the selection criteria and the corresponding significance; the authors are unaware of such a study, but the required dedicated analysis lies beyond the scope of the present paper. We will return to this issue in a future publication.

\section{Summary and Conclusions}
\label{sec:conclusion}

In this paper, we have analyzed the optimal statistical determination of the parameters of physics beyond the Standard Model, using as a specific example the production of a new heavy charged fermion $ \psi^\pm$ that can couple to the $Z$ boson and photon. We assumed for simplicity that the photon coupling is known, and allowing for both vector and axial coupling to the $Z$ with couplings $a$ and $b$ respectively. The optimal observable technique generates the minimal statistic uncertainty to which the couplings $a,\,b$ can be determined. We find that uncertainties for the case where $|a| \sim 1 $ are roughly independent of the value of $b$ and smaller than those for the quasi-axial case $ |a| \ll1 $. Find find, in addition, that, as expected, beam polarization allows for a different distinction of these couplings.

We also studied a sample model where the vector-like case ($b=0$) is realized; the model consists of an extension of the SM by a fermion isodoublet and a fermion singlet, both assumed odd under a $\zBB_2$ symmetry. In addition to providing a realization of the more general optimal observable analysis, this model contains a viable DM candidate. the presence of which can be probed at the ILC though $ \psi^\pm$ pair production followed by their decays into DM and $W$ bosons. The analysis shows that given the expected ILC luminosity (Eq.~\eqref{eq:params}) this collider will provide early evidence (or provide an early exclusion) of the model here proposed, and that this collider will be able to measure the model parameters with an accuracy very close to the optimal one. It is worth noting that the analysis is applicable to supersymmetric chargino pair production in the limit of heavy sneutrino, including our results on the optimal statistical uncertainties for the charged-lepton couplings to the $Z$ in the $a\neq0,b=0$ case.

\acknowledgments

SB acknowledges grant CRG/2019/004078 from SERB, Govt. of India and Mr. Abdur Rouf with whom the project was initiated. SJ would like to acknowledge Dr. Basabendu Barman and Dr. Sunando Patra for technical help.

\appendix

\section{Derivation of the optimal covariance matrix}
\label{sec:covmat}

Consider the observable as differential production cross section involving NP, which takes the form
\beq
\mathcal{O}=\frac{d\sigma_{\tt theo}}{d\phi} = \sum_i  c_if_i(\phi) \,,
\label{eq:obs}
\eeq
where $c_i$ denote functions of NP parameters and $f_i$ a corresponding set of functions of the phase-space coordinates $ \phi $. The goal is to provide observables with which to measure the $c_i$ optimally. 

The total cross section for the process is
\beq
\Nt =\vevof{\frac{d\sigma_{\tt theo}}{d\phi}}\,.
\label{eq:totcrss}
\eeq
where, to simplify notation, we defined the bracket $\vevof\cdots$ such that for any quantity $W$ depending on $ \phi$,
\bal
\vevof W = \int d\phi \, W(\phi)\,.
\label{eq:vev}
\end{align}

Next, if $ \lum$ is the integrated luminosity over prescribed period, the event rate $ \ert$ is given by
\beq
\ert = \lum \Nt\,.
\label{eq:totnum}
\eeq
With this definitions, the probability density function for observing $A$ events at phase-space points $ \phi_a\, (a=1,\ldots,A)$ is given by
\beq
\FFt(A;\Phi_A)   = \frac{e^{-\ert} }{A!} \prod_{a=1}^A \fft(\phi_a)\,, \quad \text{where} ~~\fft (\phi_a)= \lum \frac{d\sigma_{\tt theo}}{d\phi}(\phi_a)=\lum \mathcal{O}(\phi_a)\,;
\label{eq:fth} 
\eeq
and $ \Phi_A = (\phi_1,\ldots,\phi_A)$. Choose now a set of observables $ \op_i(\phi) $ and let
\beq
\Op_i(A;\Phi_A) = \sum_{a=1}^A \op_i(\phi_a)\,.
\label{eq:Opi}
\eeq

In the following we will need the expectation values of quantities using $\FFt$ as the probability density function. These averages involve an integration over phase-space variables and a summation over the number of events. To simplify notation we also define $\vvevof\cdots$ such that for any quantity $U$ that depends on $A$ and $ \Phi_A$,
\beq
\vvevof U = \sum_{A=0}^\infty \int d\Phi_A U(A;\Phi_A)= \sum_{A=0}^\infty \int  \prod_{c=1}^A d\phi_c U(A;\Phi_A)\,.
\label{eq:bigav}
\eeq

Now, the normalization of the probability density function ($\FFt$) is given by,

\bal
\vvevof{\FFt}=&\sum_{A=0}^\infty  \int d\Phi_A  \frac{e^{-\ert}}{A!} \prod_{a=1}^A \fft(\phi_a) \mcr
=&\sum_{A=0}^\infty \int  \frac{e^{-\ert}}{A!} \prod_{c=1}^A d\phi_c \prod_{a=1}^A \fft(\phi_a)  \mcr
=& \sum_{A=0}^\infty\int   \frac{e^{-\ert}}{A!} d\phi_1 d\phi_2 ... d\phi_A \fft(\phi_1) \fft(\phi_2) ... \fft(\phi_A)\mcr
=&\sum_{A=0}^\infty  \frac{e^{-\ert}}{A!} \int \fft(\phi_1)d\phi_1 \int \fft(\phi_2)d\phi_2... \int \fft(\phi_A)d\phi_A \mcr
=&\sum_{A=0}^\infty  \frac{e^{-\ert}}{A!} \ert^A \qquad \mcr
=&e^{-\ert}\underbrace{\sum_{A=0}^\infty  \frac{\ert^A}{A!}}_{e^{\ert}} \mcr
=&1.
\end{align}

With these preliminaries  the average of $ \Op_i$ is
\beq
\vvevof{\FFt \Op_i} = \sum_{A=1}^\infty \frac{e^{-\ert} }{A!} \ert^{A-1} A \vevof{\fft \op_i} =  \vevof{\fft \op_i} =\lum \vevof{\op_i f_j} c_j\,.
\label{eq:ftho} 
\eeq
Let then
\beq
\Gamma_i = \left(M^{-1}\right)_{ij} \Op_j\,, \quad \text{where} \quad M_{ij} = \lum \vevof{\op_i f_j}\,,
\label{eq:gammai} 
\eeq
in which case
\beq
c_i = \vvevof{\FFt \Gamma_i}\footnote{$\Gamma_i$ are the sets of function that follow the distribution $\FFt$, so that $c_i$ are the expectation values of $\Gamma_i$.}\,.
\label{eq:ci} 
\eeq
Therefore, any $ \Gamma_i $ that has the above form can be used to determine the $c_i$. The idea now is to choose the $ \Gamma_i$ that has the smallest covariance matrix.

The covariance matrix is given by
\bal
V_{ij} &= \vvevof{\FFt \Gamma_i' \Gamma_j'} =\vvevof{\FFt \Gamma_i \Gamma_j}-c_i c_j \, ; \quad \text{where}~~\Gamma_i' = \Gamma_i - c_i;
\label{eq:vij}
\end{align}

which we extremize as a function of the $\op_i $. To this end we vary $ a_i a_j V_{ij}$ (where the $ a_i$ are for arbitrary constants) as a function of the $ \op_i$:
\beq
\delta \left(\half a_i a_j V_{ij} \right) = \vvevof{\FFt (b\Op)\left(b_i \, \delta\!\Op_i - b_i \, \delta M_{ij} \, \Gamma_j\right)} =0 \,,
\eeq
where $ b_i = a_j (M^{-1})_{ji}$. Now one can show, 
\bal
\vvevof{\FFt \Op_k \, \delta\Op_i} &= \int d\Phi \sum_{A=1}^\infty \frac{e^{-\ert}}{A!} \prod_{a=1}^A \fft(\phi_a) \sum_{b=1}^A \op_k(\phi_b) \sum_{c=1}^A \delta\op_i(\phi_c)  \mcr
&= \vevof{\fft \op_k \delta\!\op_i}  + \vevof{ \fft \op_k} \vevof{\fft \delta\!\op_i}\,.
\end{align}

Similarly,

\bal
\vvevof{\FFt \Op_k \Gamma_j} &= M^{-1}_{jl} \vevof{\fft \op_k \op_l} + \vevof{\fft \op_k} c_j.
\label{eq:fthodo} 
\end{align}

where we used $ M^{-1}_{jl} \vevof{\fft \op_l} = c_j $. The extremum condition then reads

\bal
b_i b_k \bigg\{  \fft \op_k   + \vevof{ \fft \op_k} \fft   - \lum  f_j \left[ M^{-1}_{jl} \vevof{\fft \op_k \op_l} + \vevof{\fft \op_k} c_j \right] \bigg\} =0 .  
\end{align}	

Then, since $b_ib_k\neq0$ as $a$ is arbitrary and $ \lum \sum_j f_j c_j = \fft$, the solution is
\beq
\fft \op_k    = \lum  f_j  M^{-1}_{jl} \vevof{\fft \op_k \op_l}  \then \op_k = \frac{f_k}\fft  \,,
\eeq
for which the condition $ \vevof{f_k} = M_{kl} c_l  $ is satisfied and the optimal covariance matrix is then
\beq
V = \inv\lum M^{-1}\,; \qquad M_{ij} = \lum \vevof{ \frac{f_i f_j}\fft} = \vevof{ \frac{f_i f_j}{\mathcal{O}}}\,. 
\eeq

\section{Optimal Analysis with other CM energies (unpolarized beams)}
\label{sec:1tev}
\begin{figure}[htb!]
	\begin{align}
		\includegraphics[scale=0.17]{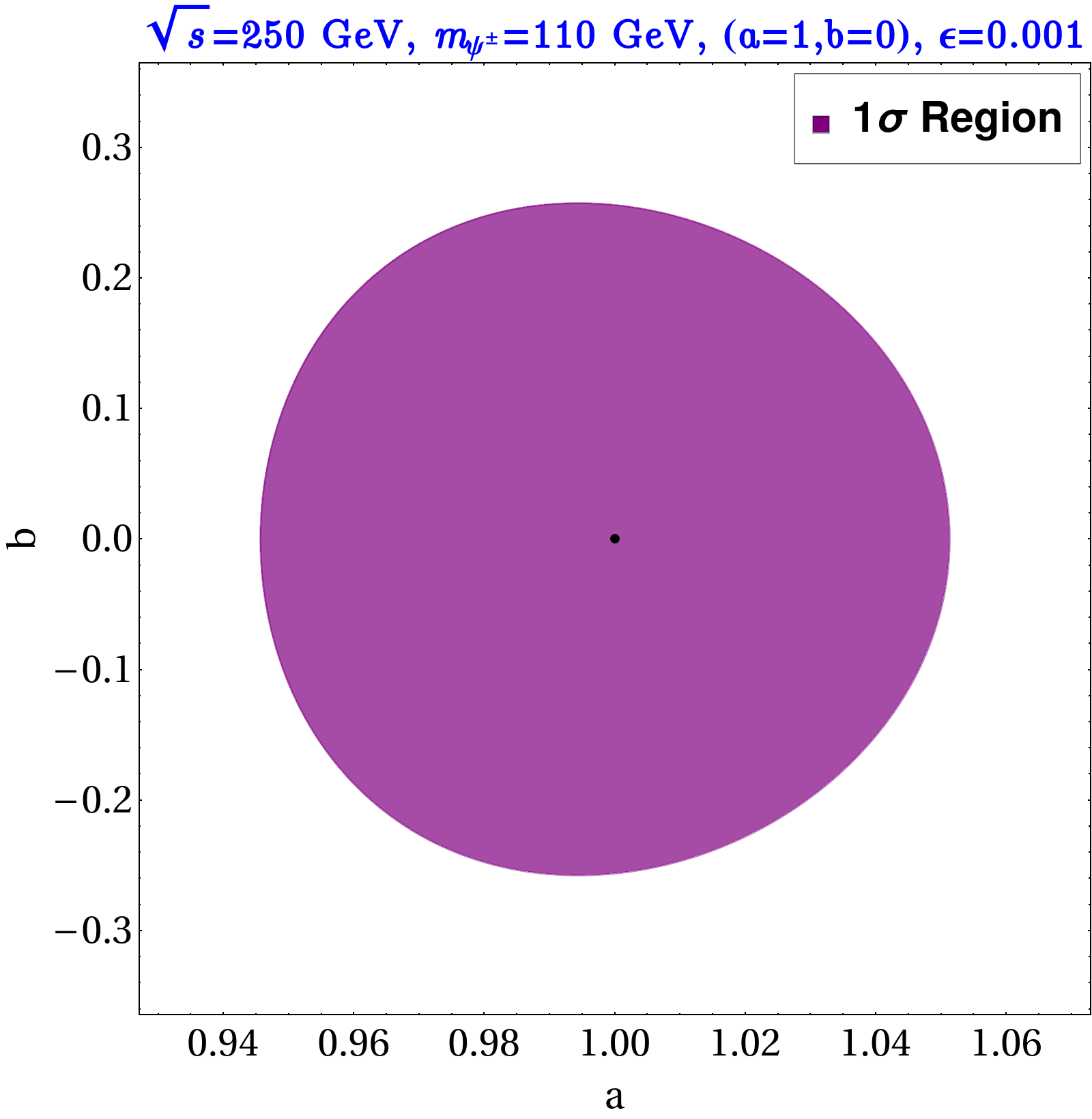} & \quad \includegraphics[scale=0.17]{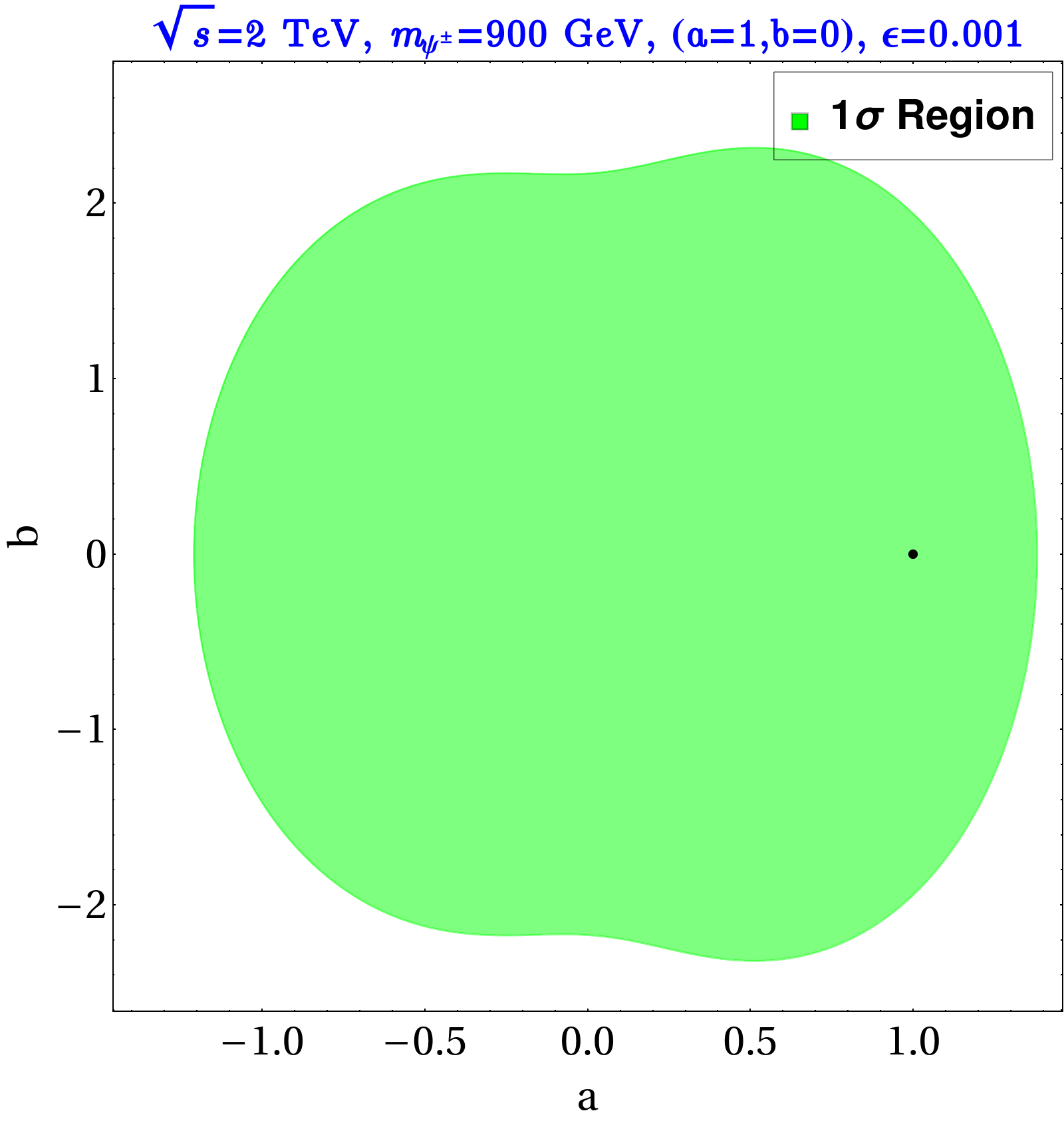} \cr
		\includegraphics[scale=0.17]{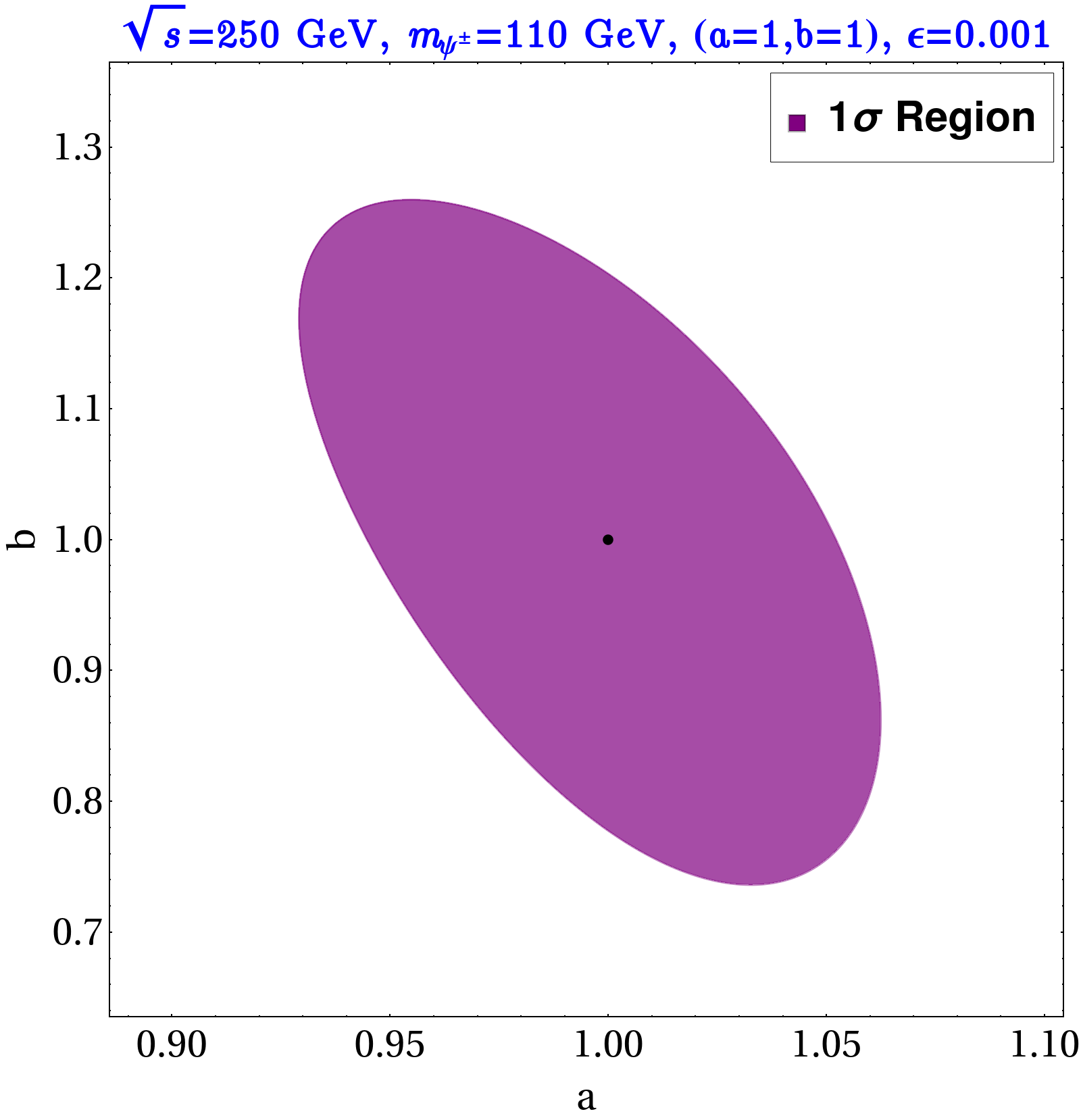} & \quad \includegraphics[scale=0.17]{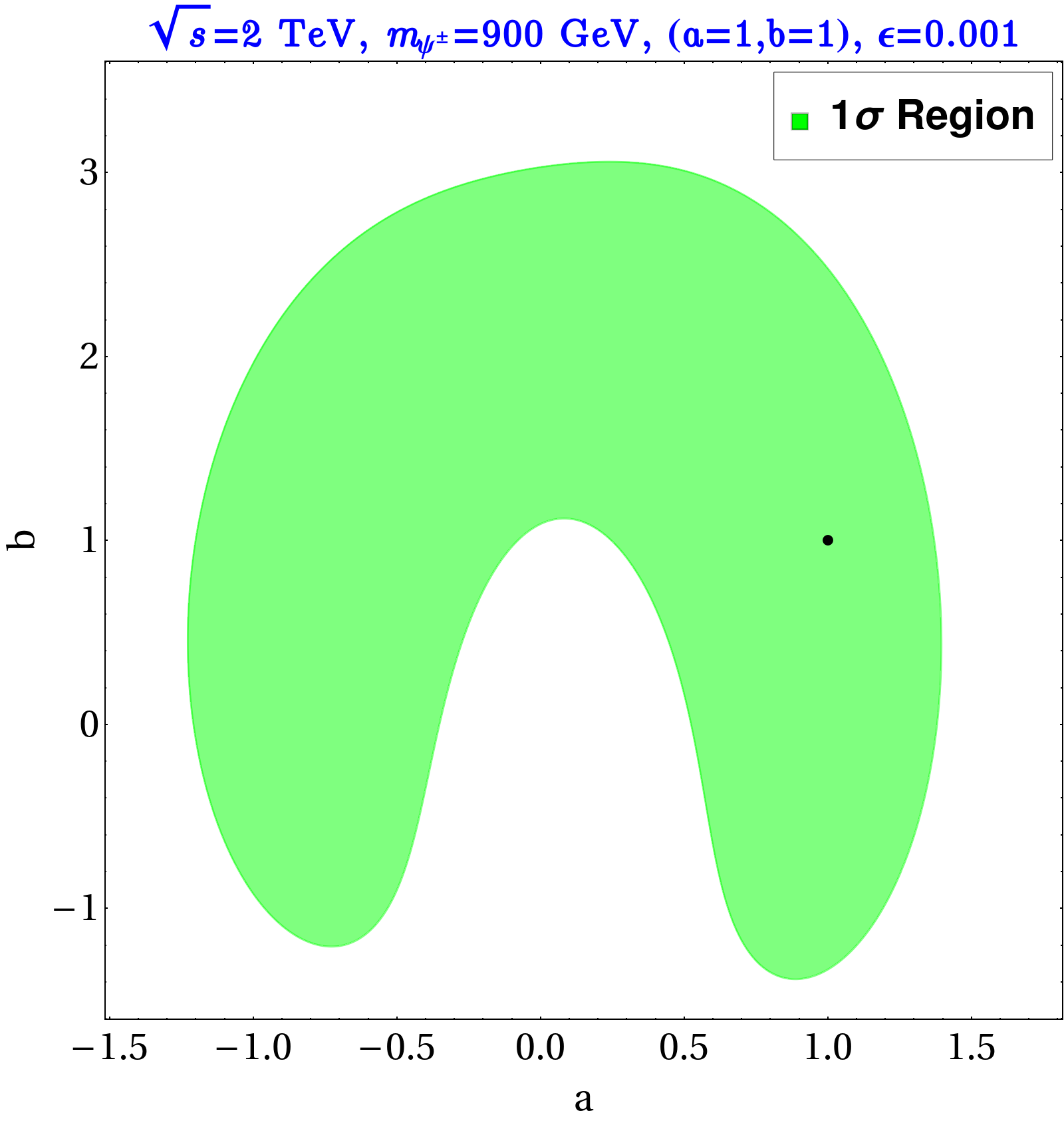} \cr
		\includegraphics[scale=0.17]{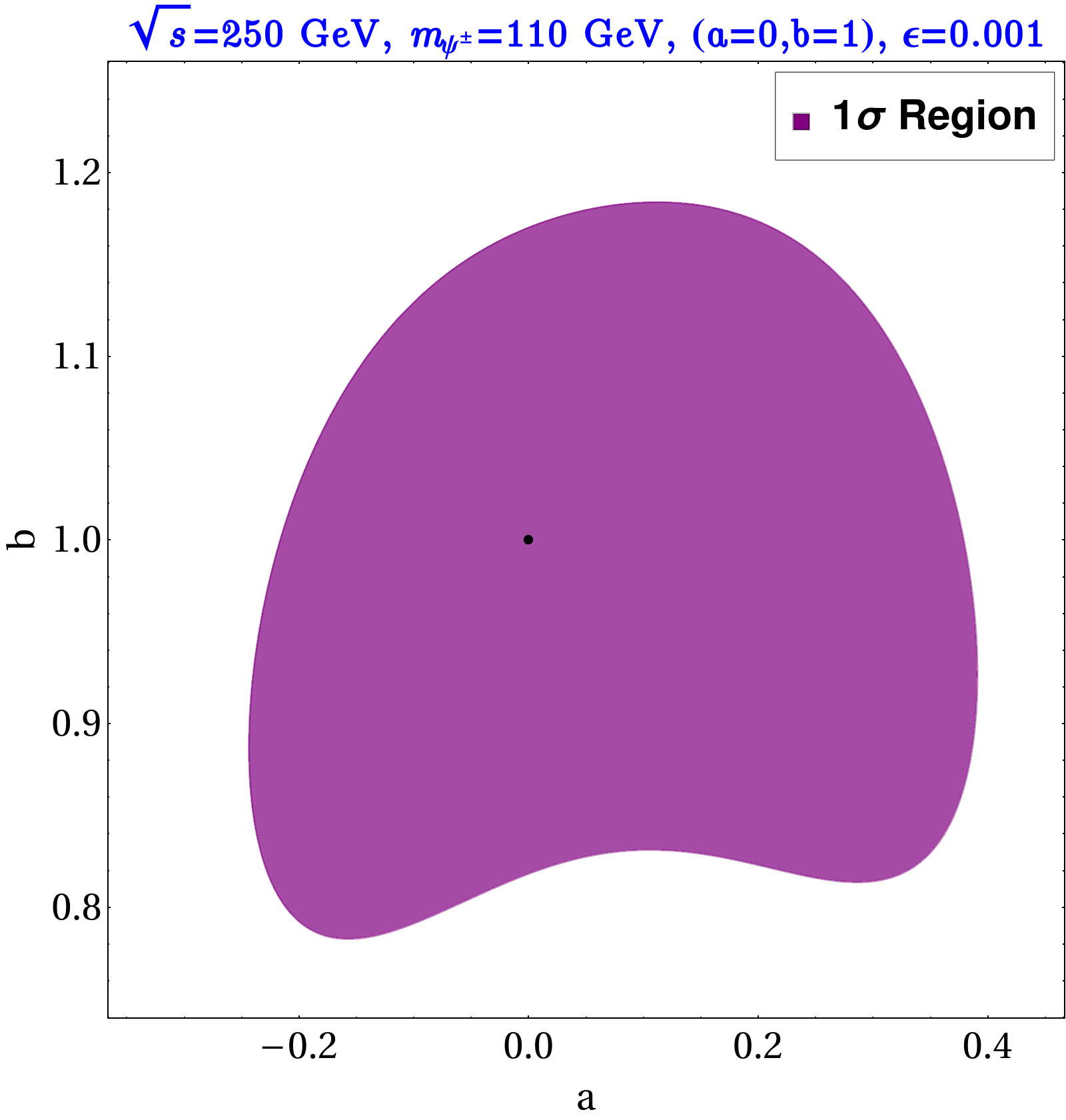} & \quad \includegraphics[scale=0.17]{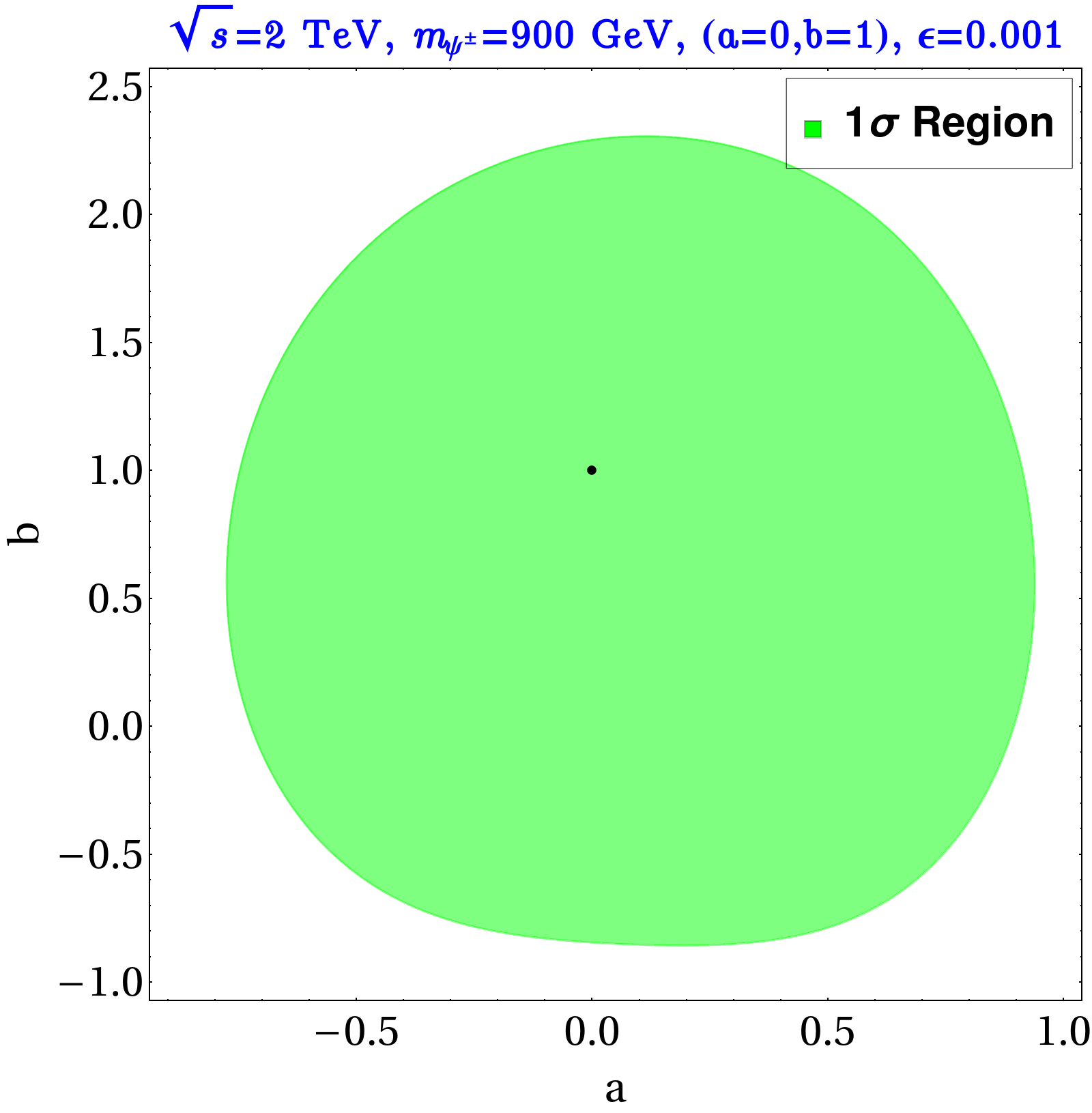} \cr
		\includegraphics[scale=0.17]{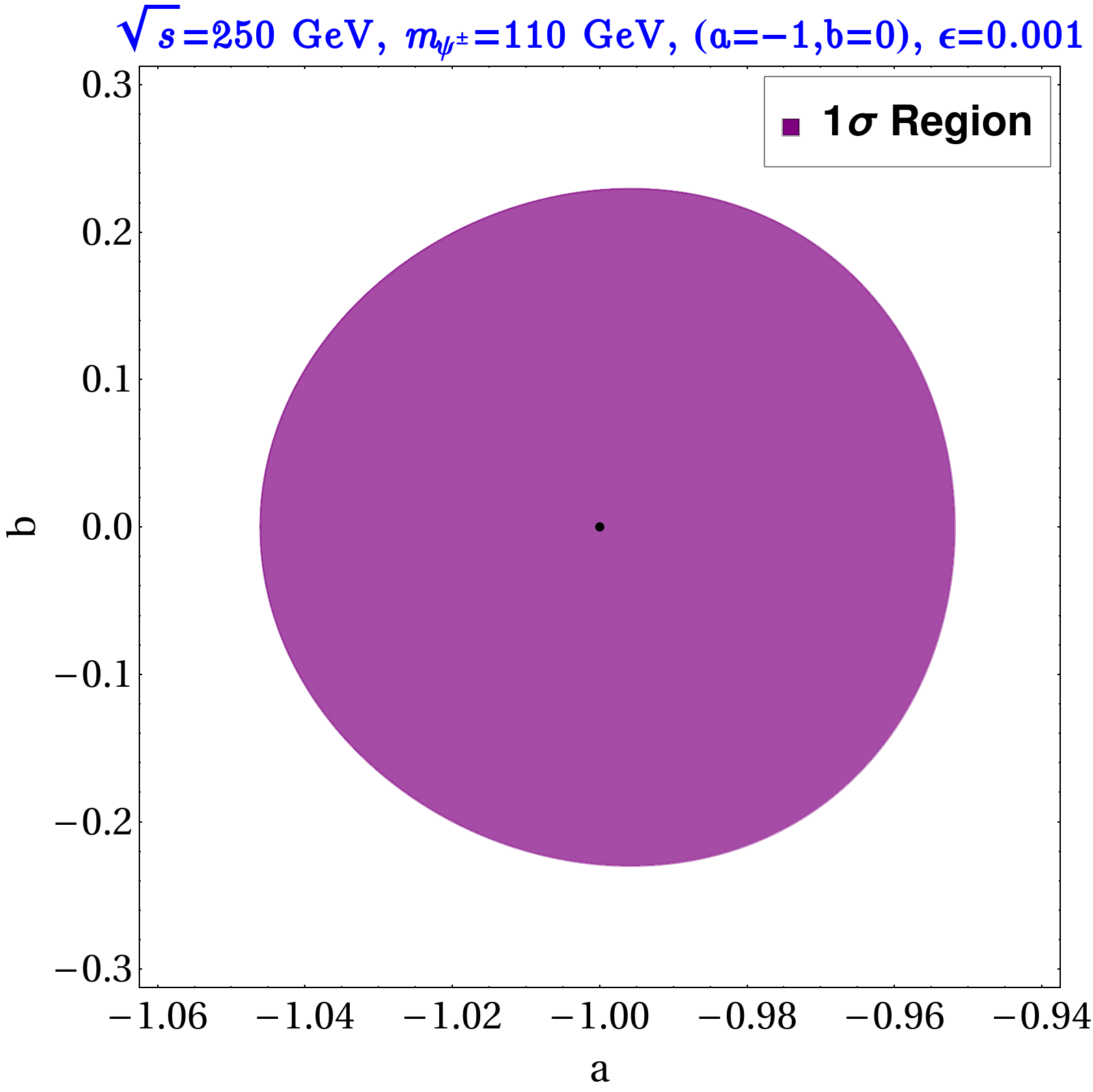}& \quad \includegraphics[scale=0.17]{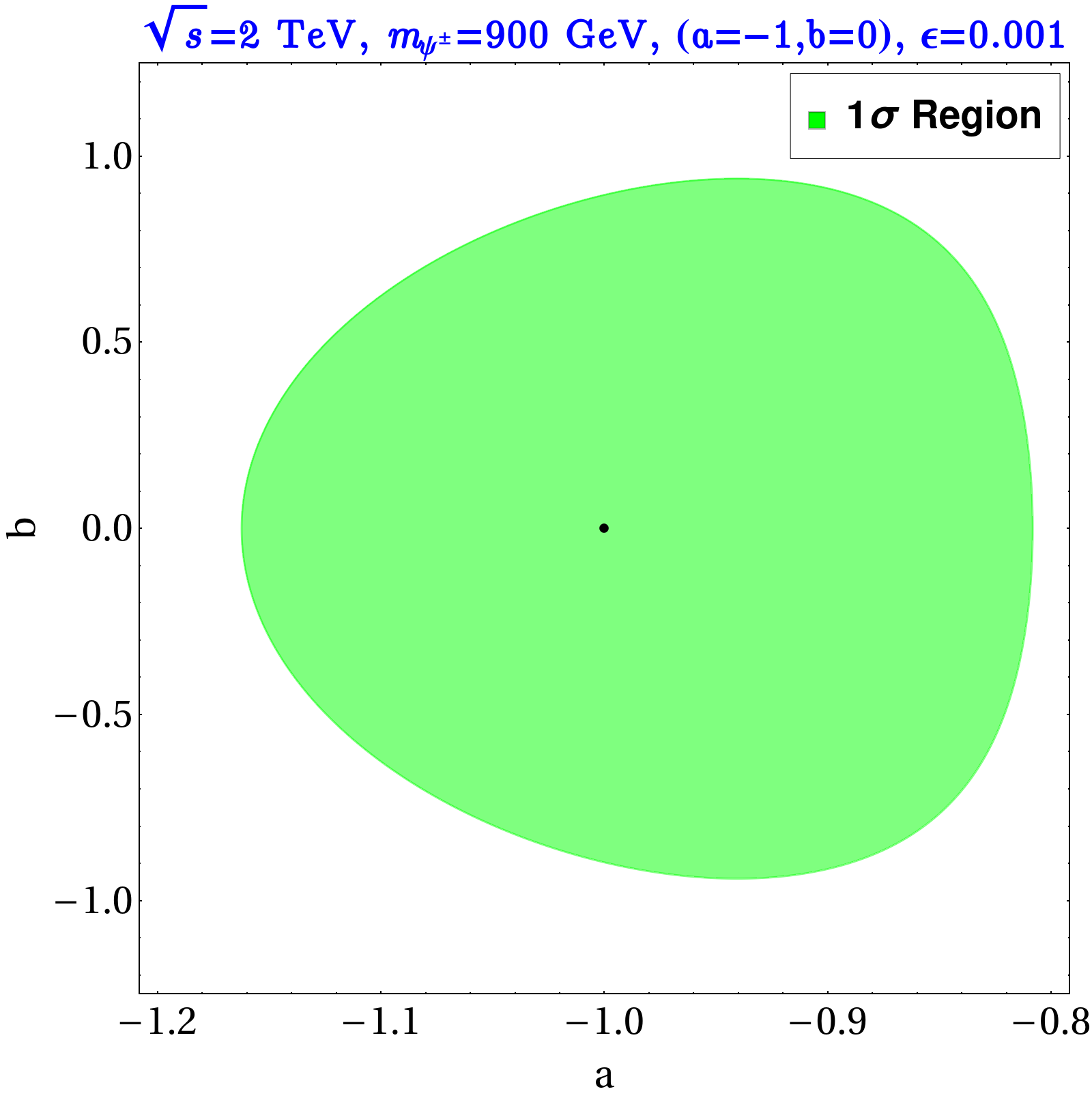}\cr
		\includegraphics[scale=0.17]{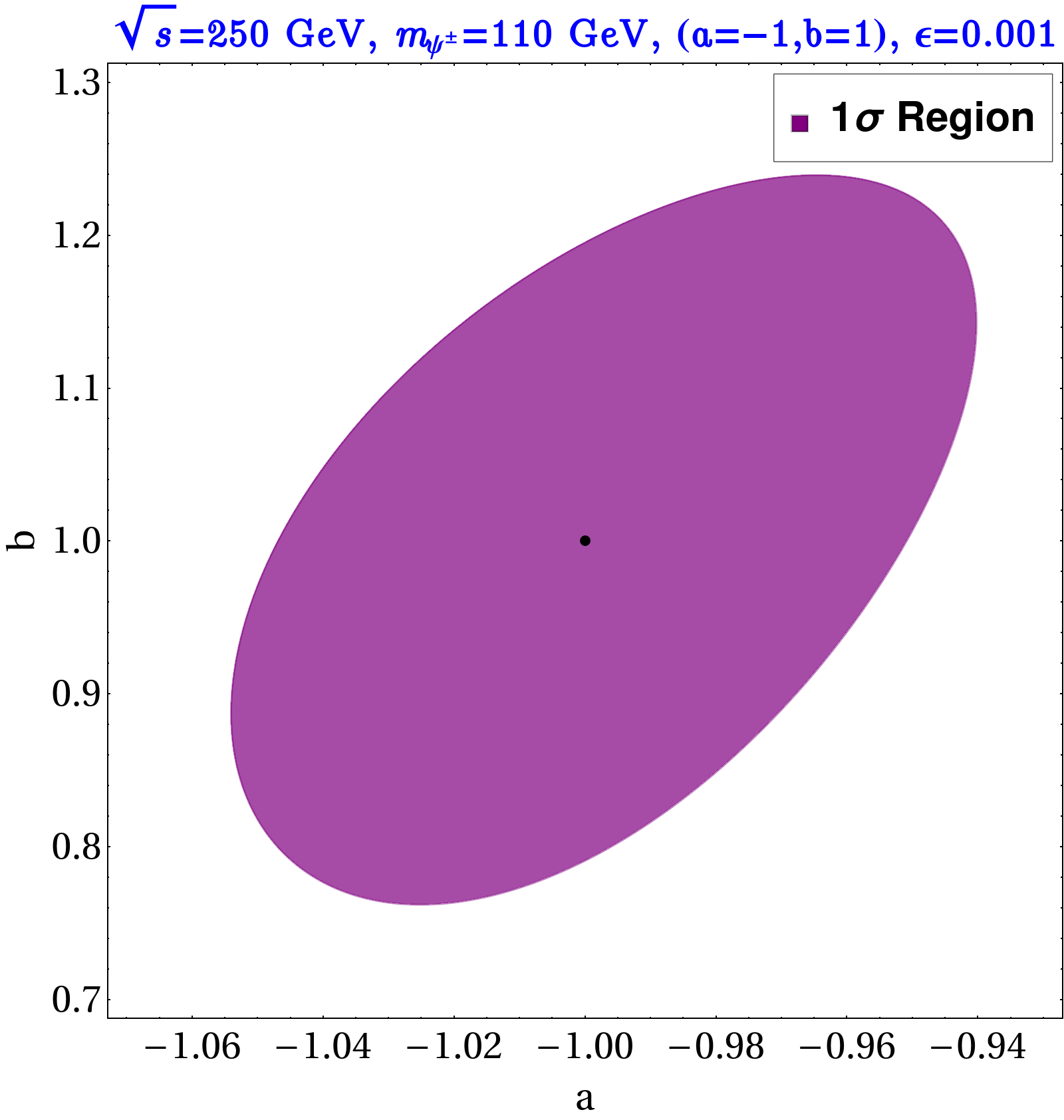}& \quad \includegraphics[scale=0.17]{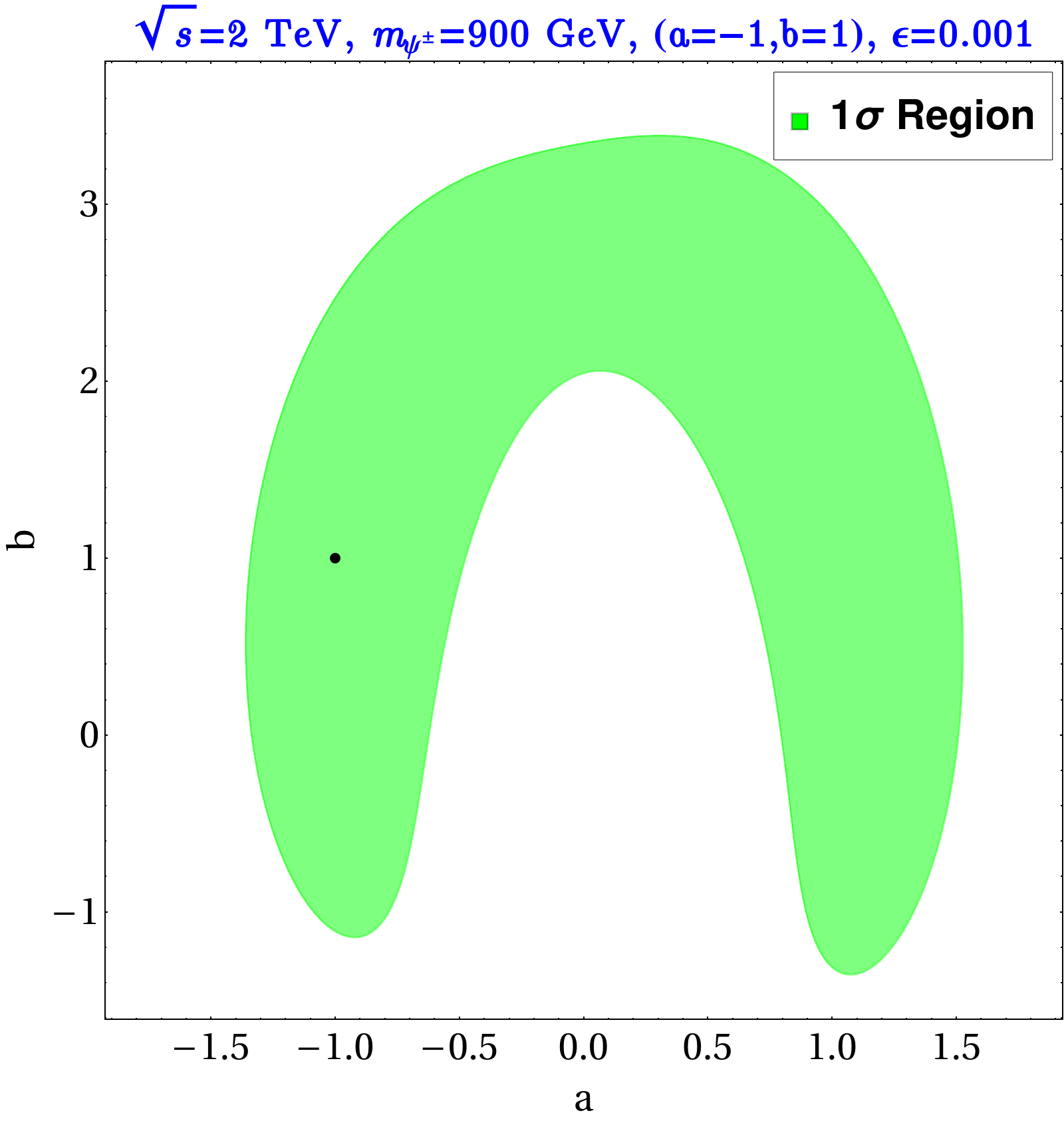}
		\nonumber
	\end{align}
	\caption{ $\chi^2\le1$ regions for unpolarized beams. Left column: $\sqrt{s}=250\,\gev,~m_{\psi^\pm} =  110$ GeV; right column: $\sqrt{s}=2\,\tev,~m_{\psi^\pm}=900$ GeV.}
	\label{fig:1sigma124-900gev}
\end{figure}

In this appendix we repeat the analysis of Sect. \ref{sec:analysis} for both lower and higher CM energies and different charged-fermion masses. The results are presented in Fig.~\ref{fig:1sigma124-900gev}:  OOT $1$-$\sigma$ regions for $\sqrt{s}=250\,\gev,~m_{\psi^\pm} =  110$ GeV (left column), and $\sqrt{s}=2\,\tev,~m_{\psi^\pm}=900$ GeV (right column). Note that as $m_{\psi^\pm}$ and $s$ increase the cross-section drops (Drell-Yann process falls like $\sim 1/\sqrt{s}$, see Eq.~\eqref{fi})
and the eigenvalues of the covariance matrix increase (since $ V^{-1} \sim M \sim 1/\sigma$, {\it cf.} Eqs.~\eqref{eq:covmat2}, \eqref{eq:mij}, and \eqref{chi2}); the $ 1$-$ \sigma$ regions are corresponding larger. The two cases where $a, b \neq 0$, the 1$\sigma$ regions are very asymmetric because the probability distribution of NP couplings are heavily distorted from the normal distribution.

\section{$\chi^2$ ellipsoids in $c_i$ plane}
\label{sec:ellipses}

\begin{figure}[htb]
$$
\includegraphics[scale=0.28]{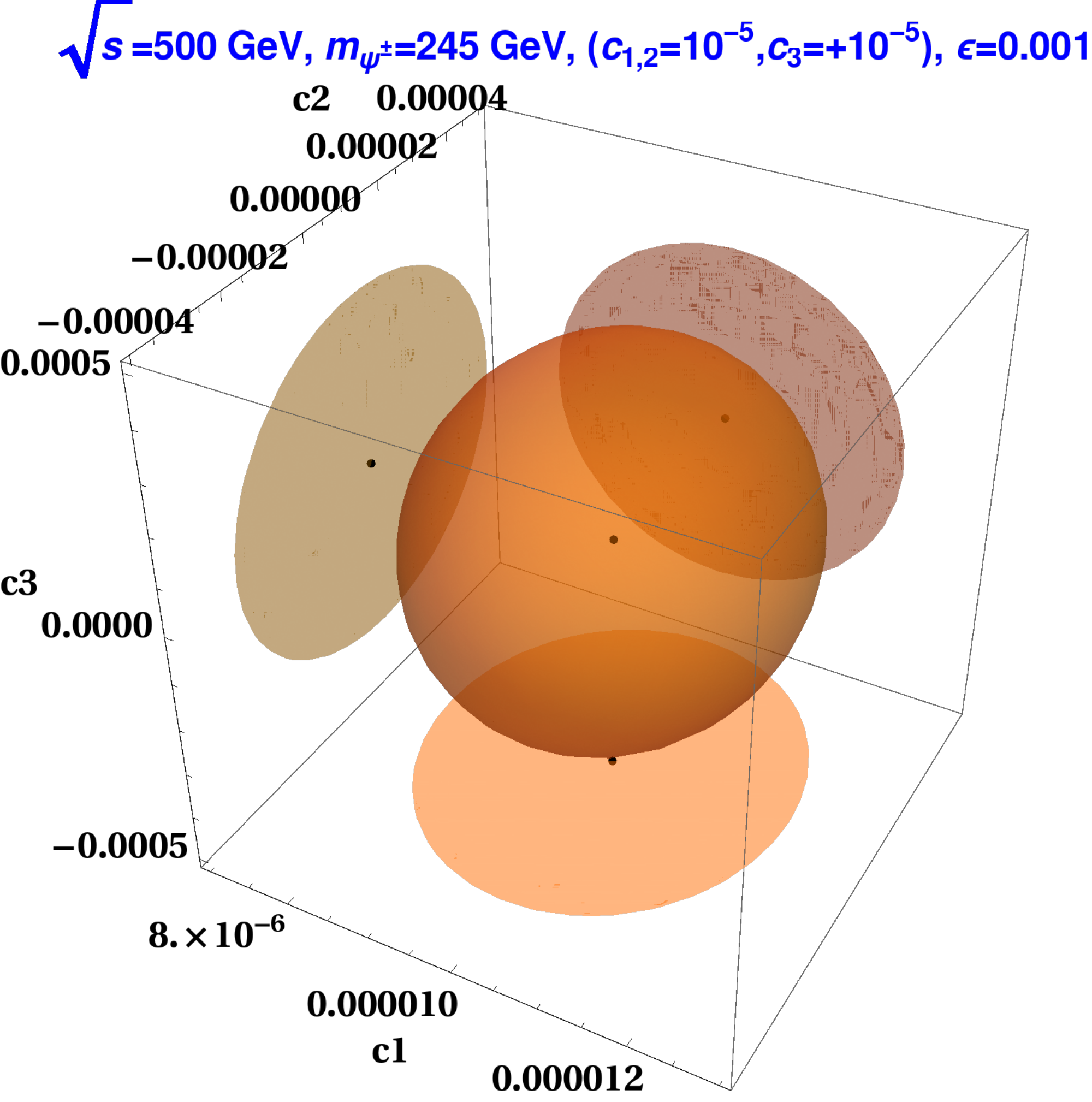} \qquad
\includegraphics[scale=0.28]{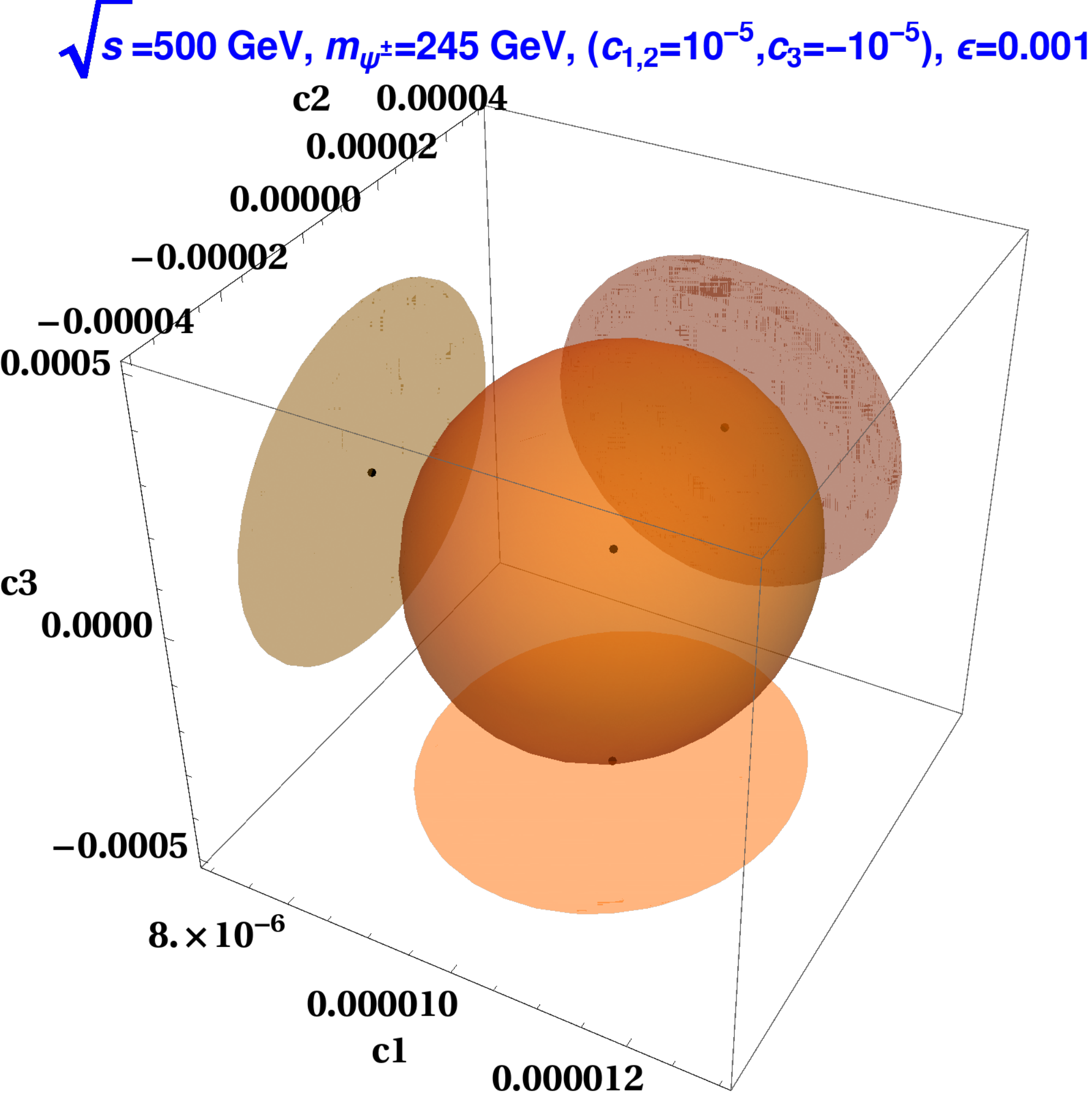}
$$
\caption{$\chi^2=1$ ellipsoids in $c_i$ space for $ c^0_{1,2} = 10^{-5},\, c^0_3 = \pm 10^{-5}$ and unpolarized beams. The 2-dimension projections are also drawn.}
\label{fig:ci3Dunpol}
\end{figure}

If the $c_i $ are taken as the NP parameters, then the $ \chi^2 = 1 $ regions become ellipsoids in this space, as illustrated in Fig.~\ref{fig:ci3Dunpol} for two examples. In general $V_0 $ in Eq.~(\ref{chi2}) is not diagonal, so that the $c_i$'s  are correlated, however, one can always choose alternative coefficients, linear combinations of the $c_i $, which are uncorrelated and whose ($1\sigma$) statistical uncertainties equal the square-root of the eigenvalues of $ V_0 $. That is, if $ c_i = R_{ij} \tilde c_i $, one can always find $R$ orthogonal which diagonalizes $V_0 $, whence Eq.~(\ref{chi2}) becomes $ \chi^2 = \epsilon \sum (\delta \tilde c_i)^2/\bar V^i_0  $, where the eigenvalues of $V_0$ are denoted by $\bar V_0^i$ and $\delta\tilde c_i $ is the statistical uncertainty of the $ \bar c_i $. An example is given in Table \ref{table:dciunpol} and corresponding 1$\sigma$ ellipsoids are shown in Fig.~\ref{fig:ci3Dunpol}.

\begin{center}
	\begin{table}[htb]
		\begin{tabular}{| c | c | c | c | c | c | c | c | c |} 
			\hline
			&
			\multicolumn{3}{|c|}{Uncertainties $\times 10^{-6}$ ($\epsilon=0.001$)} &
			\multicolumn{3}{|c|}{Uncertainties $\times 10^{-6}$ ($\epsilon=0.005$)}\\
			\hline
			Seed parameters $\times 10^{-5}$ & $ \quad|\delta\tilde c_1| \quad$ & $\quad |\delta\tilde c_2| \quad$ & $\quad |\delta\tilde c_3| \quad$ & $\quad |\delta\tilde c_1| \quad$ & $\quad |\delta\tilde c_2| \quad$ & $\quad  |\delta\tilde c_3| \quad$ \\
			\hline
			$c^0_1 = c^0_2=1;\,c_3^0= 1$ & $1.07$ & $18.40$ & $179.06$ & $0.48$ & $8.23$ & $80.07$ \\ \hline
			$c^0_1 = c_2^0=1;\,c_3^0=-1$ & $1.06$ & $18.18$ & $177.16$ & $0.47$ & $8.13$ & $79.23$ \\
			\hline
		\end{tabular}
		\caption{$1\sigma$ statistical uncertainties for the uncorrelated parameters $\tilde c_i $ (see text) for the indicated seed parameters and efficiencies; we assume unpolarized beams.}
		\label{table:dciunpol}
	\end{table}
\end{center}

\begin{figure}[htb]
		\includegraphics[scale=0.32]{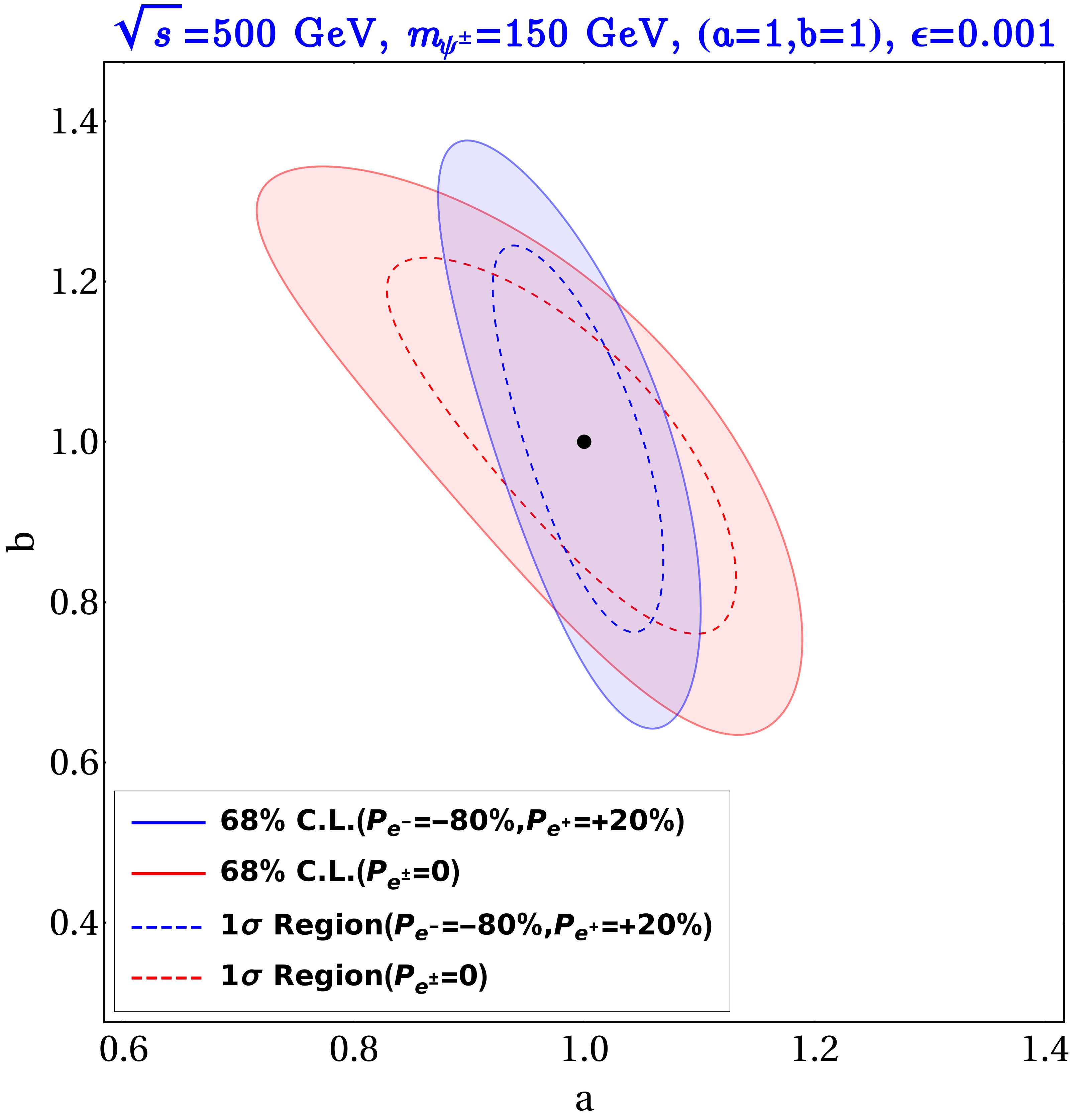}  \includegraphics[scale=0.34]{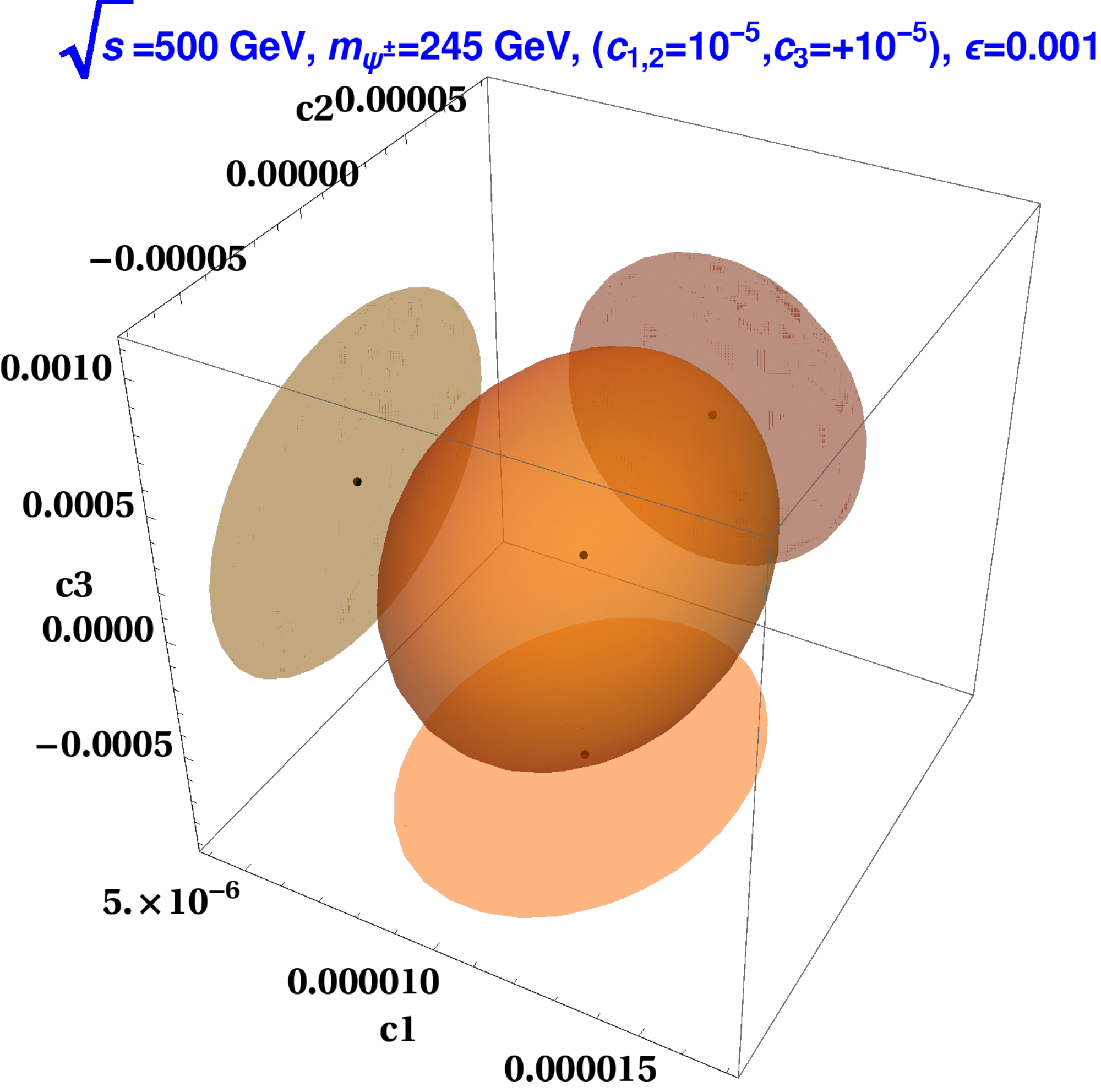} 
		\nonumber
	\caption{ 68\% C.L. in two-parameter and three-parameter distributions; $\chi^2\le2.3$ (left) for $a-b$ plane for ($a=1, b=1$) case (with dotted lines 
		denoting 1$\sigma$ surfaces) and $\chi^2\le3.5$ (right) for $c_i=10^{-5} (i=1,2,3)$.}
	\label{fig:68cl}
\end{figure}

\section{68\% C.L. in two-parameter and three-parameter distributions}
\label{sec:68cl}

As noted below Eq. \eqref{chi2} in Sect. \ref{sec:method}, when the model has $n$  normally-distributed parameters, the C.L. for $ \chi^2 \le \ell$ equals $\left(1-\Gamma(n/2,\ell/2)/\Gamma(n/2) \right)$. For example, if $n=2\, (3) $, a 68\% C.L. corresponds to $ \ell = 2.3\, (3.5)$. The 68\% C.L. contours are plotted in Fig.~\ref{fig:68cl} when $a_0=b_0=1 $ (left panel) and for $ c_i^0 = 10^{-5} $ (right panel); a comparison with the $ \chi^2\le1$ region is also provided. It is quite evident that 68\% C.L. regions are larger than 1$\sigma$ surfaces.

\bibliographystyle{JHEP}
\bibliography{ref}

\end{document}

%% file: mylatex.tex
\usepackage{bbm}  
\usepackage{amsmath}                                                

\newcommand{\nc}{\newcommand}
\nc{\beq}{\begin{equation}}  \nc{\eeq}{\end{equation}}
\nc{\bea}{\begin{eqnarray}}  \nc{\eea}{\end{eqnarray}}
\nc{\baa}{\begin{array}}     \nc{\eaa}{\end{array}}
\nc{\bit}{\begin{itemize}}   \nc{\eit}{\end{itemize}}
\nc{\ben}{\begin{enumerate}} \nc{\een}{\end{enumerate}}
\nc{\bce}{\begin{center}}    \nc{\ece}{\end{center}}
\nc{\bpm}{\begin{pmatrix}}   \nc{\epm}{\end{pmatrix}}
\nc{\bvt}{\begin{verbatim}}  \nc{\evt}{\end{verbatim}}
\nc{\bal}{\begin{align}}
\def\mcr{\nonumber\\[6pt]}

\def\half{\frac12}	

\def\to{\rightarrow}

\def\boldoverdot{\,{\raise6pt\hbox{\bf.}\!\!\!\!\>}}

\def\then{{\quad\Rightarrow\quad}}

\def\lcal{{\cal L}}
\def\mcal{{\cal M}}

\def\pp{{\bf p}}

\def\WW{{\bf W}}

\def\zBB{{\mathbbm Z}}
\def\mati{{\mathbbm1}}

\usepackage{bm}
\def\sigbf{{\bm\sigma}}		

\def\vev{vacuum expectation value}

\def\diag{\hbox{\diag}}

\def\gev{\hbox{GeV}}
\def\tev{\hbox{TeV}}

\def\vevof#1{\left\langle #1 \right\rangle}

\def\doubleundertext#1{
{\undertext{\vphantom{y}#1}}\par\nobreak\vskip-\the\baselineskip\vskip4pt%
\undertext{\hbox to 2in{}}}
\def\inbox#1{\vbox{\hrule\hbox{\vrule\kern5pt
     \vbox{\kern5pt#1\kern5pt}\kern5pt\vrule}\hrule}}
\def\sqr#1#2{{\vcenter{\hrule height.#2pt
      \hbox{\vrule width.#2pt height#1pt \kern#1pt
         \vrule width.#2pt}
      \hrule height.#2pt}}}

\def\today{\ifcase\month\or
  January\or February\or March\or April\or May\or June\or
  July\or August\or September\or October\or November\or December\fi
  \space\number\day, \number\year}
\def\pmb#1{\setbox0=\hbox{#1}%
  \kern-.025em\copy0\kern-\wd0
  \kern.05em\copy0\kern-\wd0
  \kern-.025em\raise.0433em\box0 }

\def\pmbb#1{\setbox0=\hbox{#1}%
  \kern-.02em\copy0\kern-\wd0
  \kern.04em\copy0\kern-\wd0
  \kern-.02em\raise.03464em\box0 }
\def\up#1{^{\left( #1 \right) }}
\def\inv#1{\frac1{#1}}
\def\su#1{{SU(#1)}}
\def\ui{U(1)}
\def\sumprime_#1{\setbox0=\hbox{$\scriptstyle{#1}$}
  \setbox2=\hbox{$\displaystyle{\sum}$}
  \setbox4=\hbox{${}'\mathsurround=0pt$}
  \dimen0=.5\wd0 \advance\dimen0 by-.5\wd2
  \ifdim\dimen0>0pt
  \ifdim\dimen0>\wd4 \kern\wd4 \else\kern\dimen0\fi\fi
\mathop{{\sum}'}_{\kern-\wd4 #1}}